\documentclass[11pt, A4]{article}

\usepackage{graphicx}
\usepackage{subfigure,caption}
\usepackage{amssymb}
\usepackage{amsmath}
\usepackage{authblk}
\usepackage{anysize}
\usepackage{xcolor}
\usepackage{hyperref}
\usepackage[ruled,vlined,linesnumbered,resetcount]{algorithm2e}
\usepackage{multirow}

\makeatletter
\providecommand\phantomcaption{\caption@refstepcounter\@captype}
\makeatother

\marginsize{2cm}{2cm}{2cm}{2cm}
\graphicspath{{figs/}}

\begin{document}
\title{Electrical activity of fungi: Spikes detection and complexity analysis}

\author[1]{Mohammad Mahdi Dehshibi}
\author[2]{Andrew Adamatzky}

\affil[1]{Department of Computer Science, Universitat Oberta de Catalunya, Barcelona, Spain}
\affil[2]{Unconventional Computing Laboratory, University of the West England, Bristol, UK}

\maketitle

\begin{abstract}

\noindent 
Oyster fungi \emph{Pleurotus djamor} generate actin potential like spikes of electrical potential. The trains of spikes might manifest propagation of growing mycelium in a substrate, transportation of nutrients and metabolites and communication processes in the mycelium network. The spiking activity of the mycelium networks is highly variable compared to neural activity and therefore can not be analysed by standard tools from neuroscience. We propose original techniques for detecting and classifying the spiking activity of fungi. Using these techniques, we analyse the information-theoretic complexity of the fungal electrical activity. The results can pave ways for future research on sensorial fusion and decision making of fungi.\\

\end{abstract}

\section{Introduction} \label{sec:1}

Extracellular (EC) recordings of action potentials have been widely used to record and measure neural activity in a number of species. The broad functionality of this method has been shown for studying neural activity in several applications, ranging from single nerve fibres in invertebrate sensory organs to cortical neurons involved in cognition, learning and memory \cite{fyhn2004spatial,quiroga2009explicit,trainito2019extracellular}. 

Excitation is an essential property of all living creatures, from bacteria~\cite{masi2015electrical},  Protists~\cite{eckert1979ionic,hansma1979sodium,bingley1966membrane}, fungi~\cite{mcgillviray1987transhyphal} and plants~\cite{trebacz2006electrical,fromm2007electrical,zimmermann2013electrical} to vertebrates~\cite{hodgkin1952quantitative,aidley1998physiology,nelson2012excitable,davidenko1992stationary}. Waves of excitation could be also found in various physical~\cite{kittel1958excitation,tsoi1998excitation,slonczewski1999excitation,gorbunov1987excitation}, chemical~\cite{belousov1959periodic, zhabotinsky1964periodic,zhabotinsky2007belousov} and 
social systems~\cite{farkas2002social,farkas2003human}. When recorded with differential electrodes, a propagating excitation wave is manifested by spike.

In our recent studies \cite{adamatzky2018towards,adamatzky2019plant}, we demonstrated that the oyster fungi \emph{Pleurotus djamor} generate action potential like impulses of electrical potential. We observed trains of the spontaneous spikes\footnote{Calling the spikes spontaneous means that they are not invoked by an intentional external stimulation. Otherwise, the spikes indeed reflect physiological and morphological processes ongoing in mycelial networks.} with two types of activity, \textit{i.e.}, high-frequency (period 2.6~min) and low-frequency (period 14~min). Appropriate utilisation of this information is, however, subject to the accurate extraction of the EC spike waveform, separating it from the background activity of neighbouring cells, and sorting the characteristics.

\begin{figure}[!htb]
    \centering
    \subfigure[]{\includegraphics[width=0.7\textwidth]{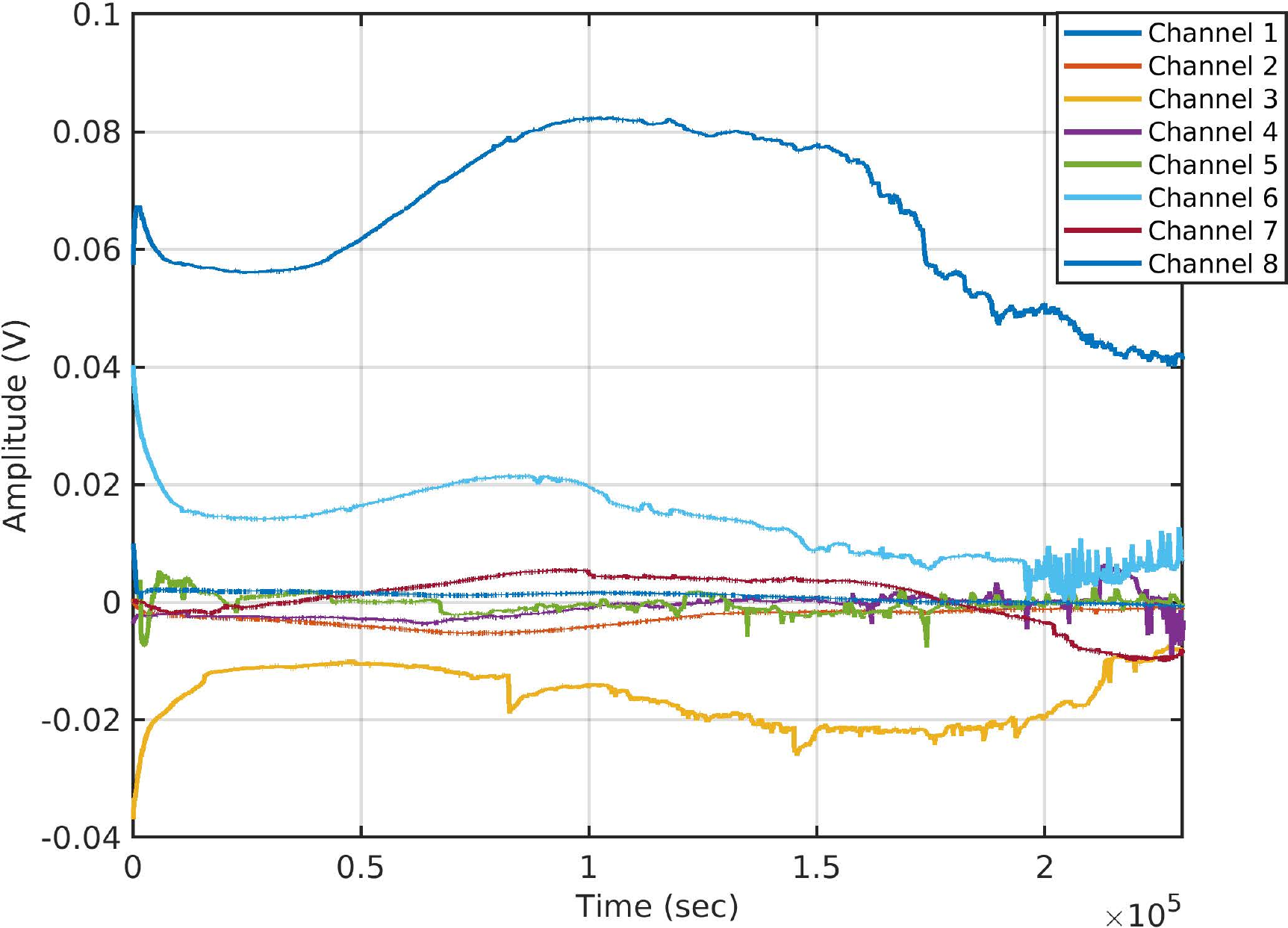}}
    \subfigure[]{\includegraphics[width=0.48\textwidth]{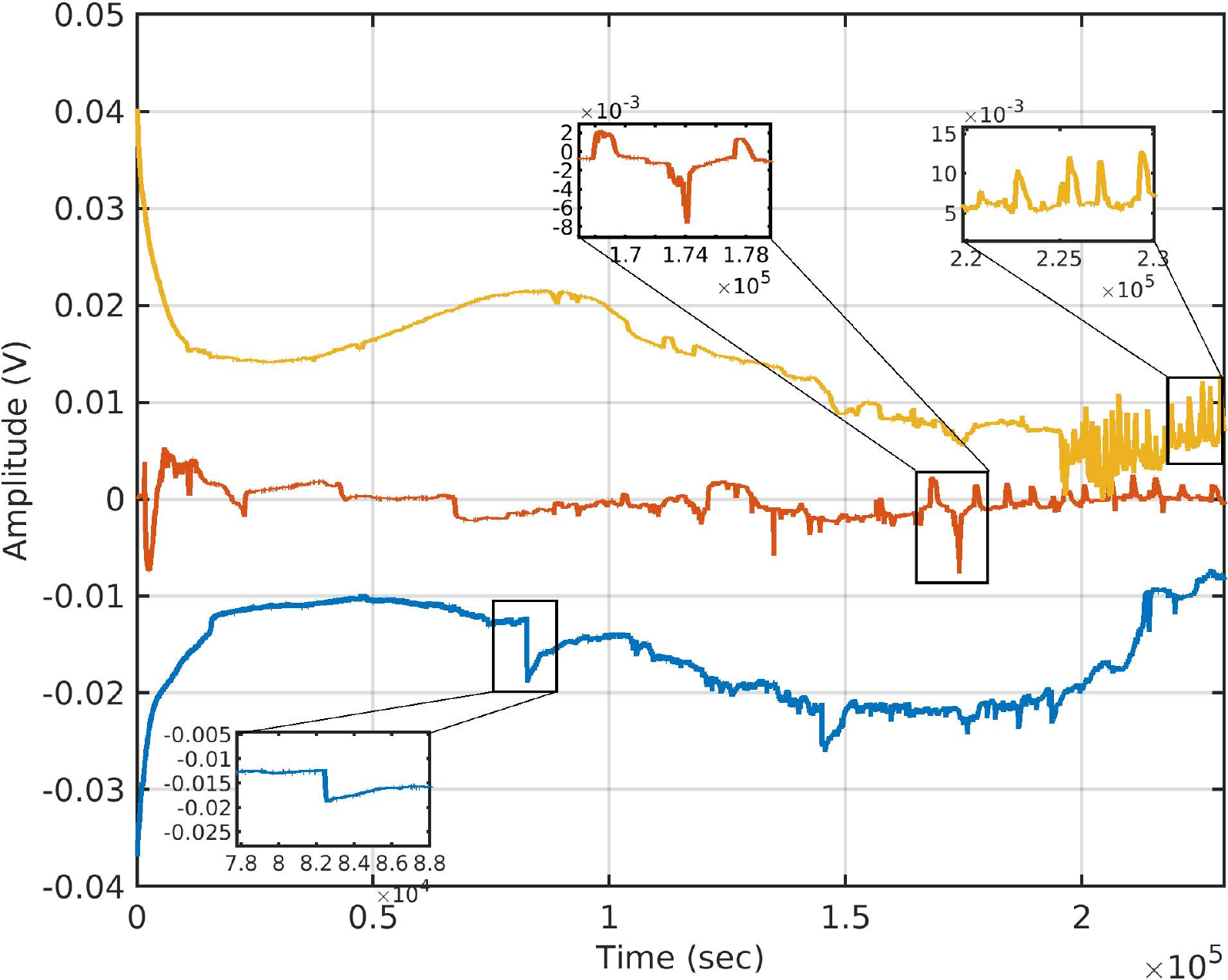}}
    \subfigure[]{\includegraphics[width=0.48\textwidth]{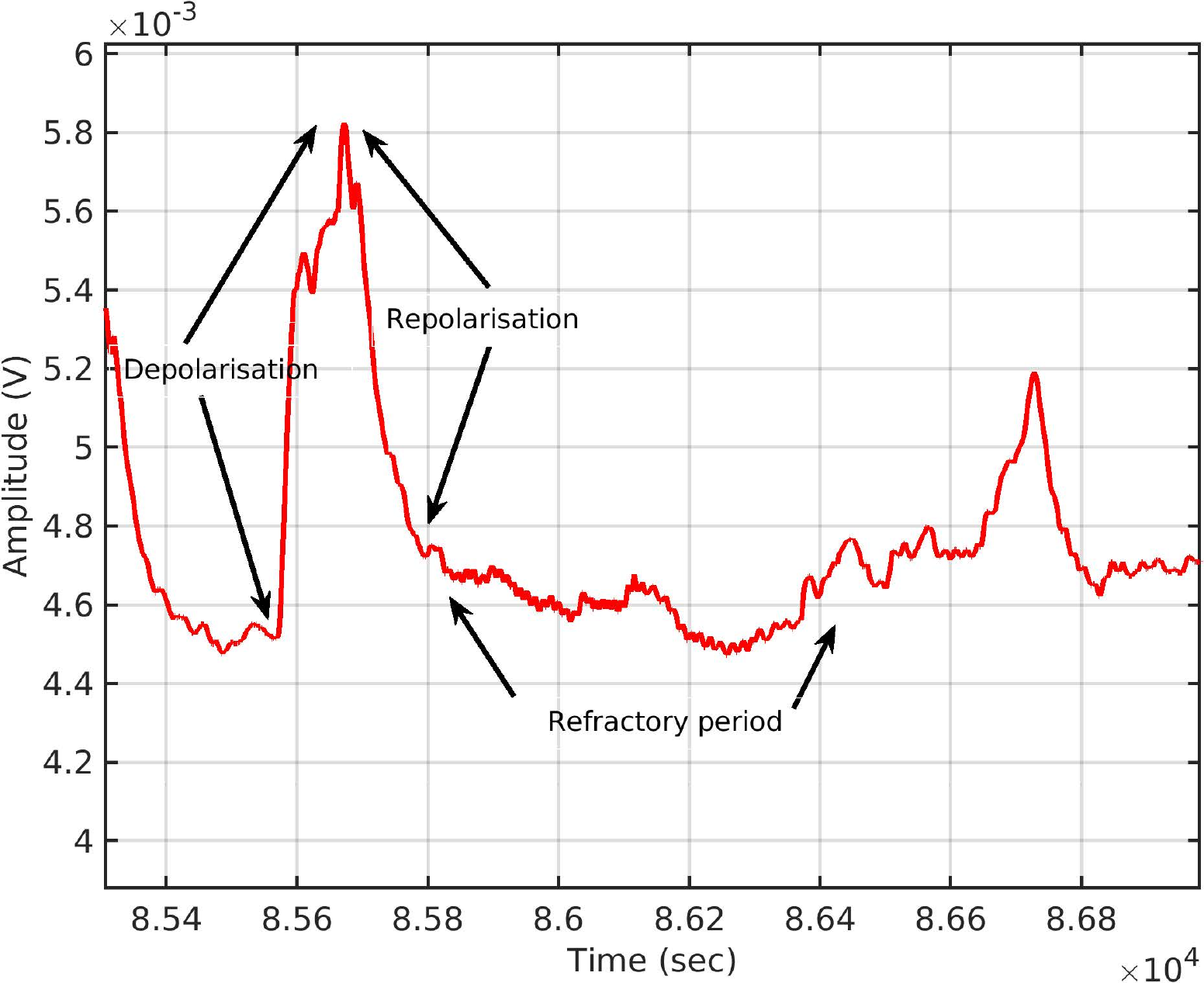}}
    \caption{The electrical activity of the mycelium of the grey oyster fungi. (a) Example of a dynamics of electrical potentials recorded from eight channels of the same cluster during 63 hours. (b) Three channels are zoomed in the inserts to show the rich combination of slow (hours) drift of base electrical potential combined with relatively fast (minutes) oscillations of the potential. (c) All `classical’ parts of a spike, \textit{i.e.},  depolarisation, repolarisation and refractory period, can be found in this exemplar spike. This spike has a period of 220~s, from base-level potential to refractory-like period, and refractory period of 840~s. The depolarisation and repolarisation rates are 0.03 and 0.009~mV/s, respectively.} \label{fig:1}
\end{figure}

Lack of an algorithmic framework for exhaustive characterisation of the electrical activity of a substrate colonised by mycelium of oyster fungi \emph{Pleurotus djamor} motivated us to develop this framework to extract spike patterns, quantify the diversity of spiking events, and measure the complexity of fungal electrical communication. We evidenced the spiking activity of the mycelium (see an example in Fig.~\ref{fig:1}), which will enable us to develop an experimental prototype of fungi-based information processing devices.

We evaluated the proposed framework in comparison to the existing, in neuroscience, techniques of spike detection~\cite{nenadic2004spike,shimazaki2010kernel}, and observed considerable improvement in extracting spike activity periods. Evaluation of the proposed method for detecting spikes events compared to the determined spikes' arrival time by an expert shows true-positive and false-positive rates of 76\% and 16\%, respectively. We found that the average dominant duration of an action-potential-like spike is 402 sec. The spikes' amplitude varies from 0.5~mV to 6~mV and depends on the location of the electrical activity source (the position of electrodes). We observed that the Kolmogorov complexity of fungal spiking varies from 11$\times 10^{-4}$ to 57$\times 10^{-4}$. This might indicate mycelium sub-networks in different parts of the substrate have been transmitting different information to other parts of the mycelium network, \textit{i.e.}, more extended propagation of excitation wave corresponds to higher values of complexity.

The rest of this paper is organised as follows. Sect.~\ref{sec:2} presents the experimental setup. The details of the proposed methods for spike detection are explained in Sect.~\ref{sec:3}. Experimental results and complexity analysis are presented in Sect.~\ref{sec:4}. Finally, the discussion is given in Sect.~\ref{sec:5}.

\section{Experimental set-up} \label{sec:2}

\begin{figure}[!htb]
    \centering
    \subfigure[]{\includegraphics[width=0.45\textwidth]{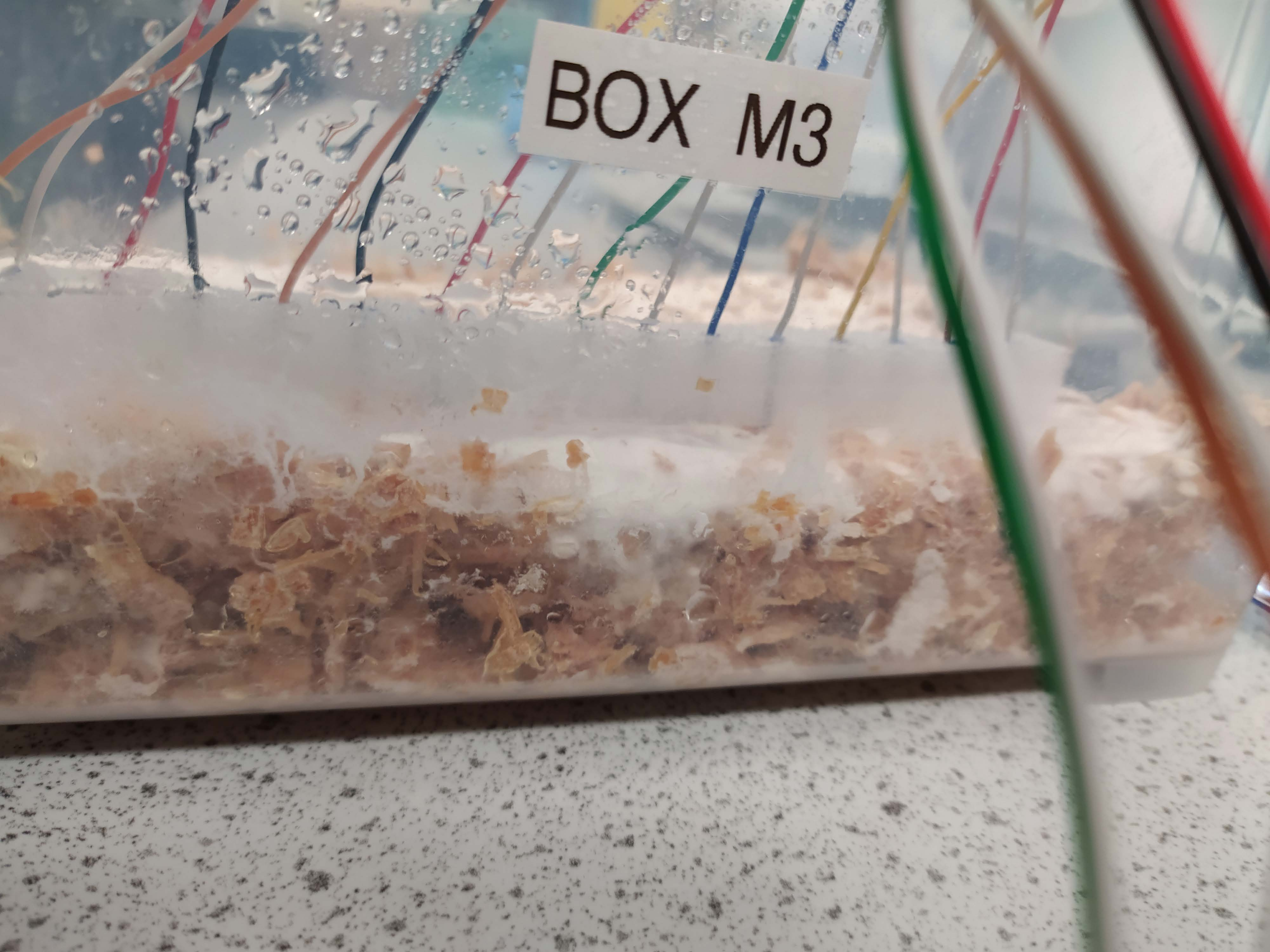}}
    \subfigure[]{\includegraphics[width=0.45\textwidth]{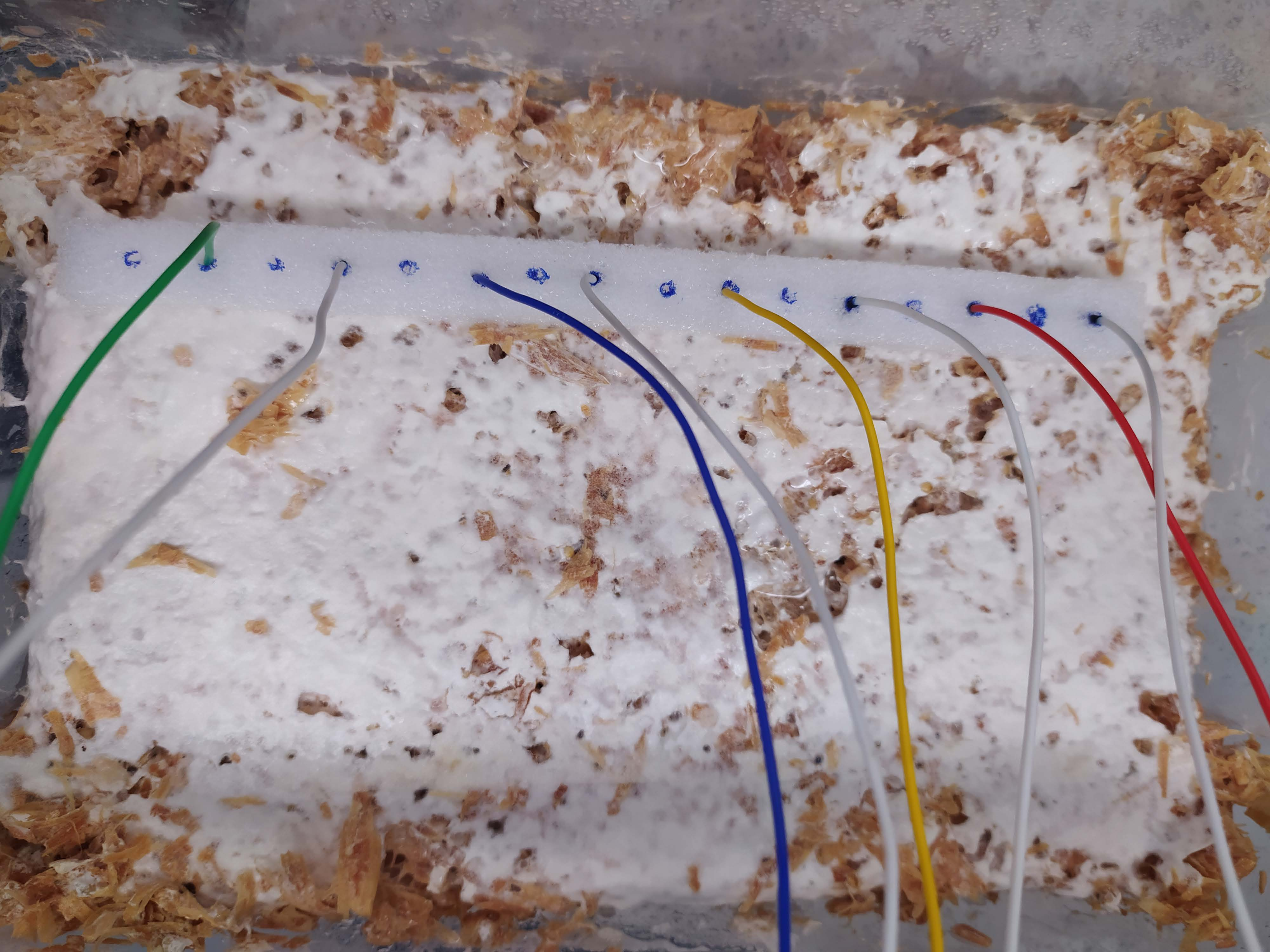}}
    \subfigure[]{\includegraphics[width=0.45\textwidth]{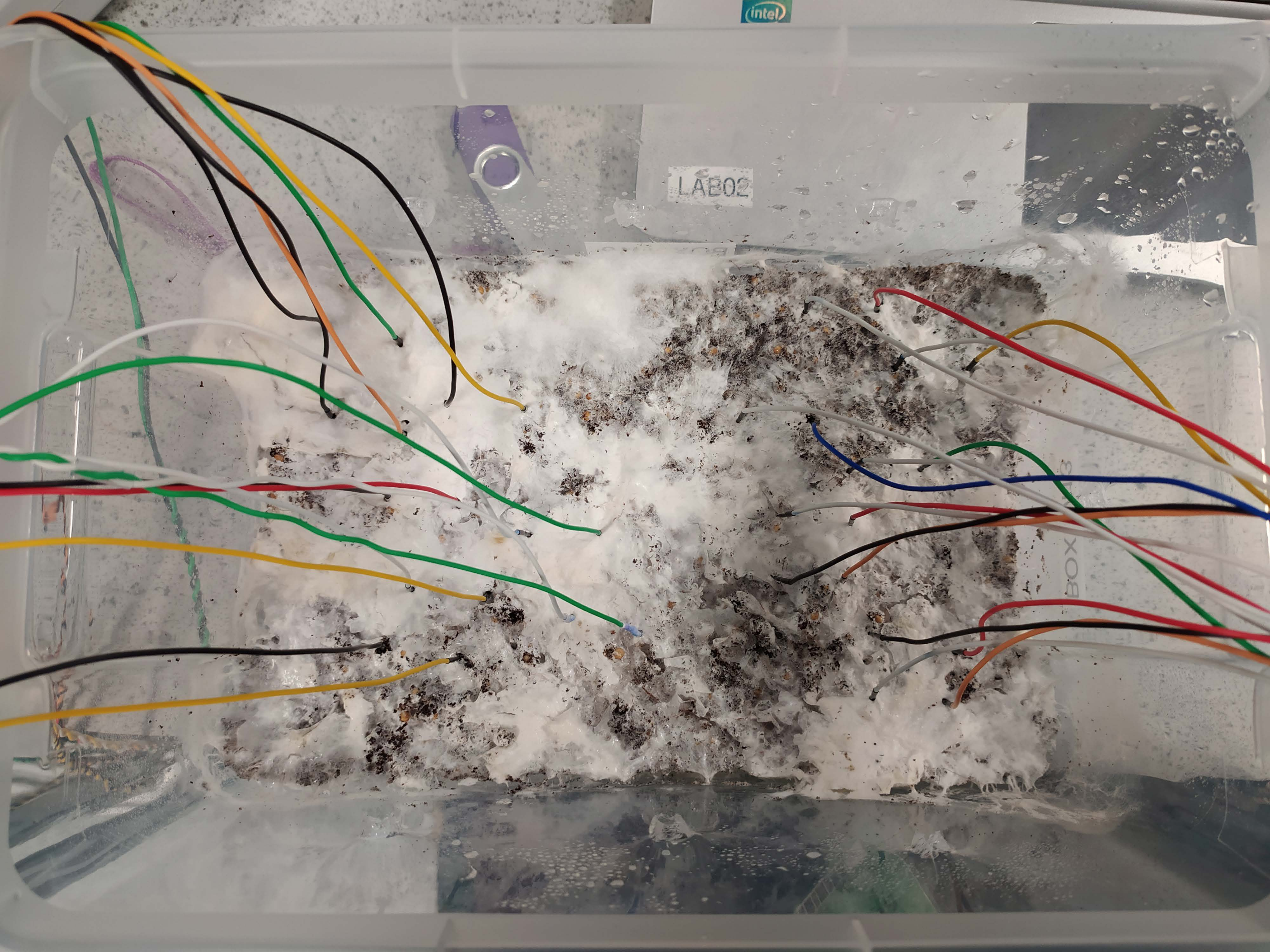}}
    \caption{Three examples of the experimental set-up with (a) in lines placement of electrodes (1~cm distance), (b) in lines placement of electrodes (2~cm distance), (c) random electrode placement.} \label{fig:2}
\end{figure}

A wood shavings  substrate was colonised by the mycelium of the grey oyster fungi, \emph{Pleurotus ostreatus} (Ann Miller's Speciality Mushrooms Ltd, UK). The substrate was placed in a hydroponic growing tent with a silver Mylar lightproof inner lining (Green Box Tents, UK).  Figure~\ref{fig:2} shows three examples of the experimental set-up.

We inserted pairs of iridium-coated stainless steel sub-dermal needle electrodes (Spes Medica SRL, Italy), with twisted cables into the colonised substrate to obtain electrical activity. Using an ADC-24 (Pico Technology, UK) high-resolution data logger with a 24-bit A/D converter, galvanic isolation and software-selectable sample rates all contribute to a superior noise-free resolution. We recorded electrical activity one sample per second, where the minimum and maximum logging times were 60.04 and 93.45 hours, respectively. During the recording, the logger makes as many measurements as possible (typically up to 600 per second) and saves the average value. We set the acquisition voltage range to 156~mV with an offset accuracy of 9~$\mu$V at 1~Hz to maintain a gain error of 0.1\%. Each electrode pair was considered independently with the noise-free resolution of 17 bits and conversion time of 60~ms. In our experiments, electrode pairs were arranged in one of two configurations: random placement or in lines. Distance between electrodes was 1-2~cm. In each cluster, we recorded 5–16 electrode pairs (channels) simultaneously.

\section{Proposed method} \label{sec:3}

A spike event can be formally defined as an extracellular signal that exceeds a simple amplitude threshold and passes through a subsequent pair of user-specified time-voltage boxes. The spike, which includes depolarisation, repolarisation, and refractory periods, reflects physiological and morphological processes ongoing in mycelial networks. To extract spike events, we proposed an unsupervised  method which consists of three major steps.

In the first step, we split the whole recording period, $F(t)$, into $k$ chunks, $f_{k}(t)$, with respect to the signal's transitions. To determine the transitions, we estimated the state levels of the signal by its histogram and identified all regions that cross the upper-state boundary of the low state and the lower-state boundary of the high state. Then, we calculated scale-to-frequency conversions of the analytic signal in each chunk using Morse wavelet basis~\cite{lilly2012generalized}. To assess the presence of spike-like events, we scaled the wavelet coefficients at each frequency and obtained the sum of the scales that were less than the threshold defined in Algorithm~\ref{alg:1}. Finally, we selected regions of interest (ROI) enclosed between a consecutive local minimum and maximum whose lengths were greater than 30~sec.

In the second step, we calculated the envelopes of the analytic signal using spline interpolation over local maximums. To determine the analytical signal, we first applied the discrete approximation of Laplace’s differential operator to $f_{k}(t)$ to obtain a finite sequence of equally-spaced samples. Then, we converted this finite sequence into a same-length sequence of equally-spaced samples of the discrete-time Fourier transform. From the average signal envelope, we extracted regions that fall in a consecutive local minimum and maximum. These regions created constraints that observing them led to the identification of spike events.

In the third step, we preserved ROIs from the first step where satisfied constraints obtained in the second step. The signal envelope could guide wavelet decomposition in an unsupervised way to cluster signal into the spike, pseudo-spike, and background activity of neighbouring cells. We detailed the proposed method in the following sub-sections.

\subsection{Slicing fungi electrical activity}

To split the fungi electrical activity, $F(t)$, with a length of $t$ second into $k$ chunks $f_{k}(t), 1 \leq k \leq t-1$, we used signal transitions that compose each pulse. To determine the transitions, we estimated the state levels of $F(t)$ by a histogram method~\cite{ieee2003}. Then, we identified all regions that cross the upper-state boundary of the low state and the lower-state boundary of the high state. To estimate the states of the signal, we followed the following steps.

\begin{figure}[!htb]
    \centering
    \subfigure[]{\includegraphics[width=0.48\textwidth]{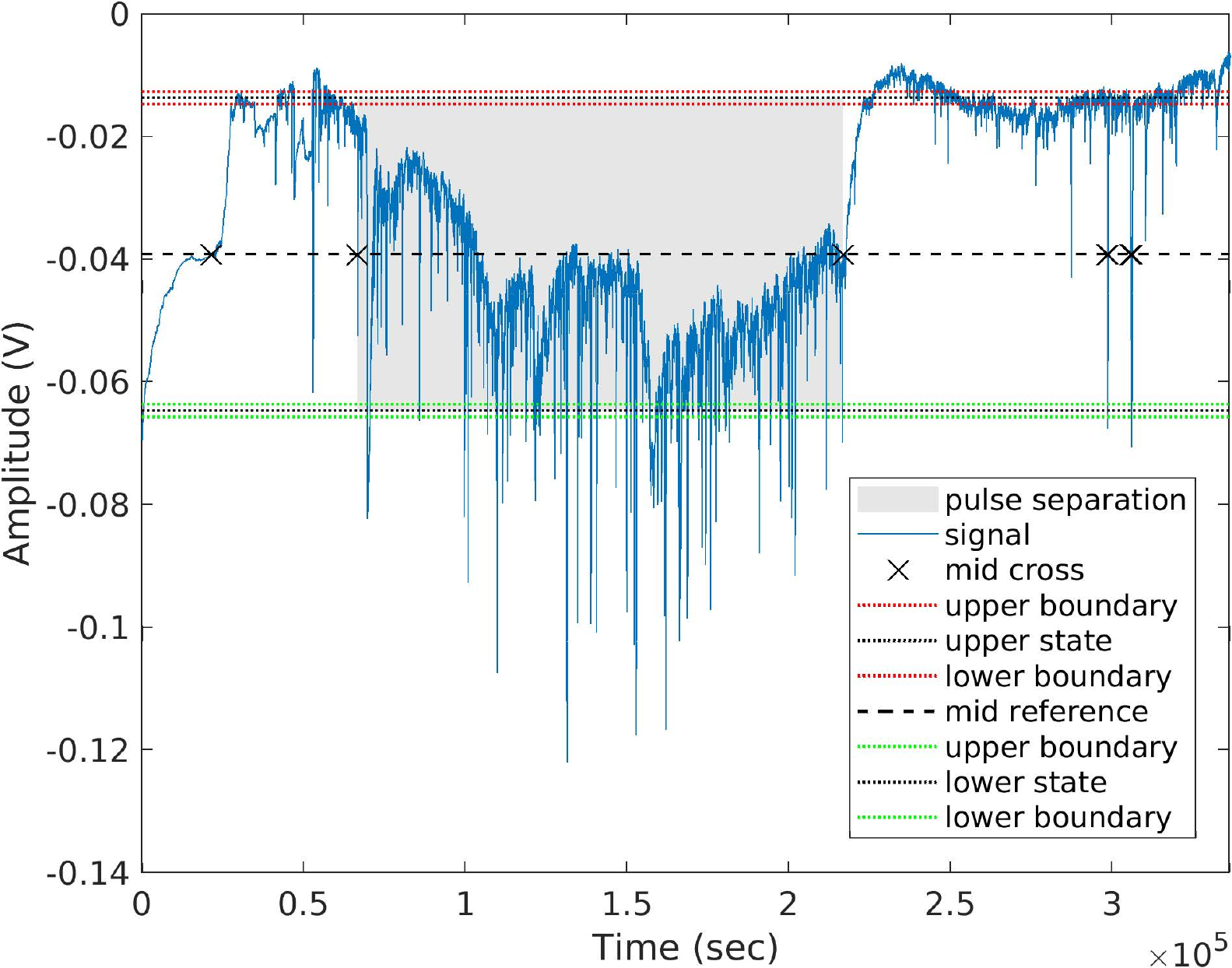}}
    \subfigure[]{\includegraphics[width=0.48\textwidth]{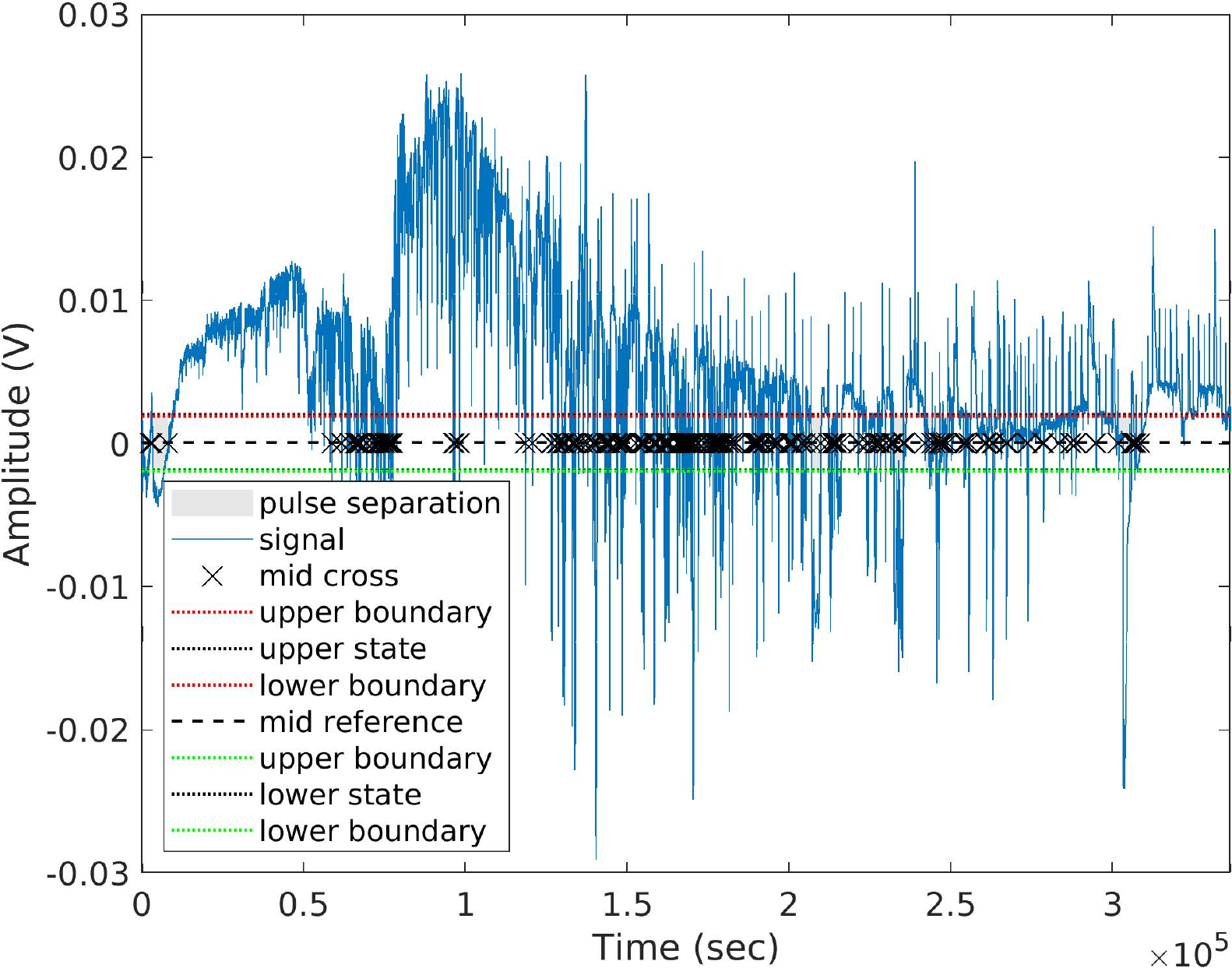}}
    \caption{Slicing electrical potential recordings for two channels.} \label{fig:0}
\end{figure}

\begin{enumerate}
    \item Determining the minimum, maximum and range of amplitudes.
    \item Sorting amplitude values into the histogram bins and determining the bin width by dividing the amplitude range to the number of bins.
    \item Identifying the lowest- and highest-indexed histogram bins, $hb_{low}$, $hb_{high}$, with non-zero counts. 
    \item Dividing the histogram into two sub-histograms, where the indices of the lower and upper histogram bins are $hb_{low} \leq hb \leq \frac{1}{2}(hb_{high}-hb_{low})$ and $hb_{low}+\frac{1}{2}(hb_{high}-hb_{low}) \leq hb \leq hb_{high}$, respectively.
    \item Calculating the mean of the lower and upper histogram to compute the state levels.
\end{enumerate}
Each chunk is then enclosed between the last negative-going transitions of every positive-polarity pulse and the next positive-going transition. Figure~\ref{fig:0} shows slicing results for two channels.

\subsection{Detecting time-localised events by Morse-based wavelets}

The electrical activity of the mycelium exhibits modulated behaviour with variation in amplitude and frequency over time. This feature hints that the signal can be analysed with analytic wavelets, which are naturally grouped into even or cosine-like and odd or sine-like pairs, allowing them to capture phase variability. A wavelet $\psi(t)$ is a finite energy function which projects the $f(t)$ onto a family of time-scale waveforms by translation and scaling. The Morse wavelet, $\psi_{\beta,\gamma}(t)$, is an analytic wavelet whose Fourier transforms is supported only on the positive real axis \cite{lilly2012generalized,lilly2017element}. This wavelet is defined in the frequency domain for $\beta \geq 0$ and $\gamma > 0$ using Eq.~\ref{eq:1}

\begin{align}
    \psi_{\beta,\gamma}(t) = \frac{1}{2\pi}\,\int_{-\infty}^{\infty}\,\Psi_{\beta,\gamma}(\omega)\,e^{i\omega t}\,\rm{d}\omega, \nonumber \\
    \Psi_{\beta,\gamma}(\omega) \equiv a_{\beta,\gamma}\,\omega^{\beta}\,e^{-\omega^{\gamma}}\,\times\,\begin{cases} 1 & \omega > 0 \\ \frac{1}{2} & \omega = 0 \\ 0 & \omega < 0 \end{cases}.
    \label{eq:1}
\end{align}
where $\omega$ is the angular frequency and $a_{\beta,\gamma} \equiv 2 \left ( \frac{e\gamma}{\beta} \right)^{\frac{1}{\gamma}}$ is the amplitude coefficient used as a real-valued normalised constant. Here, $e$ is Euler’s number, $\beta$ characterises the low-frequency behaviour, and $\gamma$ defines the high-frequency decay. We can rewrite Eq.~\ref{eq:1} in the Fourier domain, parameterised by $\beta$ and $\gamma$ as Eq.~\ref{eq:2}.

\begin{align}
    \phi_{\beta,\gamma}(\tau,s) \equiv \int_{-\infty}^{\infty}\frac{1}{s}\,\psi^{*}_{\beta,\gamma}(\frac{t-\tau}{s})f(t)\,\rm{d}t = \frac{1}{2\pi}\,\int_{-\infty}^{\infty}\,e^{i\omega \tau}\, \Psi^{*}_{\beta,\gamma}(s\omega)F(\omega)\,\rm{d}\omega.
    \label{eq:2}
\end{align}
where $F(\omega)$ is the Fourier transform of $f(t)$, and $*$ denotes the complex conjugate. When $\Psi^{*}_{\beta,\gamma}(\omega)$ is real-valued, the conjugation may be omitted. The scale variable $s$ causes stretching or compression of the wavelet in time. In order to reflect the energy of $f(t)$ and normalise the time-domain wavelets to preserve constant energy, $\frac{1}{\sqrt{s}}$ is usually used. However, we used $\frac{1}{s}$ instead, since we describe time-localised signals by the amplitude. To recover the time-domain representation, we can use the inverse Fourier transform by $f(t) = \frac{1}{2\pi}\,\int_{-\infty}^{\infty}\,e^{i\omega t}\,F(\omega)\,\rm{d}\omega$ and $\psi_{\beta,\gamma}(t) = \int_{-\infty}^{\infty}\,e^{i\omega t}\,\rm{d}t = 2\pi\delta(\omega)$, where $\delta(\omega)$ is the Dirac delta function.

\begin{figure}[!htbp]
    \centering
    \subfigure[]{\includegraphics[width=0.48\textwidth]{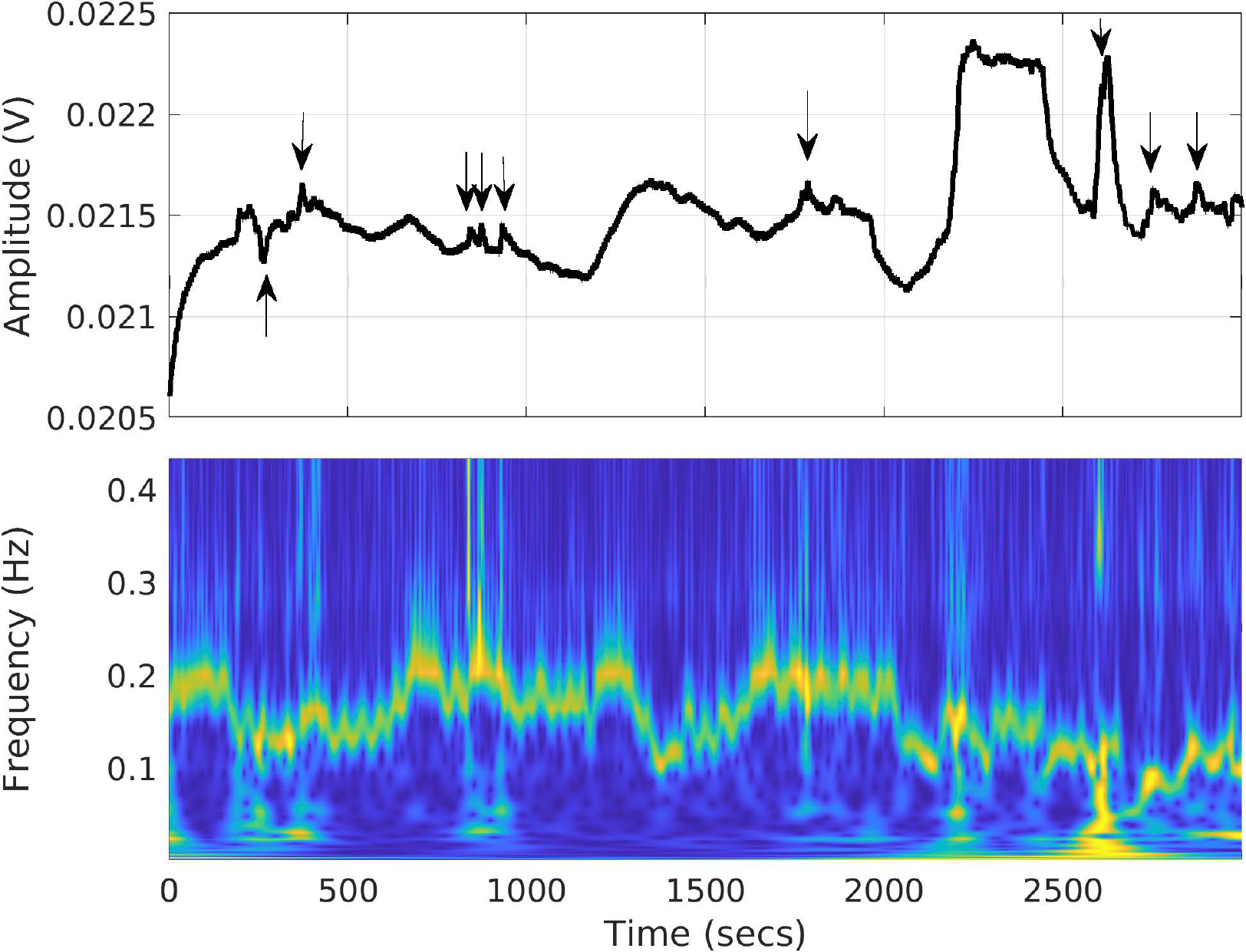}}
    \subfigure[]{\includegraphics[width=0.48\textwidth]{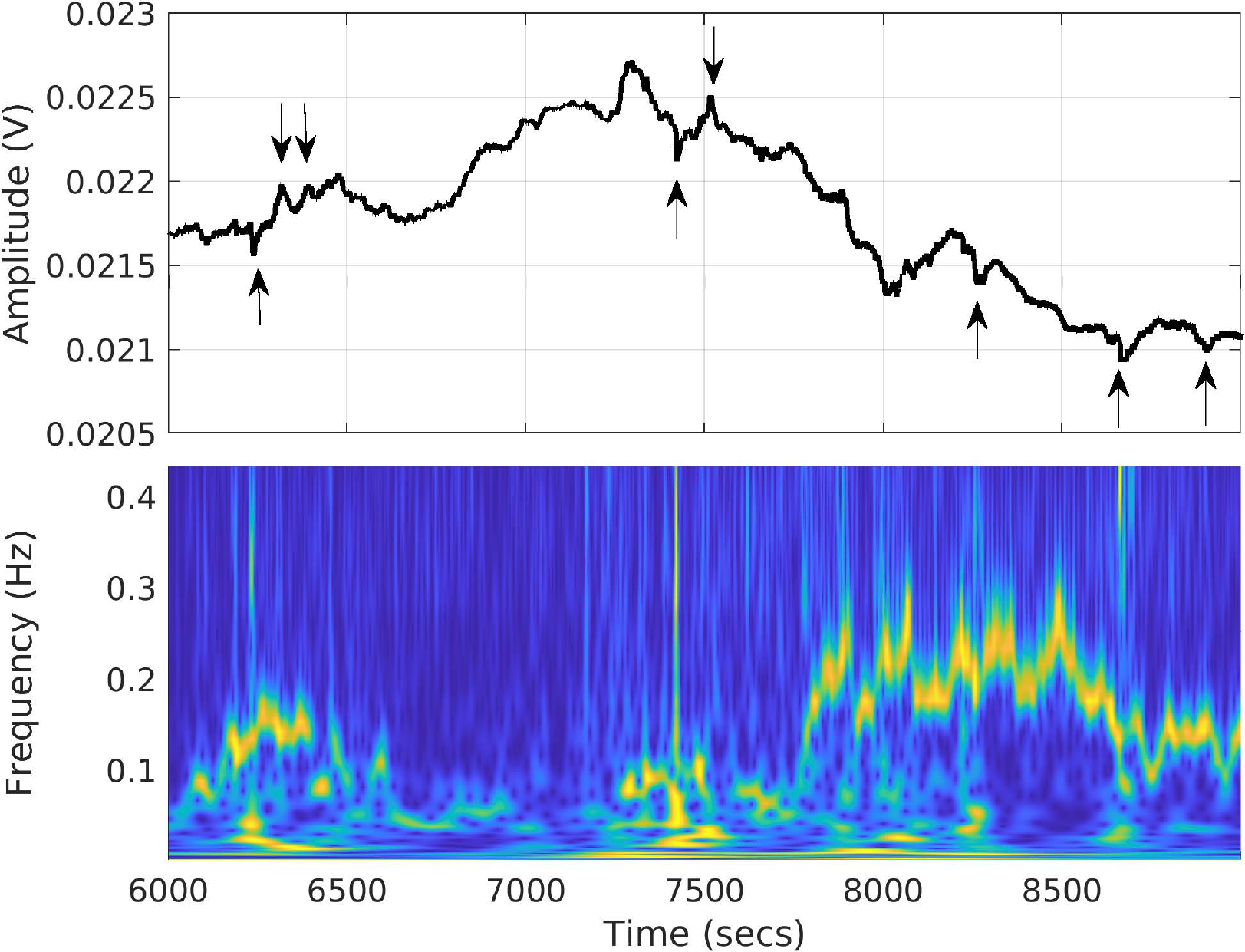}}
    \caption{Annotated spikes by the expert over with the Morse wavelet scalogram for (a) $Slice_1$ and (b) $Slice_2$. We added black arrows to point to the spike identified by the expert. The scalogram is plotted as a function of time and frequency in which the maximum absolute value at each frequency is used to normalise coefficient. Frequency axis is displayed on a linear scale.} \label{fig:3}
\end{figure}

The representation of Morse wavelets can be more oscillatory when both $\beta$ and $\gamma$ increase, and more localised with impulses when these parameters decrease. On the other hand, increasing $\beta$ and keeping $\gamma$ fixed broaden the central portion of the wavelet and increase the long-time decay rate. Whereas, increasing $\gamma$ by keeping $\beta$ constant expands the wavelet envelope without affecting the long-time decay rate. Following explanations given in~\cite{lilly2008higher}, we set the symmetry parameter $\gamma$ to 3 and the time-bandwidth product $P^{2} = \beta\gamma$ to 60. We also used $L1$ normalisation to have an equal magnitude in the wavelets when we have equal amplitude oscillatory components at different scales. Figure \ref{fig:3} shows two randomly selected 3000-second chunks of the fungi electrical activity (namely, $Slice_1$ and $Slice_2$) with their Morse wavelet scalograms. 

\begin{figure}[!htb]
    \centering
    \subfigure[]{\includegraphics[width=0.48\textwidth]{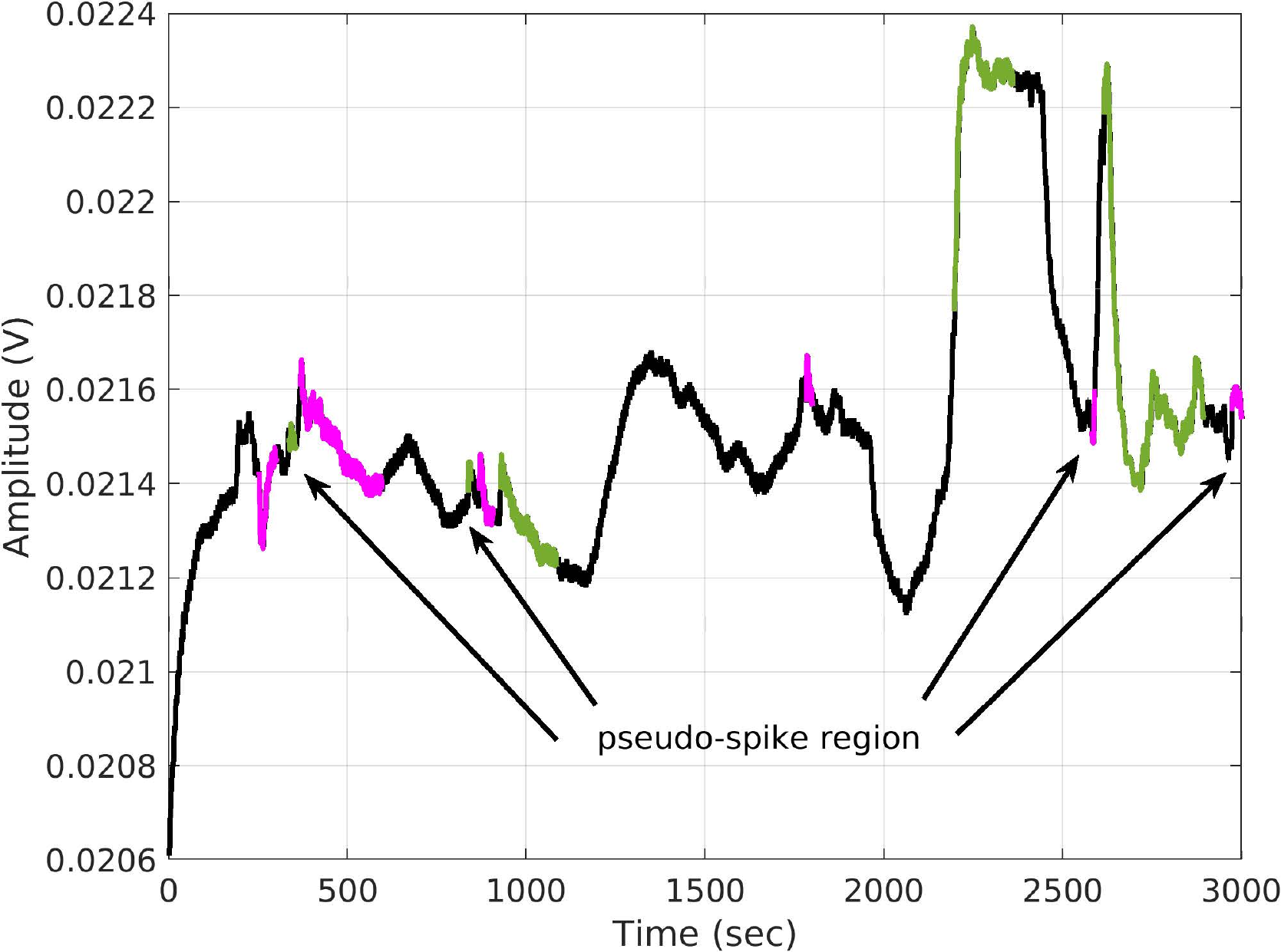}}
    \subfigure[]{\includegraphics[width=0.48\textwidth]{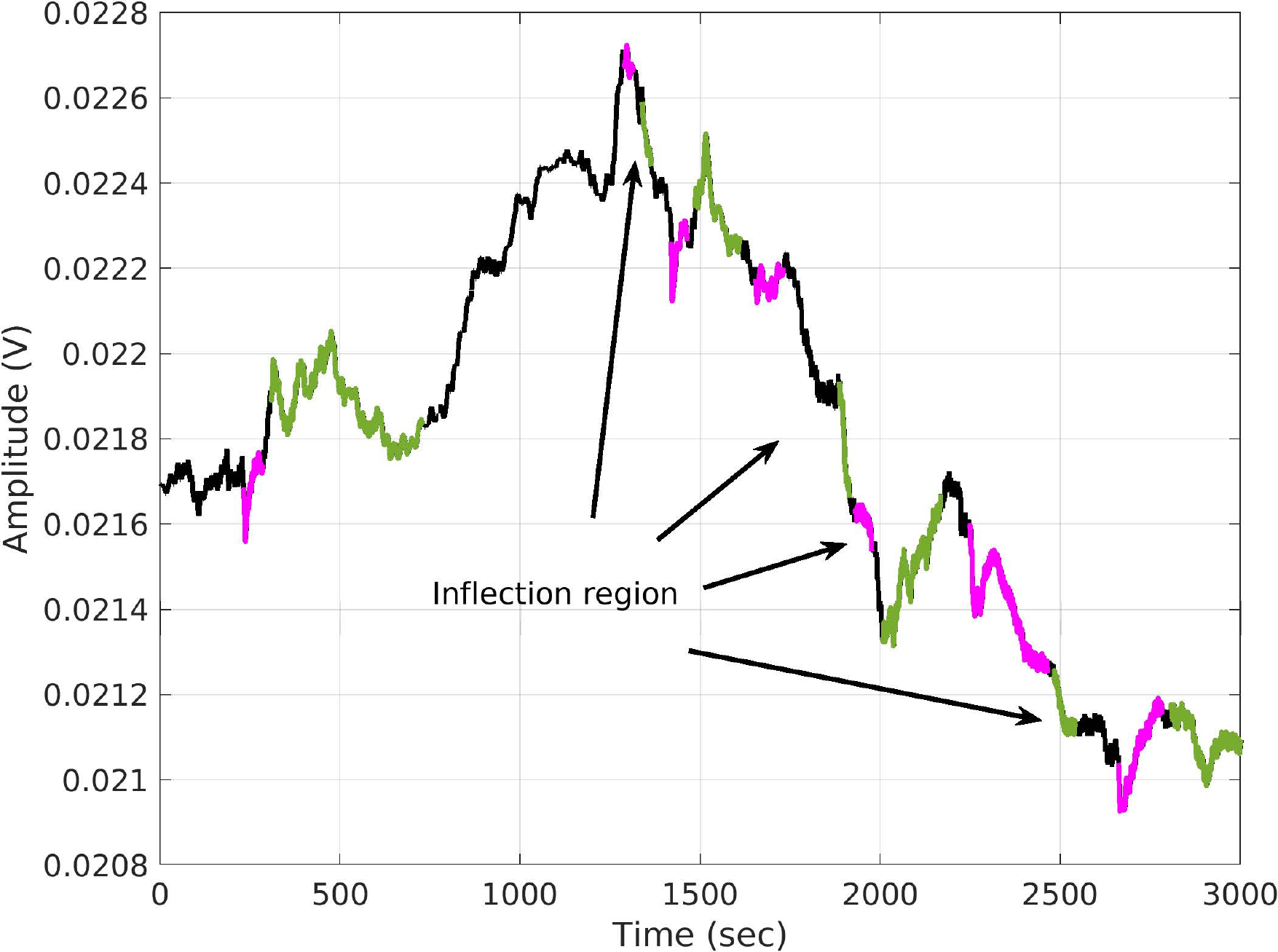}}
    \caption{(a,b) Identified local maxima and minima over $g_{\beta,\gamma}(\tau,s)$ in (a) $Slice_1$ and (b) $Slice_2$. The second-row of the plot is the inverse of the first row; therefore, the marked maximums are identical to the local minima. (c,d) Candidate regions of interest which are alternately coloured purple and green to ease visual tracking.} \label{fig:4}
\end{figure}

We observed that using the maximum absolute value at each frequency (level) to normalise coefficients can help in identifying events that may contain spikes. Hence, we proposed to use Eq.~\ref{eq:3} for normalisation and subsequently set all zero entries to 1.

\begin{align}
    \kappa_{\beta,\gamma}(\tau,s) = |\phi_{\beta,\gamma}(\tau,s)|^\intercal, \nonumber \\
    g_{\beta,\gamma}(\tau,s) = \left ( \eta \times \frac{\kappa_{\beta,\gamma}(\tau,s) - \min_{s}(\kappa_{\beta,\gamma}(\tau,s))}{\max_{s}(\kappa_{\beta,\gamma}(\tau,s))} \right )^\intercal.
    \label{eq:3}
\end{align}
where $|\bullet|$ and $(\bullet)^\intercal$ return the absolute value and the matrix transpose, respectively. Here, $\eta$ is a scaling factor that we empirically set it to 240. We used $g_{\beta,\gamma}(\tau,s)$ in Algorithm~\ref{alg:1} to extract candidate ROIs, which are shown in Fig.~\ref{fig:4}.

\RestyleAlgo{ruled}
\SetAlgoNoLine
\LinesNumbered
\begin{algorithm}[!htb]
\SetKw{KwBy}{by}
\SetKwFunction{Lmin}{LocalMinimum}\SetKwFunction{Lmax}{LocalMaximum}
\SetKwInOut{Input}{Input}\SetKwInOut{Output}{Output}
\Input{$g_{\beta,\gamma}(\tau,s)$ -- Scaled wavelets coefficients.}
\Output{$\mathcal{B}$ -- set of candidate regions.}
\BlankLine
\Begin{
    $\epsilon = 0.05 \times (\max(g_{\beta,\gamma}(\tau,s))-\min(g_{\beta,\gamma}(\tau,s)))$\;
    $max_{g} \gets$ set of all \Lmax{$g_{\beta,\gamma}(\tau,s), \epsilon$}\;
    \tcp{\Lmax{} returns $\tau^{*}$ if $\forall \tau \in (\tau^{*} \pm \epsilon), ~ g_{\beta,\gamma}(\tau^{*},s) \geq g_{\beta,\gamma}(\tau,s)$.}
    $min_{g} \gets$ set of all \Lmin{$g_{\beta,\gamma}(\tau,s), \epsilon$}\;
    $\mathcal{U} \gets \mathbf{sort}(min_{g} \bigcup max_{g})$\;
    $n = \mathbf{card}(\mathcal{U})$\;
    \tcp{$\mathbf{card}(A)$ returns number of entries in $A$.}
    \If{$n \equiv 1 \;(\bmod\; 2)$}{
        slack $\gets \mathbf{mean}($difference of two consecutive entries$)$\; 
        Add $\mathbf{min}(\mathcal{U}_n+\mathrm{slack},\tau)$ to $\mathcal{U}$\;
        $n = n+1$\;
    }
    $\mathcal{B} \gets (\mathcal{U}_{i},\mathcal{U}_{i+1}), ~ \forall i \in \{1,3,\cdots,n-1\}$ 
}
\BlankLine
\Return{$\mathcal{B}$}
\caption{Detecting candidate regions for time-localised events.}
\label{alg:1}
\end{algorithm}

\begin{figure}[!htb]
    \centering
    \subfigure[]{\includegraphics[width=0.48\textwidth]{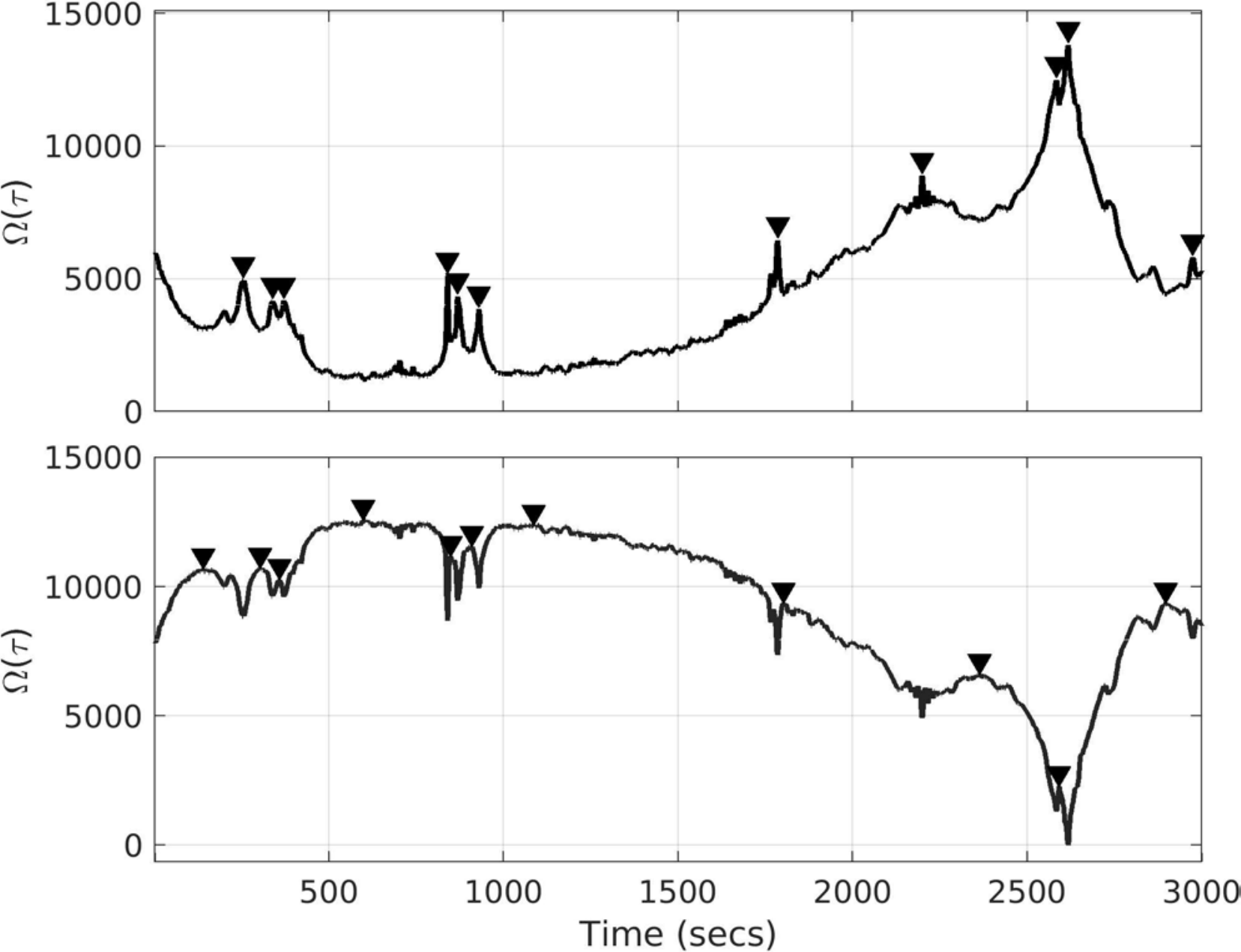}}
    \subfigure[]{\includegraphics[width=0.49\textwidth]{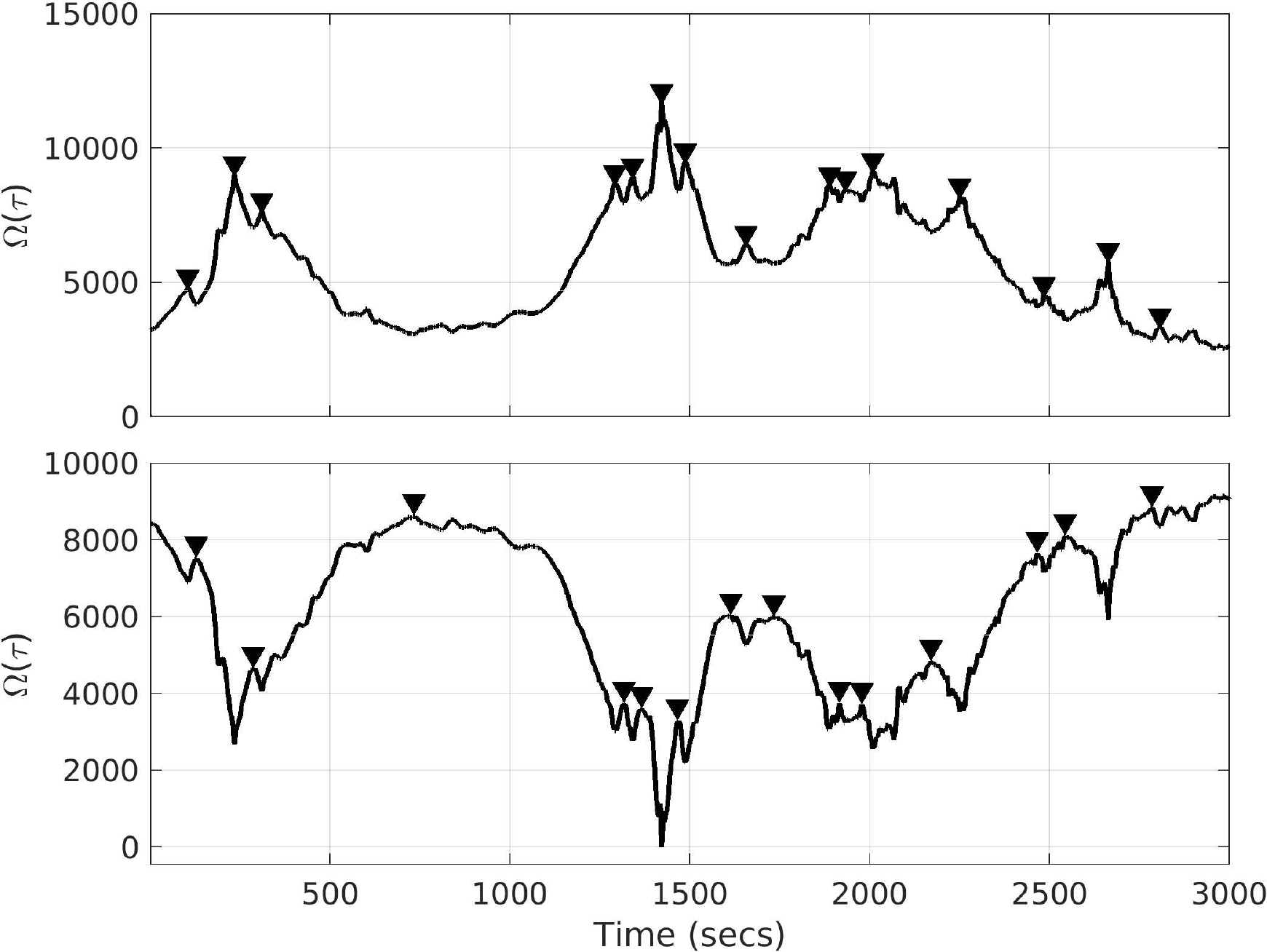}}
    \vspace{-2\baselineskip}
    \caption*{}
\end{figure}

\begin{figure}[!htb]
    \centering
    \subfigure[]{\includegraphics[width=0.48\textwidth]{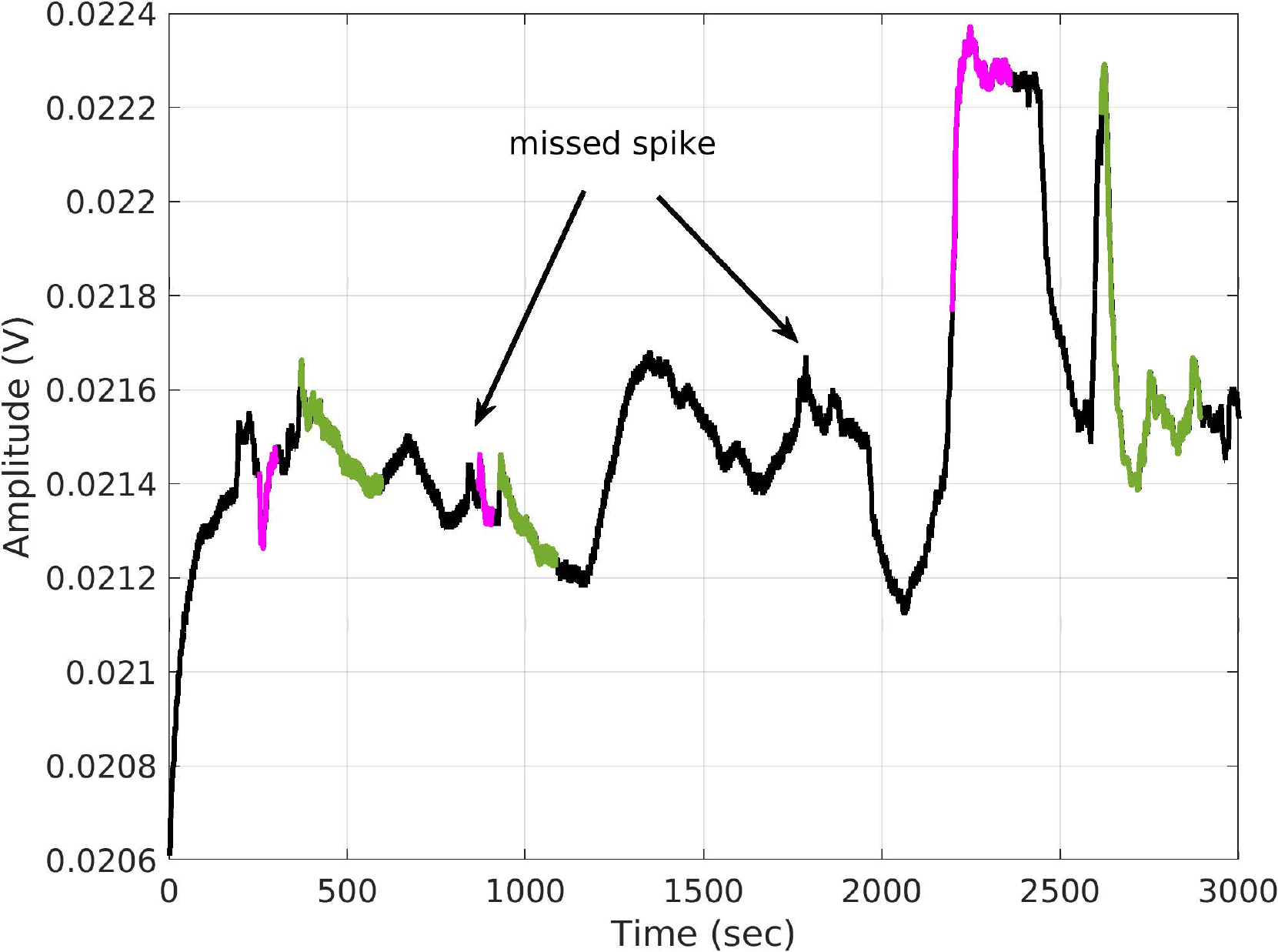}}
    \subfigure[]{\includegraphics[width=0.48\textwidth]{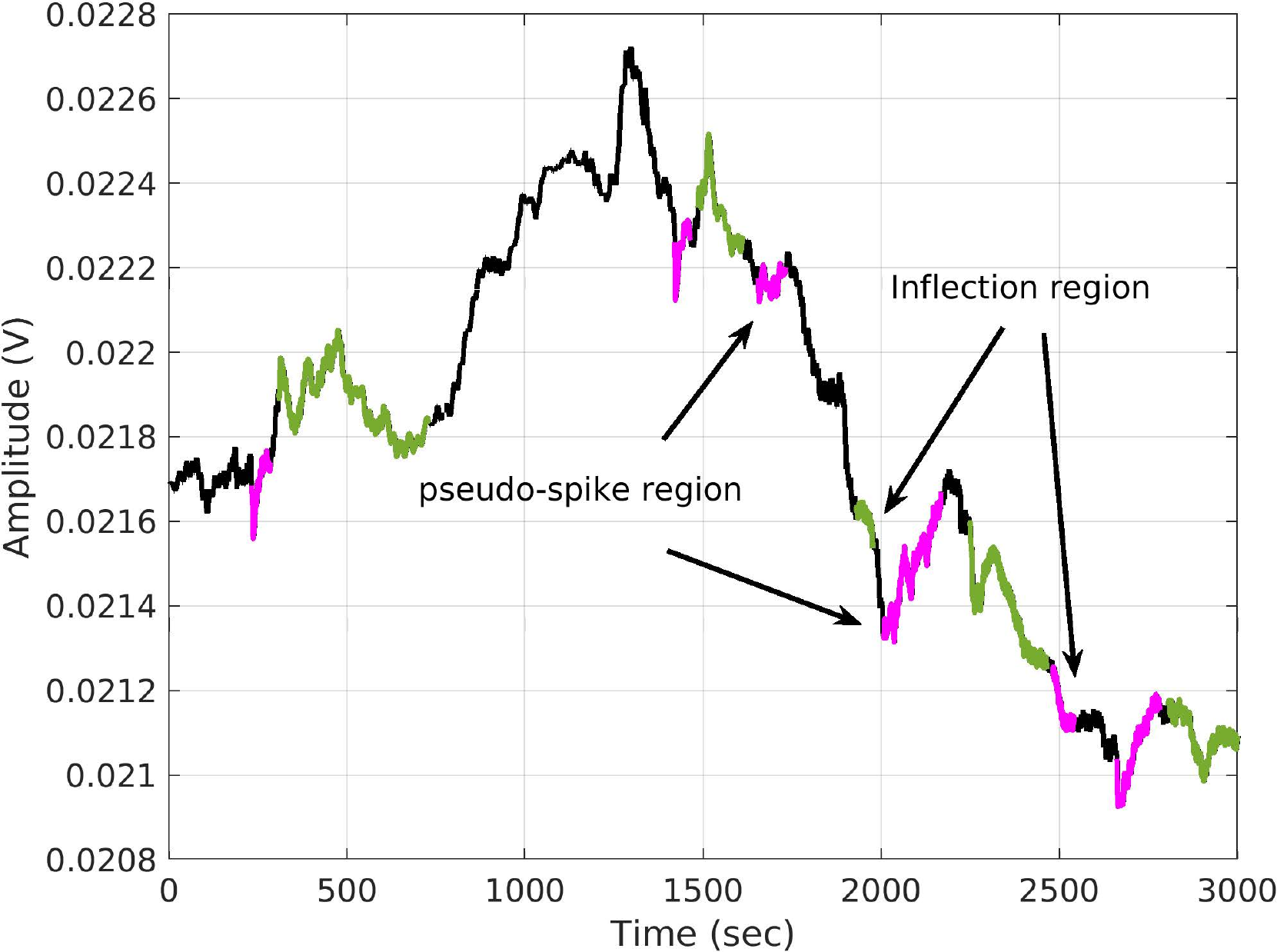}}
    \caption{Results of applying Algorithm~\ref{alg:2} to (a) $Slice_1$ and (b) $Slice_2$. Two spike events are missed in $Slice_1$. Two pseudo-spike and two inflection regions still remain in $Slice_2$.} \label{fig:5}
\end{figure}

As shown in Fig.~\ref{fig:4}(c,d), some of the detected regions are either too short\footnote{We observed in our previous studies \cite{adamatzky2018towards,adamatzky2019plant} that minimum spike length was 5 mins.} or lack repolarisation and depolarisation periods that should be removed from $\mathcal{B}$. We proposed Algorithm \ref{alg:2} to remove these regions, which we called them \emph{pseudo-spike} and \emph{inflection} regions, respectively. Figure~\ref{fig:5} shows the results.

\RestyleAlgo{ruled}
\SetAlgoNoLine
\LinesNumbered
\begin{algorithm}[!htb]
\SetKwFunction{Lmin}{isLocalMinimum}\SetKwFunction{Lmax}{isLocalMaximum}
\SetKwInOut{Input}{Input}\SetKwInOut{Output}{Output}
\Input{$\mathcal{B}$ --- set of ROI, \textit{i.e.}, Algorithm~\ref{alg:1} output,\\
        $f$ --- Electrical potential.}
\Output{$\mathcal{C}$ --- set of wavelet-based ROIs,\\
        $\mathcal{D}$ --- set of \emph{pseudo-spike} and \emph{inflection} regions.}
\BlankLine
\Begin{
	\For{$i = 1$ \KwTo $\mathbf{card}(\mathcal{B})$}{
		$lb \leftarrow \mathcal{B}(i,1)$\;
		$ub \leftarrow \mathcal{B}(i,2)$\;
		\If{$(ub-lb) > 30$}{
		    $chunk = f[lb \cdots ub]$\;
		    $minima = \mathbf{min}(\Lmin{chunk})$\;
		    \tcp{\Lmin{} and \Lmax{} use spline interpolation in locating local extreme \cite{hall1976optimal}.}
		    $maxima = \mathbf{max}(\Lmax{chunk})$\;
		    \eIf{$f(minima) < \mathbf{min}(f(lb),f(ub))~\mathbf{or}~f(maxima) > \mathbf{max}(f(lb),f(ub))$}{
		        $\mathcal{C} \leftarrow [lb,~ub]$\;
		    }{
		        $\mathcal{D} \leftarrow [lb,~ub]$\;
		    }
		}
	}
}
\BlankLine
\Return{$\mathcal{C}$, $\mathcal{D}$}
\caption{Excluding pseudo-spike and inflation regions form candidate ROI.}
\label{alg:2}
\end{algorithm}

Applying Algorithm~\ref{alg:2} led to the loss of two spikes in $Slice_1$ (see Fig.~\ref{fig:5}(a)) and failure to remove two \emph{pseudo-spike} and two \emph{inflection} regions in $Slice_2$ (see Fig.~\ref{fig:5}(b)). We found that assessing the analytic signal by its envelope can increase the accuracy of spike detection.

\subsection{Analytical signal envelope for locating spike pattern}

To obtain the signal envelope, $\xi$, we calculated the magnitude of its analytic signal. The analytic signal is found using the discrete Fourier transform as implemented in Hilbert transform. To intensify effective peaks in the signal and, specifically, \emph{inflection} regions ineffective, we calculated the second numerical derivation of the signal as $L = \frac{\partial^2 f}{4\partial t^2}$.

A frequency-domain approach to approximately generate a discrete-time analytic signal is proposed in~\cite{marple1999computing}. In this approach, the negative frequency half of each spectral period is set to zero, resulting in a periodic one-sided spectrum. The specific procedures for creating a complex-valued $N$-point ($N$ is even) discrete-time analytic signal $F(\omega)$ from a real-valued $N$-point discrete time signal $L[n]$ are as follows:

\begin{enumerate}
    \item Compute the $N$-point discrete-time Fourier transform using $F(\omega) = T \sum_{n=0}^{N-1}L[n]e^{-i2\pi\omega Tn}$, where $|\omega| \leq 1/2T$~Hz and $L[n]$ for $0 \leq n \leq N-1$ is obtained by sampling a band-limited real-valued continuous-time signal $L(nT) = L[n]$ at periodic time intervals of $T$ seconds to prevent aliasing.
    \item Form the $N$-point one-sided discrete-time analytic signal transform:
    \begin{equation}
        Z[m] = \begin{cases}
            F[0], & \text{ for } m = 0\\ 
            2F[m], & \text{ for } 1 \leq m \leq\frac{N}{2}-1 \\ 
            F[\frac{N}{2}], & \text{ for } m = \frac{N}{2}\\ 
            0, & \text{ for } \frac{N}{2}+1 \leq m \leq N-1.
    \end{cases}
    \end{equation}
    \item  Compute the $N$-point inverse discrete-time Fourier transform to obtain the complex discrete-time analytic signal of same sample rate as the original $L[n]$
    \begin{equation}
        z[n] = \frac{1}{NT}\sum_{m=0}^{N-1}Z[m]e^{\frac{i2\pi mn}{N}}
    \end{equation}
\end{enumerate}

Obtaining analytic signal in this way can satisfy two properties: (1) The real part is identical to the original discrete-time sequence; (2) the real and imaginary components are orthogonal. Calculating the magnitude of this analytic signal yields signal envelope, $\xi[n]$, containing the upper, $\xi_{H}[n]$, and lower, $\xi_{L}[n]$, envelopes of $L[n]$ (Eq. \ref{eq:6}).

\begin{equation}
    \xi[n] = \left |z[n] \right |
    \label{eq:6}
\end{equation}

Envelopes are determined using spline interpolation over local maxima separated by at least $n_p = 60$ samples. We considered $n_{p} = 60$ since we did not witness in our previous studies \cite{adamatzky2018towards,adamatzky2019plant} fungal spikes of electrical potential shorter than 60 seconds\footnote{This threshold can be changed with respect to the context of experiments.}. We proposed Algorithm~\ref{alg:3} to locate candidate regions using signal envelope.

\RestyleAlgo{ruled}
\SetAlgoNoLine
\LinesNumbered
\begin{algorithm}[!htb]
\SetKwFunction{Lmin}{isLocalMinimum}\SetKwFunction{Lmax}{isLocalMaximum}
\SetKwInOut{Input}{Input}\SetKwInOut{Output}{Output}
\Input{$\xi[n]$ --- Envelope of signal $L[t]$,\\
        $n_p =60$ --- Minimum distance between two consecutive local extreme.}
\Output{$\mathcal{R}$ --- set of envelope-based ROIs.}
\BlankLine
\Begin{
    $\xi_{M}[n] = \left (\xi_{H}[n]+\xi_{L}[n]\right )/2$\;
    $[val_{min}, ind_{min}] =$ \Lmin{$\xi_{M}[n],n_p$}\;
    $[val_{max}, ind_{max}] =$ \Lmax{$\xi_{M}[n],n_p$}\;
    \tcp{\Lmin{} and \Lmax{} locate local minimum and maximum, respectively.}
    $j \gets$ index of the first local maximum whose value is greater than the value of the first local minimum\;
    \For{$i = 1$ \KwTo $\mathbf{card}(ind_{min})$}{
		\If{$j \leq \mathbf{card}(ind_{max})$}{
		    $\Delta \gets val_{max}(j) - val_{min}(i)$\;
		    Add $\left (ind_{min}(i), ind_{max}(j), \Delta \right )$ to $\mathcal{R}$\;
		    $j \gets j+1$\;
		}
	}
	\tcp{$\mathcal{R}$ has $j$ rows and $3$ columns, as $\mathcal{R}_{1}$, $\mathcal{R}_{2}$, and $\mathcal{R}_{3}$.}
	\BlankLine
	$\rho = \mathbf{mean}(\mathcal{R}_{3}) - \mathbf{std}(\mathcal{R}_{3})$\;
	\tcp{$\mathbf{mean}()$ and $\mathbf{std}()$ calculate the mean and standard deviation, respectively.}
	Remove the $k^{th}$ entry from $\mathcal{R}$ where $\mathcal{R}_{3}(k) < \rho$ -- see Fig.~\ref{fig:6}(b)\;
}
\BlankLine
\Return{$\mathcal{R}$}
\caption{Detecting candidate spike region from signal envelope.}
\label{alg:3}
\end{algorithm}
\clearpage
\begin{figure}[!htb]
    \centering
    \subfigure[]{\includegraphics[width=0.31\textwidth]{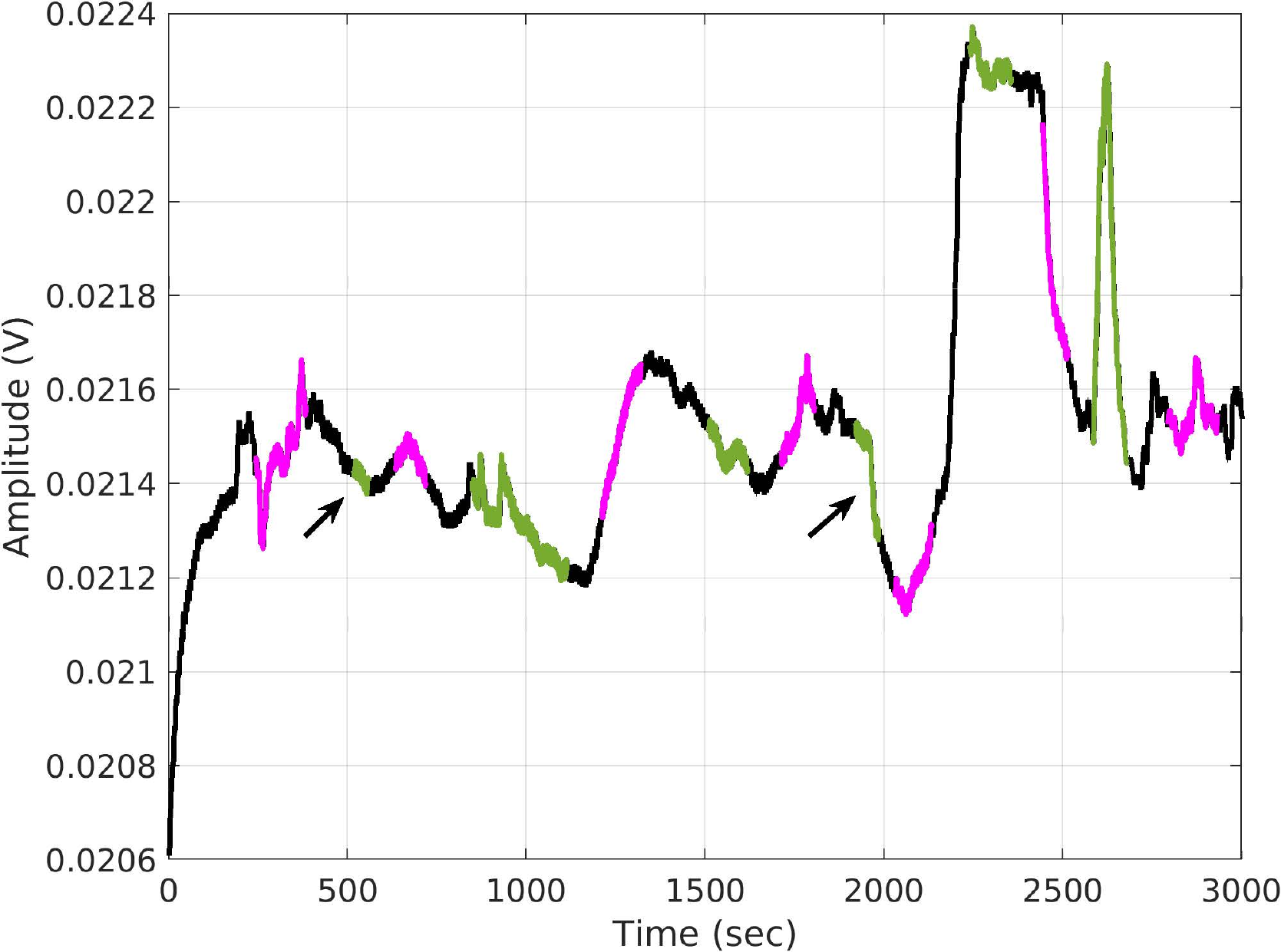}}
    \subfigure[]{\includegraphics[width=0.31\textwidth]{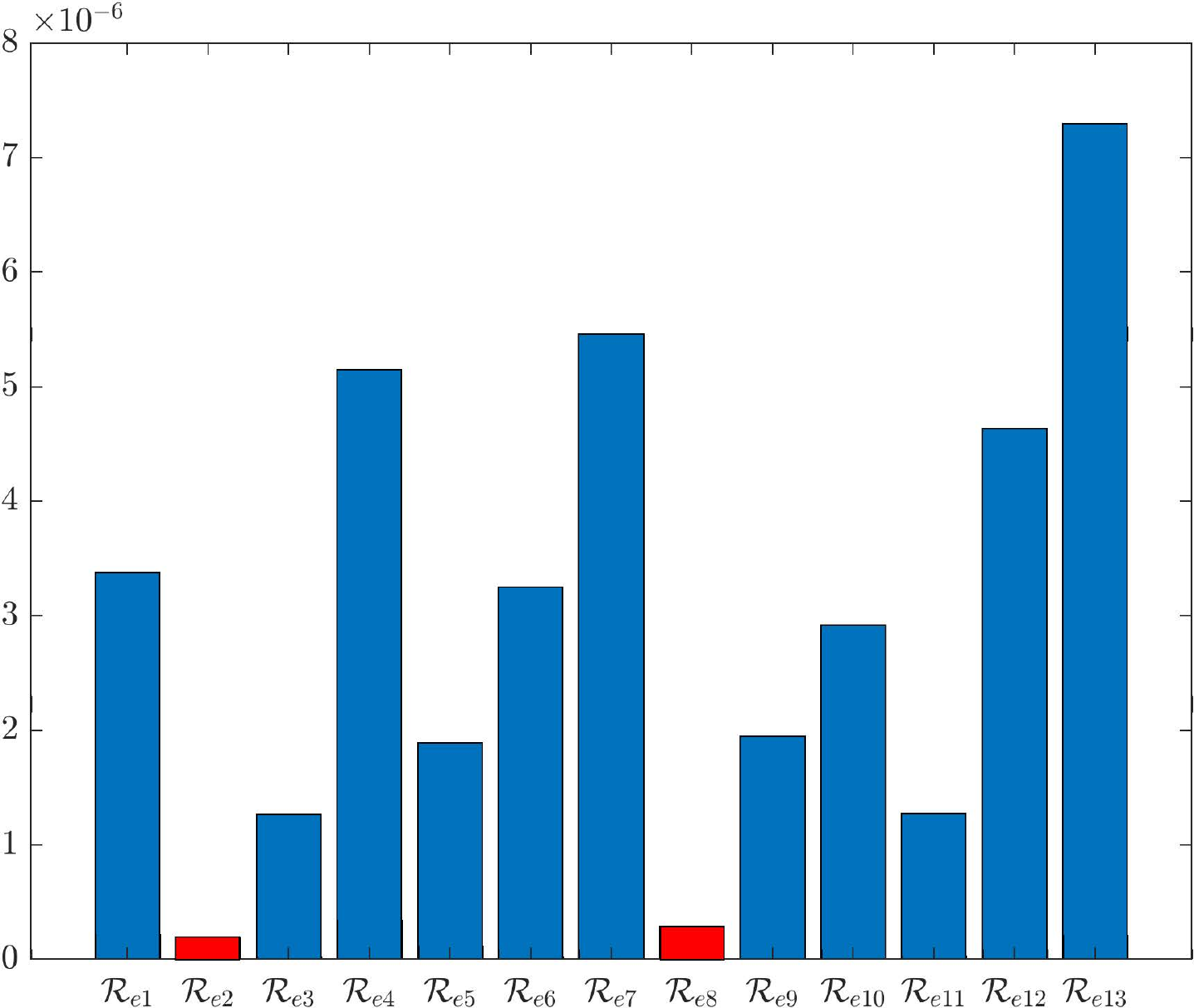}}
    \subfigure[]{\includegraphics[width=0.31\textwidth]{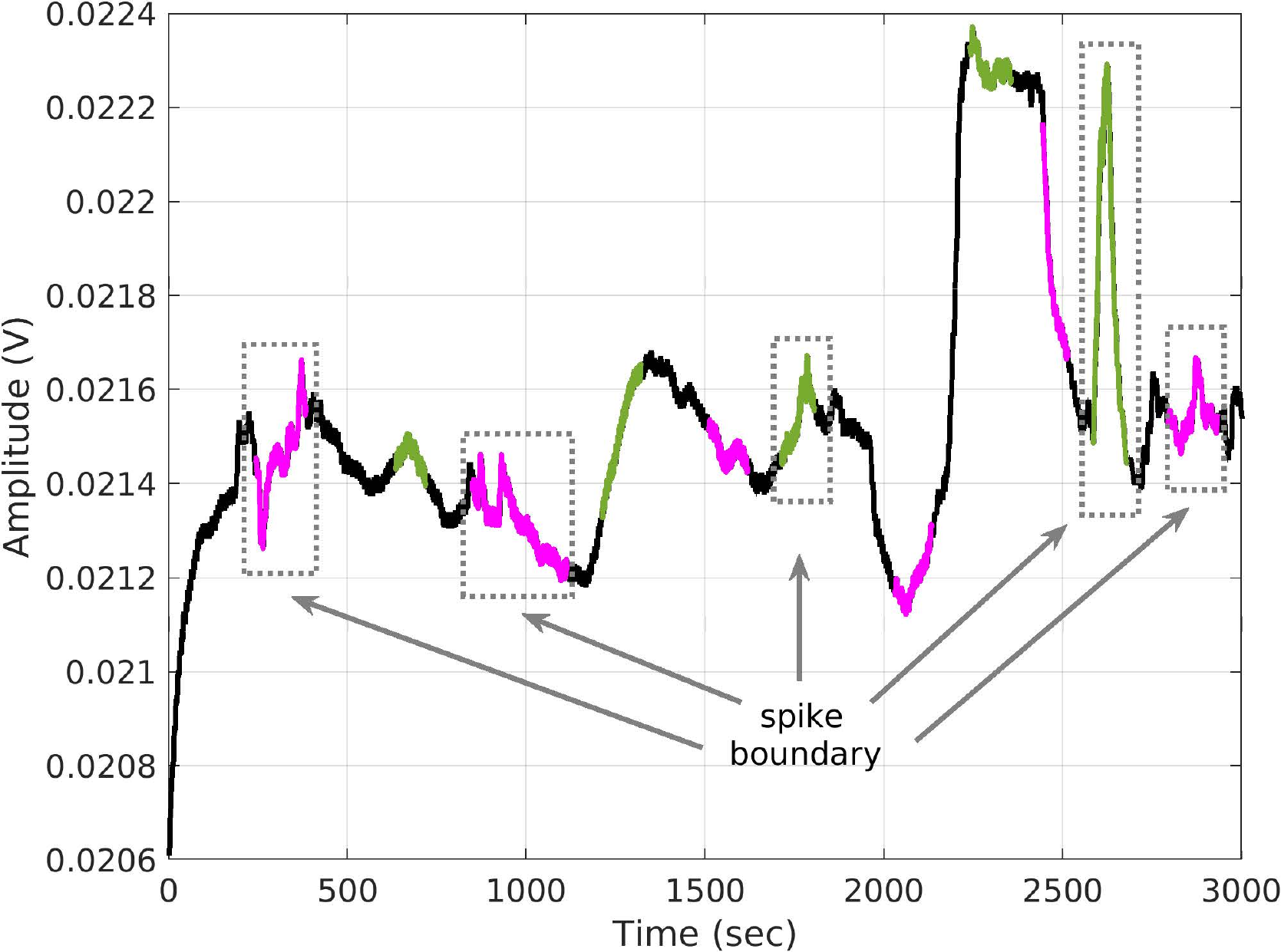}}
    \subfigure[]{\includegraphics[width=0.31\textwidth]{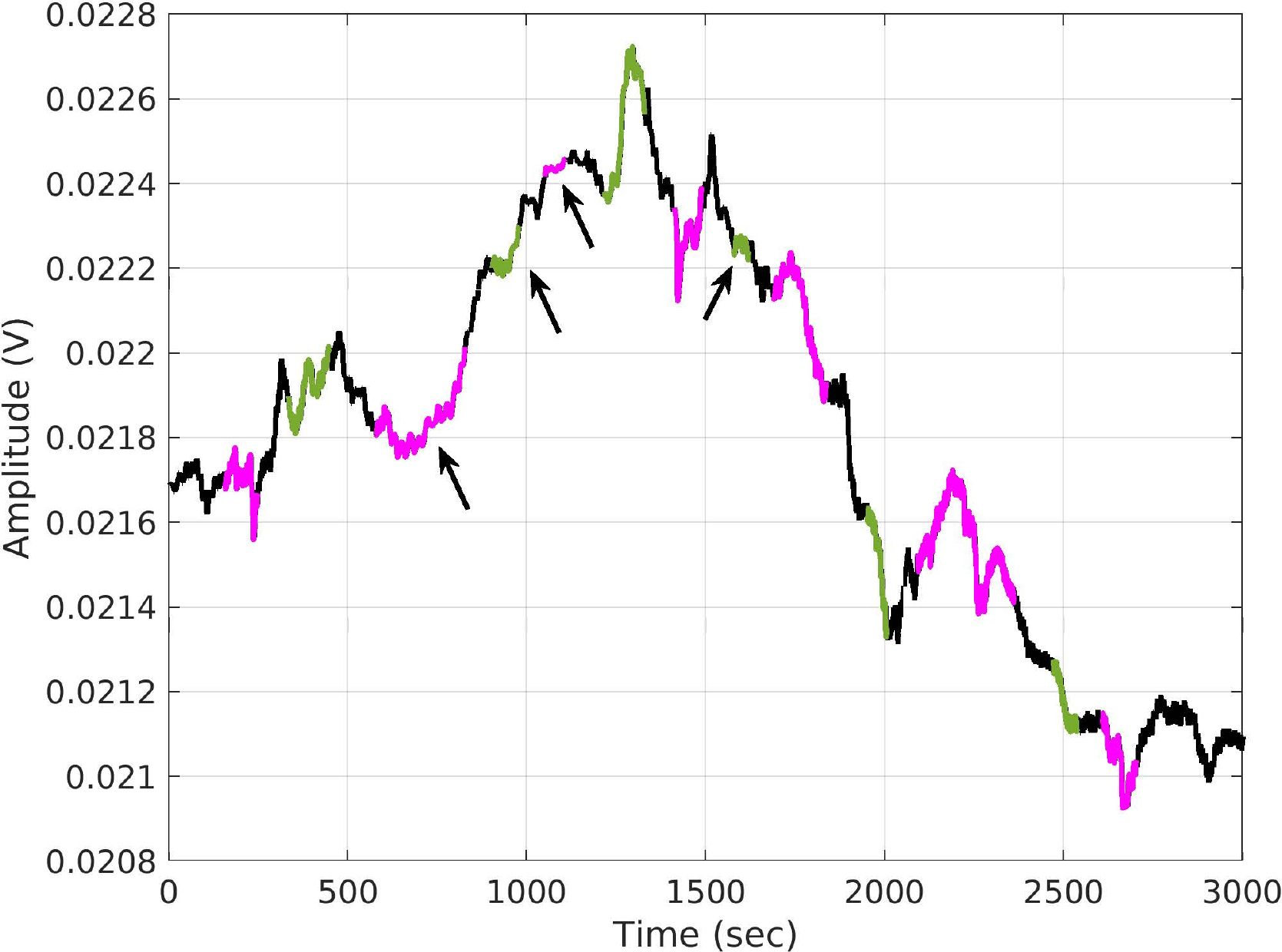}}
    \subfigure[]{\includegraphics[width=0.31\textwidth]{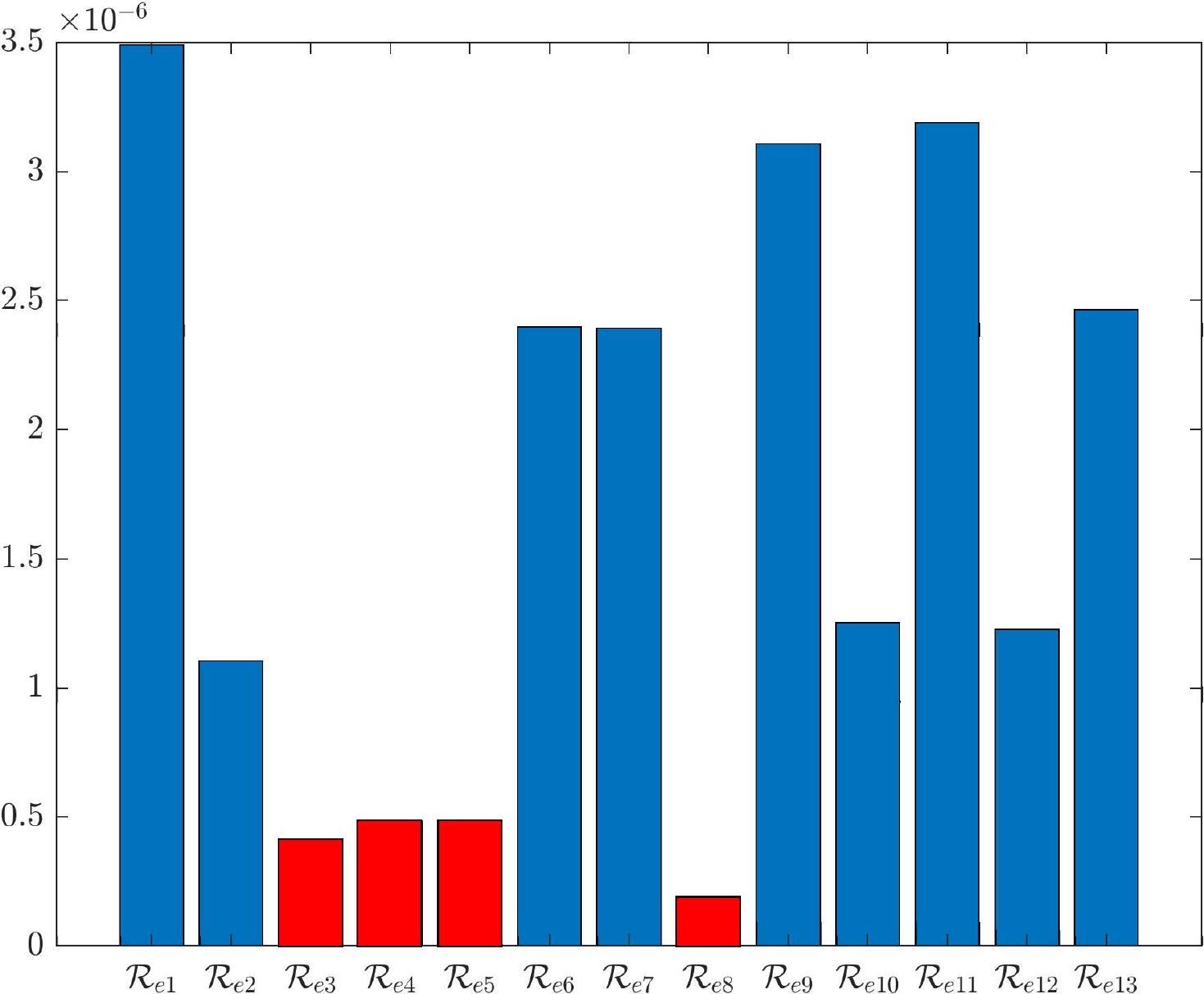}}
    \subfigure[]{\includegraphics[width=0.31\textwidth]{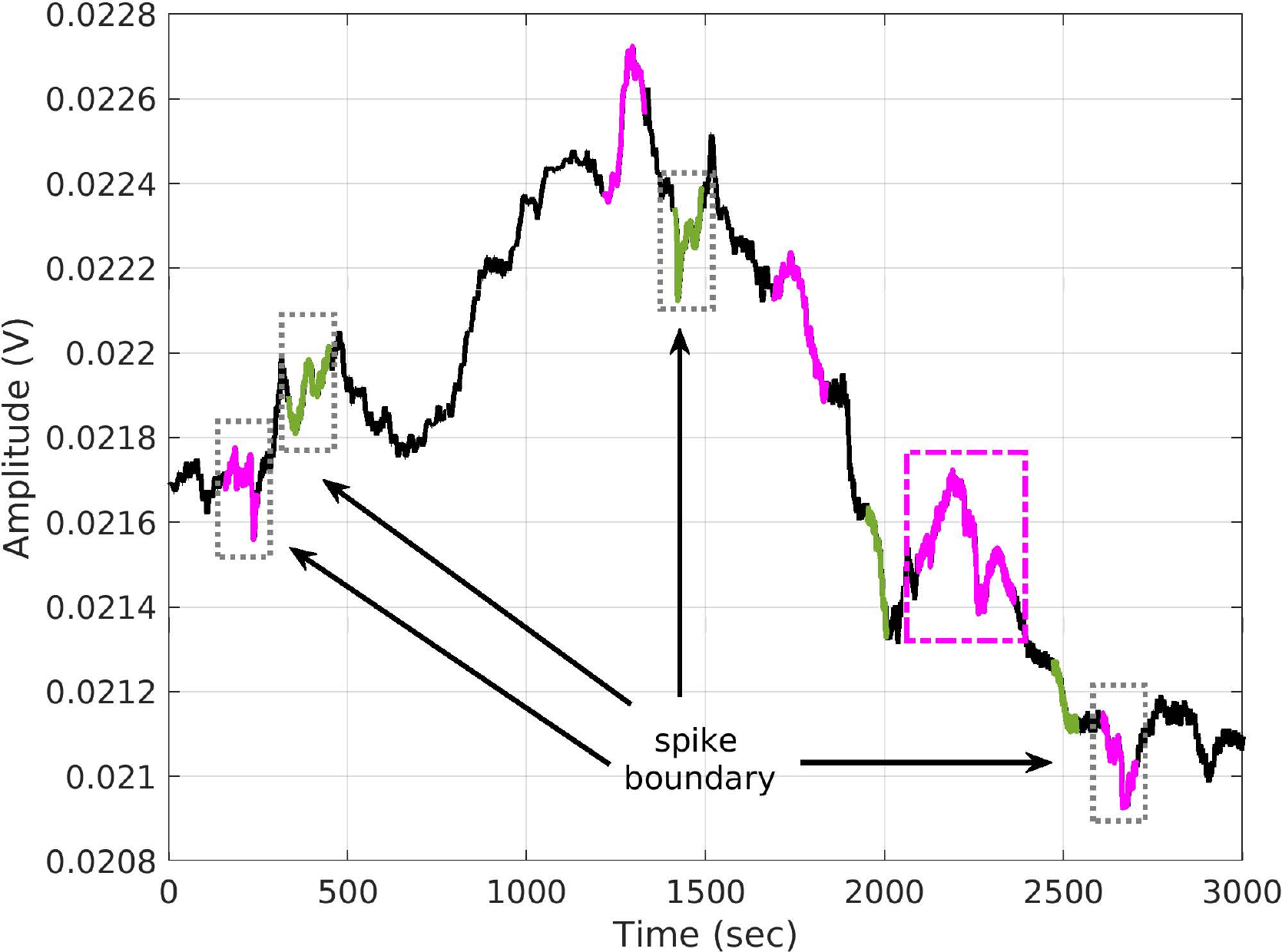}}
    \caption{Results of applying Algorithm~\ref{alg:3} to $Slice_1$ (first row) and $Slice_2$ (second row). (a,d) Candidate regions by finding local minima and maxima of the analytic signal envelope. The regions with arrows are also highlighted in red on the bar chart. (b,e) The absolute prominence difference of consecutive local minimum and maximum. Regions that do not satisfy $\mathcal{R}_{3}(k) < \rho$ are coloured in red. (c,f) Regions of Interest in $\mathcal{R}$. Gray rectangle with dash edge shows the correct spike, including repolarisation, depolarisation, and refractory periods. The purple dashed rectangle shows the region whose refractory period attached to a \emph{pseudo-spike} event.} \label{fig:6}
\end{figure}

Figure~\ref{fig:6}(a,d) shows candidate regions in $\mathcal{R}$ before applying Step 13. At this stage, although $\mathcal{R}$ includes regions that do not observe the spike definition (pointed by arrow in plot), the correctly identified spikes are consonant with our findings in~\cite{adamatzky2018towards,adamatzky2018spiking}. To eliminate non-spike regions, which are highlighted in red in Fig.~\ref{fig:6}(b,e), we applied Steps 13 and 14. Nevertheless, the output of Algorithm~\ref{alg:3} (see Fig~.\ref{fig:6}(c,f)) still contains regions that either belong to \emph{pseudo-spike}/\emph{inflection} regions or their refractory periods attached to a \emph{pseudo-spike} region. 

\begin{figure}[!htb]
    \centering
    \subfigure[]{\includegraphics[width=0.48\textwidth]{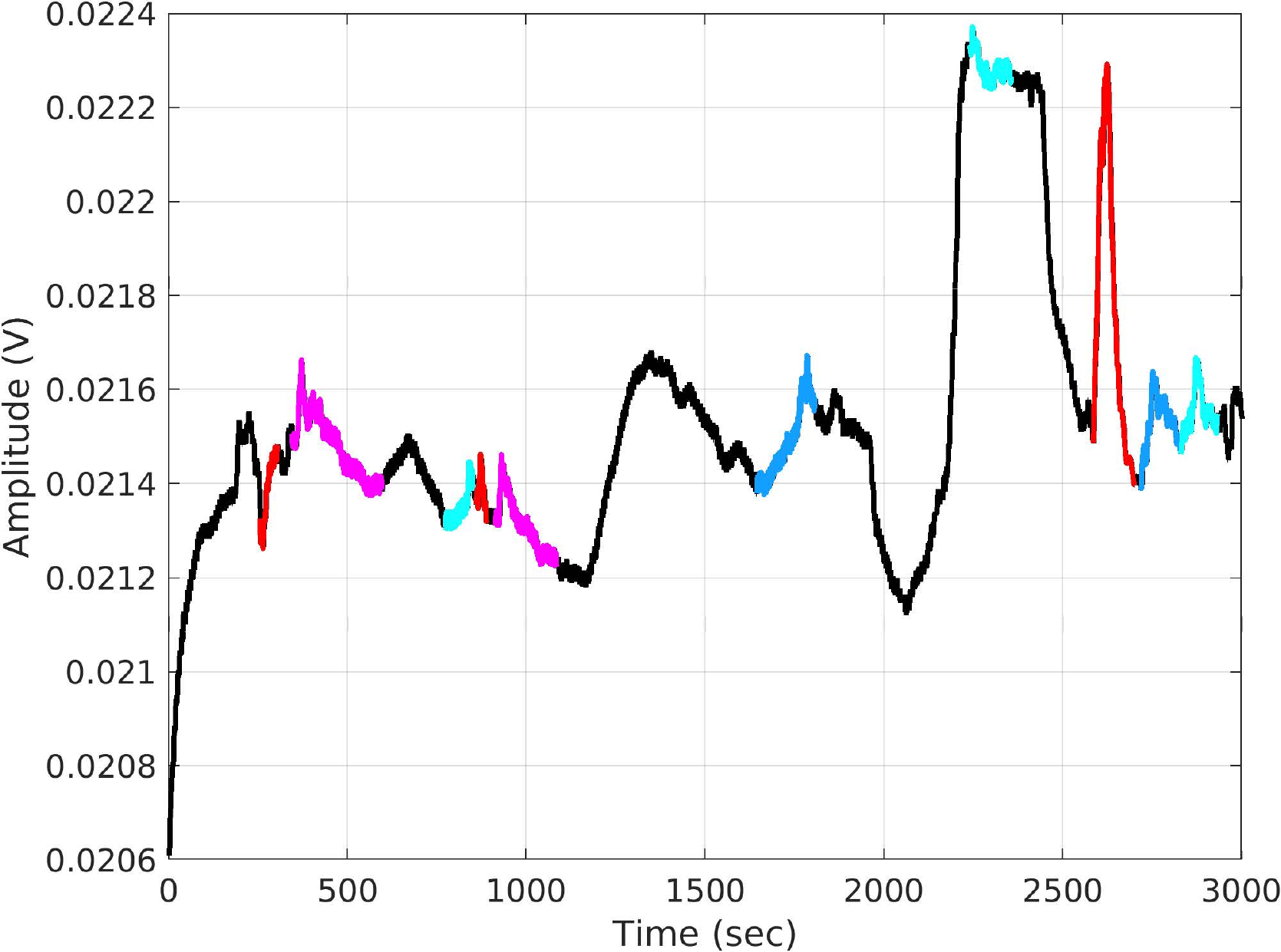}}
    \subfigure[]{\includegraphics[width=0.48\textwidth]{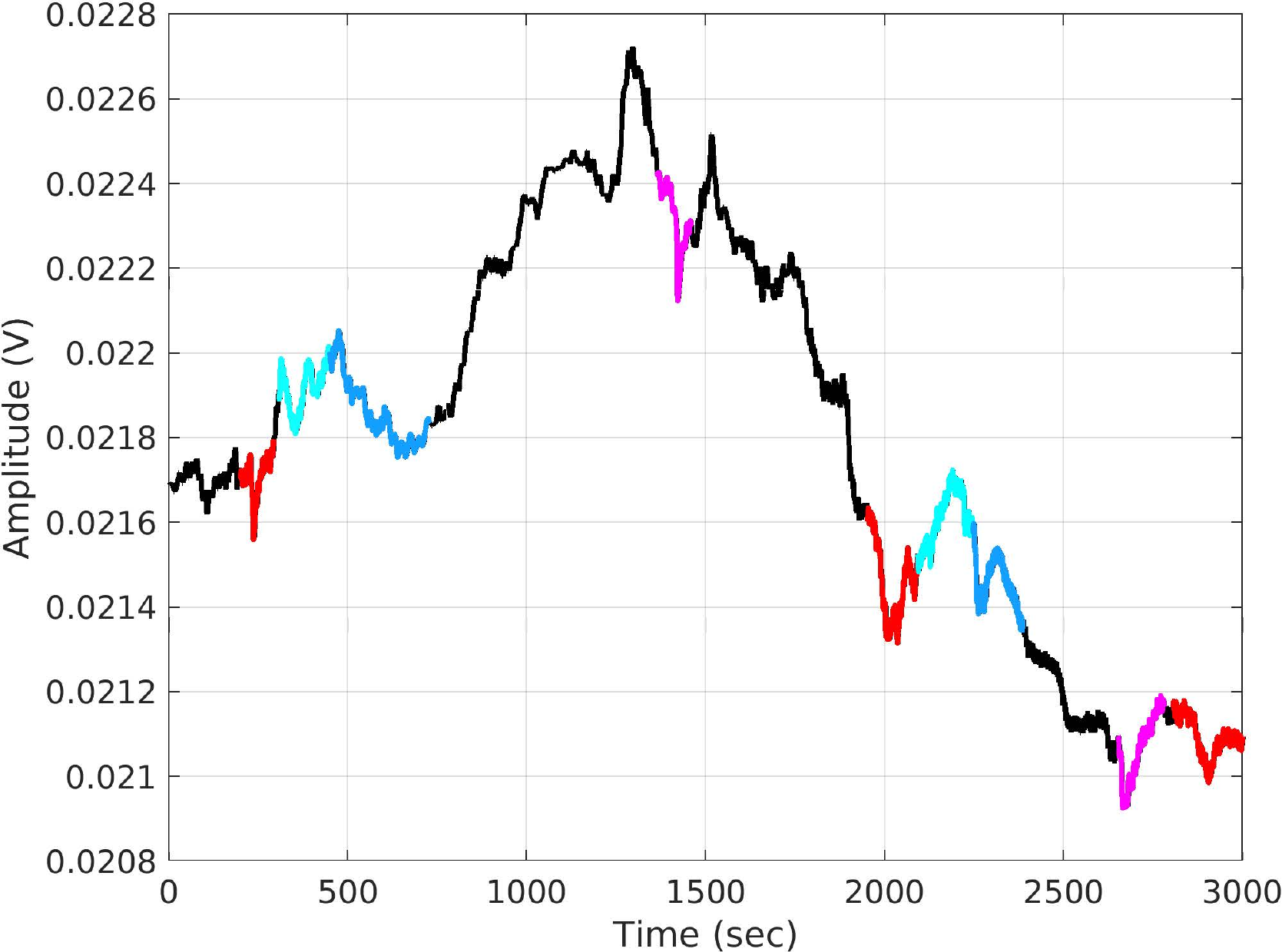}}
    \caption{Results of applying Algorithm~\ref{alg:4} to (a) $Slice_1$ and (b) $Slice_2$. We alternatively used red/purple for colouring spike events and blue/cyan for colouring \emph{pseudo-spike} events.} \label{fig:7}
\end{figure}

To resolve issues in Algorithms~\ref{alg:2} and \ref{alg:3}, we proposed Algorithm~\ref{alg:4} in which regions belong to $(\mathcal{C} \cup \mathcal{D})$ are used in updating $\mathcal{R}$. If any ROI in $\mathcal{R}$ is a subset of $(\mathcal{C} \cup \mathcal{D})$, it is added to the spike event set, $\mathcal{F}_{s}$, with an updated length. If any ROI in $(\mathcal{C} \cup \mathcal{D})$ is a subset of $\mathcal{R}$, it is added to the \emph{pseudo-spike} set, $\mathcal{F}_{p}$. In a case of having intersection without observing subset condition, we split the concatenation of ROIs from the intersection point into two slices. Then, the slice with the minimum length is added to $\mathcal{F}_{p}$. Finally, regions with a length of fewer than 60 seconds are removed from $\mathcal{F}_{s}$ and $\mathcal{F}_{p}$. Figure~\ref{fig:7} shows results of applying Algorithm~\ref{alg:4}.

\RestyleAlgo{ruled}
\SetAlgoNoLine
\LinesNumbered
\begin{algorithm}[!htb]
\SetKwFunction{Int}{intersect}
\SetKwInOut{Input}{Input}\SetKwInOut{Output}{Output}
\Input{$\mathcal{C}, \mathcal{D}, \mathcal{R}$ --- Regions of interest.}
\Output{$\mathcal{F}_{s}, \mathcal{F}_{p}$ --- Fungi \emph{spike} and \emph{pseudo-spike} events, respectively.}
\BlankLine
\Begin{
    \ForEach{$r_{e} \in \mathcal{R}$}{
        $chunk_{e} \leftarrow [r_{e}^{1} \cdots r_{e}^{2}]$\;
        \ForEach{$r_{w} \in (\mathcal{C} \cup \mathcal{D})$}{
            $chunk_{w} \leftarrow [r_{w}^{1} \cdots r_{w}^{2}]$\;
            \Switch{$chunk_{w},chunk_{e}$}{
                \Case{$chunk_{e} \subset chunk_{w}$}{
                    $chunk_{w}(end) = chunk_{e}(end)$\;
                    $\mathcal{F}_{s} \leftarrow chunk_{w}$\;
                }
                \Case{$chunk_{w} \subset chunk_{e}$}{
                    $\mathcal{F}_{p} \leftarrow chunk_{w}$\;
                }
                \Case{\Int{$chunk_{w}, chunk_{e}$}}{
                \tcp{\Int{} checks if two chunks have an intersection point.}
                    \textbf{Split} the concatenation of $chunk_{w}$ and $chunk_{e}$ from intersection point into two \emph{sub-Chunks}\;
                    $\mathcal{F}_{p} \leftarrow$ \emph{sub-Chunks}\;
                }
            }
	    }
	}
	\BlankLine
    \ForEach{$r \in (\mathcal{F}_{s} \cup \mathcal{F}_{p})$}{
        \textbf{Remove} $r$ if $|r|< 60$\;
    }
}
\BlankLine
\Return{$\mathcal{F}_{s}, \mathcal{F}_{p}$}
\caption{Extracting fungi \emph{spike} and \emph{pseudo-spike} events.}
\label{alg:4}
\end{algorithm}

\section{Experimental results} \label{sec:4}

This section comprises of objective and complexity analyses. In the objective analysis, we showed the efficiency of the spike event detection method in comparison with the existing, in neuroscience, techniques of spike detection~\cite{nenadic2004spike,shimazaki2010kernel} and the expert opinion in locating spikes' arrival time. In the complexity analysis, we selected complexity measures used in previous studies ~\cite{taghipour2016complexity,adamatzky2019exploring,casali2013theoretically,schartner2017increased} to quantify activity patterns that are spatio-temporally integrated and differentiated. 

\subsection{Objective analysis}

Various methods have been proposed to detect and sort spike events in EC recordings~\cite{quiroga2004unsupervised,nenadic2004spike,obeid2004evaluation,wilson2002spike,gotman1991state,wilson1999spike,franke2010online,racz2020spike,wang2020novel,sablok2020interictal,liu2020robust}. However, only a few of these methods do not require auxiliary information like the construction of templates and the supervised setting of thresholds to detect and sort spike events~\cite{nenadic2004spike,shimazaki2010kernel}. Nenadic and Burdick \cite{nenadic2004spike} developed an unsupervised method to detect and localise spikes in noisy neural recordings. This method benefits from continues wavelet transform. They applied multi-scale decomposition of the signal using `bior1.3,' `bior1.5,' `Haar,' or `db2'  wavelet basis. To assess the presence of spikes, they separated the signal and noise at each scale and performed Bayesian hypothesis testing. Finally, they combined decisions at different scales to estimate the arrival times of individual spikes. 

\begin{figure}[!htb]
    \centering
    \subfigure[]{\includegraphics[width=0.48\textwidth]{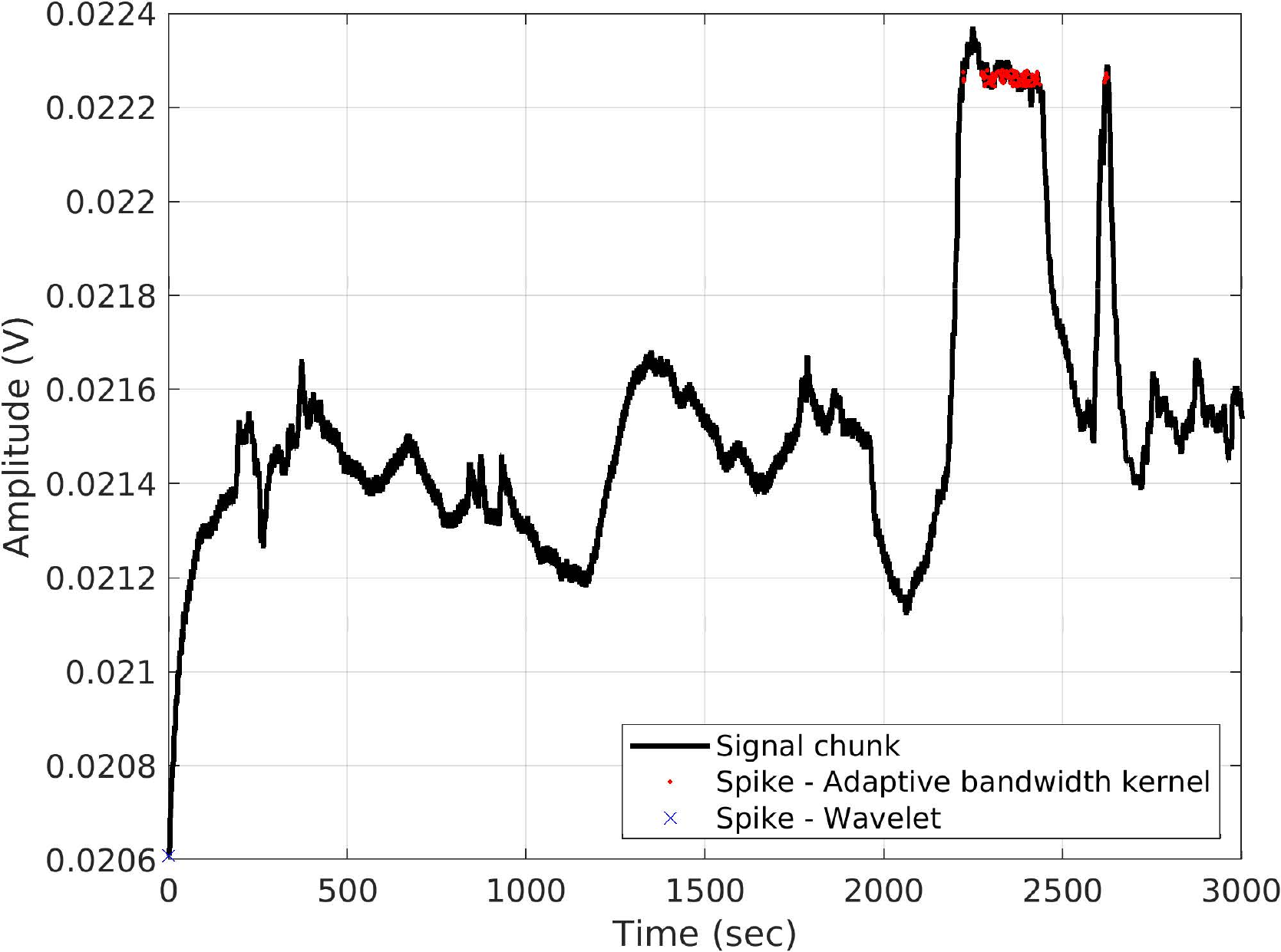}}
    \subfigure[]{\includegraphics[width=0.48\textwidth]{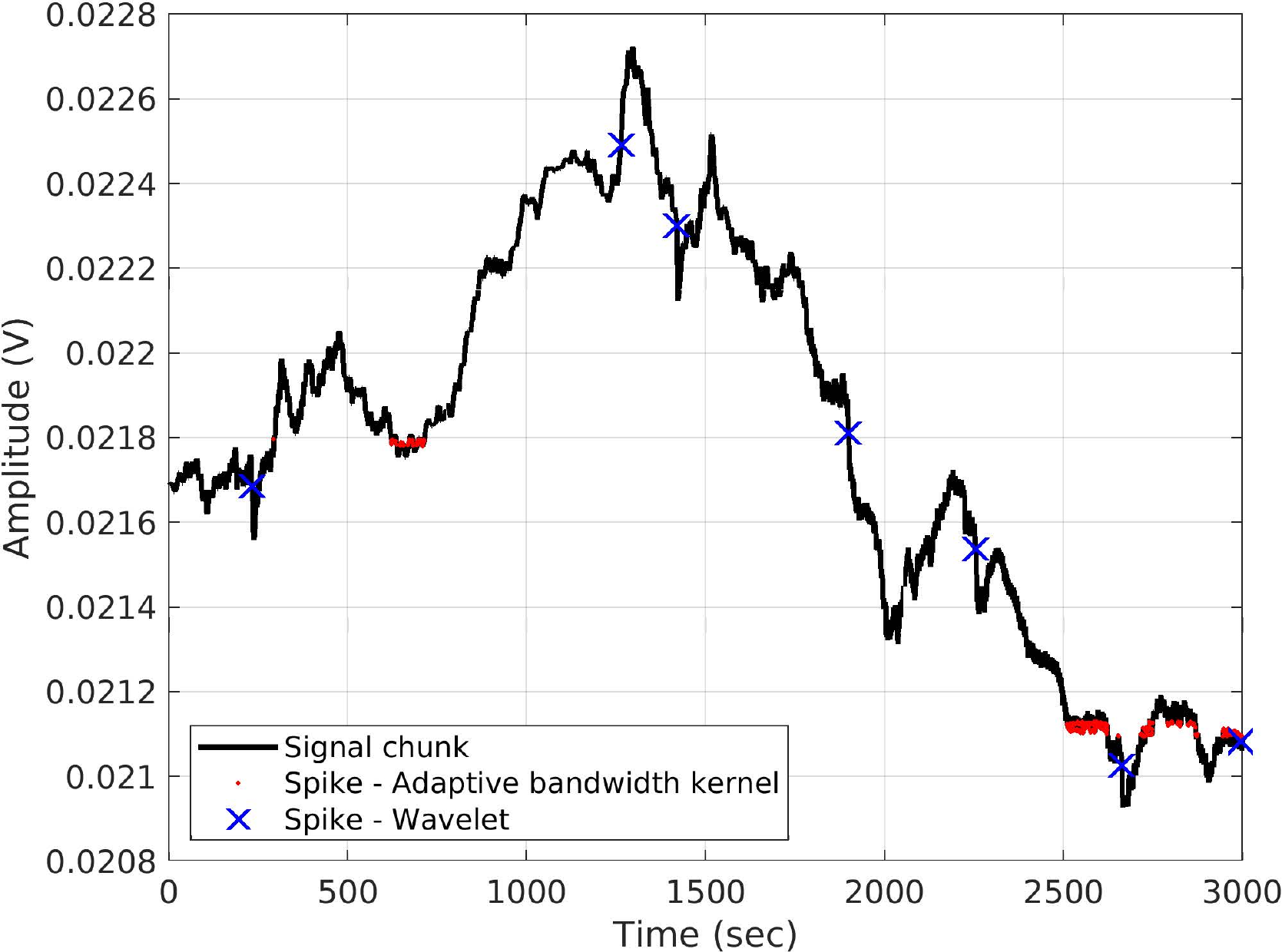}}
    \caption{Results of applying proposed algorithms in \cite{nenadic2004spike,shimazaki2010kernel} to (a) $Slice_1$ and (b) $Slice_2$. Note that the wavelet-based method can only locate spike arrival time. The kernel bandwidth optimisation can, however, extract the spike region.} \label{fig:8}
\end{figure}

Shimazaki and Shinomoto~\cite{shimazaki2010kernel} proposed an optimisation technique for selecting the bin width of the time-histogram. This optimisation minimised the mean integrated square in the kernel density estimation. This method benefited from variable kernel width, which allowed grasping non-stationary phenomena, and stiffness constant to avoid possible overfitting due to excessive freedom in the bandwidth variability. The estimated bandwidth was then used to filter spike event regions from the signal. Figure~\ref{fig:8} shows the results of applying these methods to two chunks with a length of 3000 seconds.

Both methods could not correctly detect all spike events that were located by the expert. Wavelet-based method could locate three spikes in Fig.~\ref{fig:8}(b) without detecting any spike in Fig.~\ref{fig:8}(a). The adaptive bandwidth kernel-based method could detect one spike in Fig.~\ref{fig:8}(b) and one pseudo-spike in Fig.~\ref{fig:8}(a) and (b). While our proposed method wrongly introduced one spike event in Fig.~\ref{fig:7}(a) and had a total-error (wrong- or non-detection) of three in Fig.~\ref{fig:7}(b). 

\begin{figure}[!htb]
    \centering
    \subfigure[]{\includegraphics[width=0.75\textwidth]{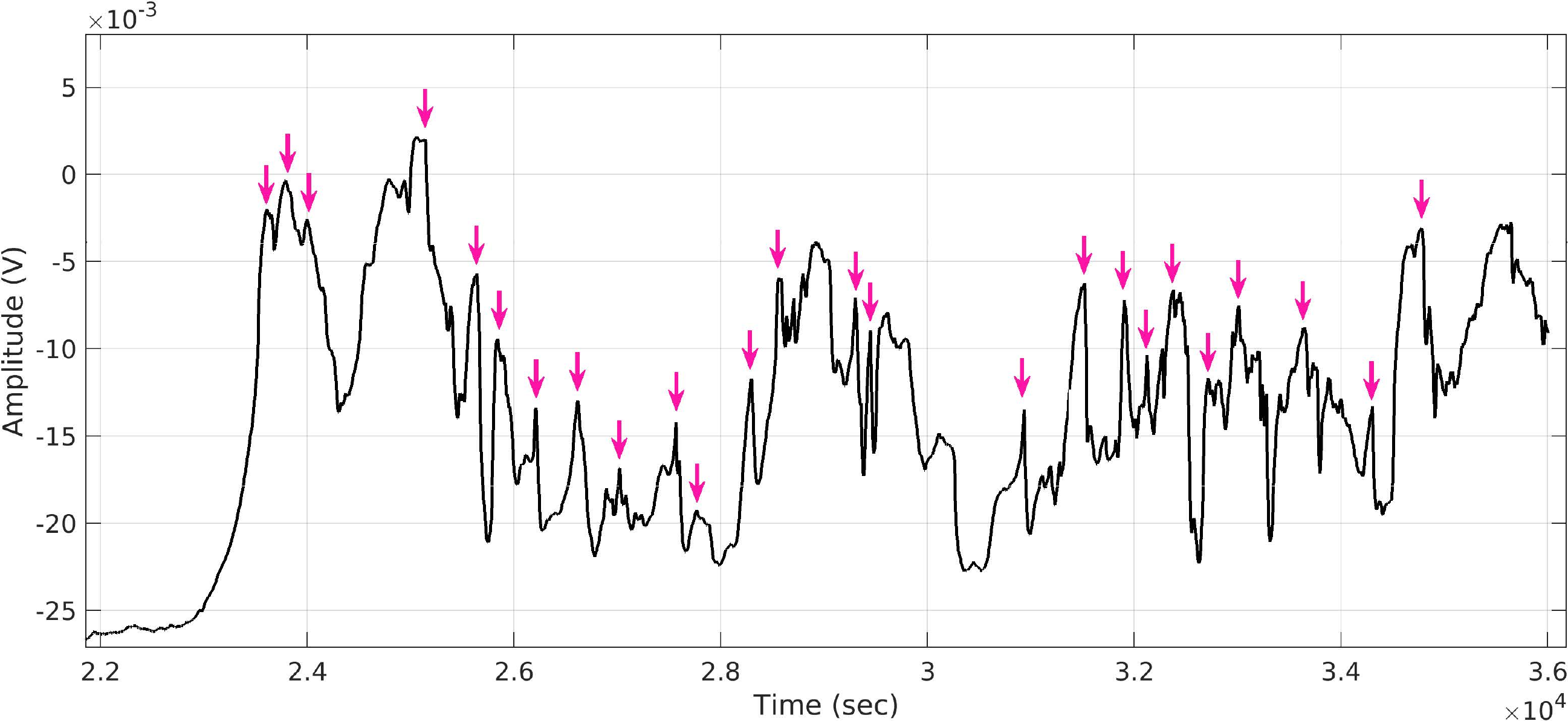}}
    \subfigure[]{\includegraphics[width=0.75\textwidth]{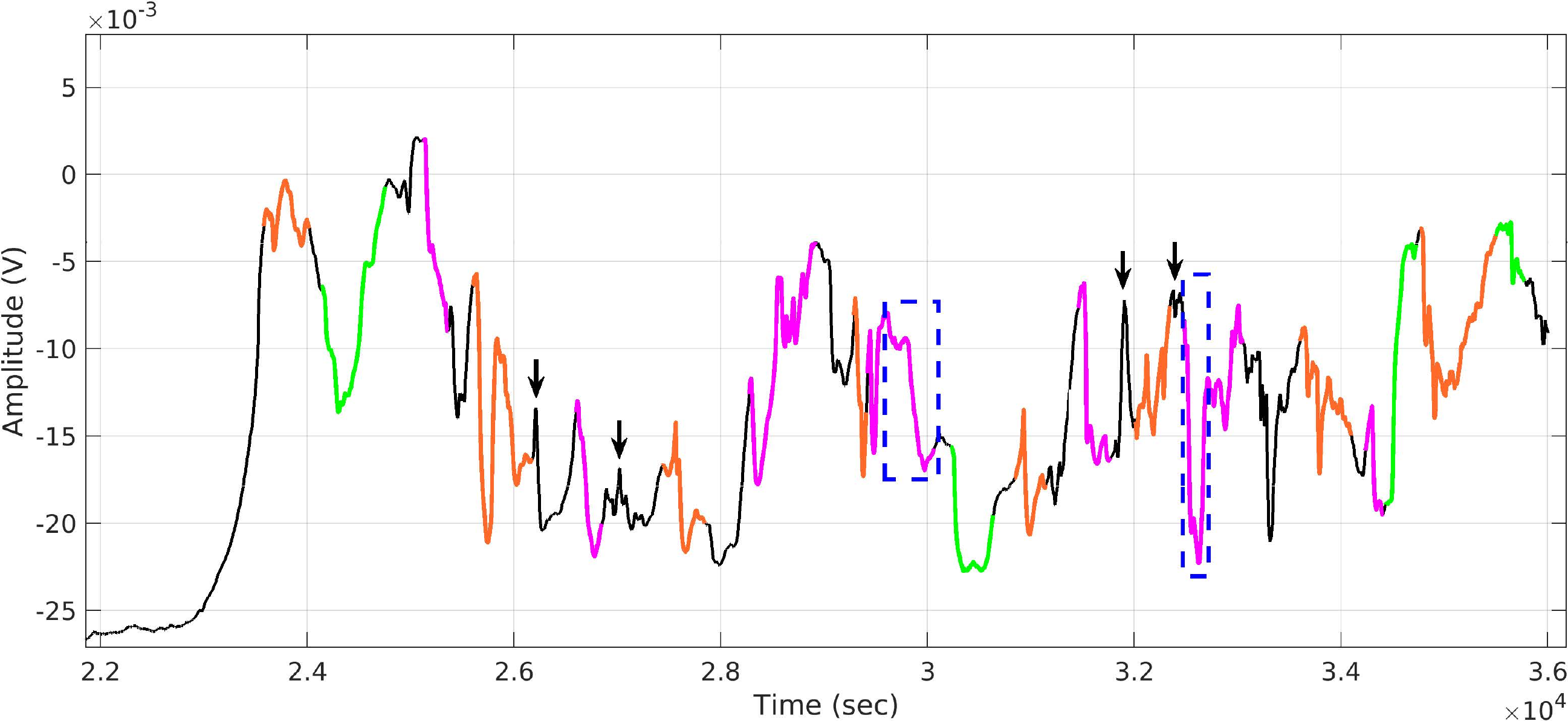}}
    \caption{(a) Spike arrival time located by the expert. Here we used augmented pink arrow to point to these spikes. (b) Spike regions extracted by the proposed method. Spike regions are alternatively coloured in orange and violet. The green areas point to \emph{pseudo-spike} regions that are mistaken for spikes. Blue rectangles with dash edge show overestimated refractory periods. We used black arrows to point to the missed spikes.} \label{fig:15}
\end{figure}

We also compared the proposed method with the expert opinion on a randomly selected 36,000-second chunk, \textit{i.e.}, 10 hours of electrical activity recordings. In this quantitative comparison, the proposed method could correctly locate 21 spikes, introduce four pseudo-spike events, overestimate two refractory periods; resulting in the true-positive and false-positive rates of 76\% and 16\%, respectively. Figure~\ref{fig:15}(a) shows located spikes by the expert and Fig.~\ref{fig:15}(b) indicates the results of the proposed spike detection method.

\begin{figure}[!htb]
    \centering
    \subfigure[]{\includegraphics[width=0.32\textwidth]{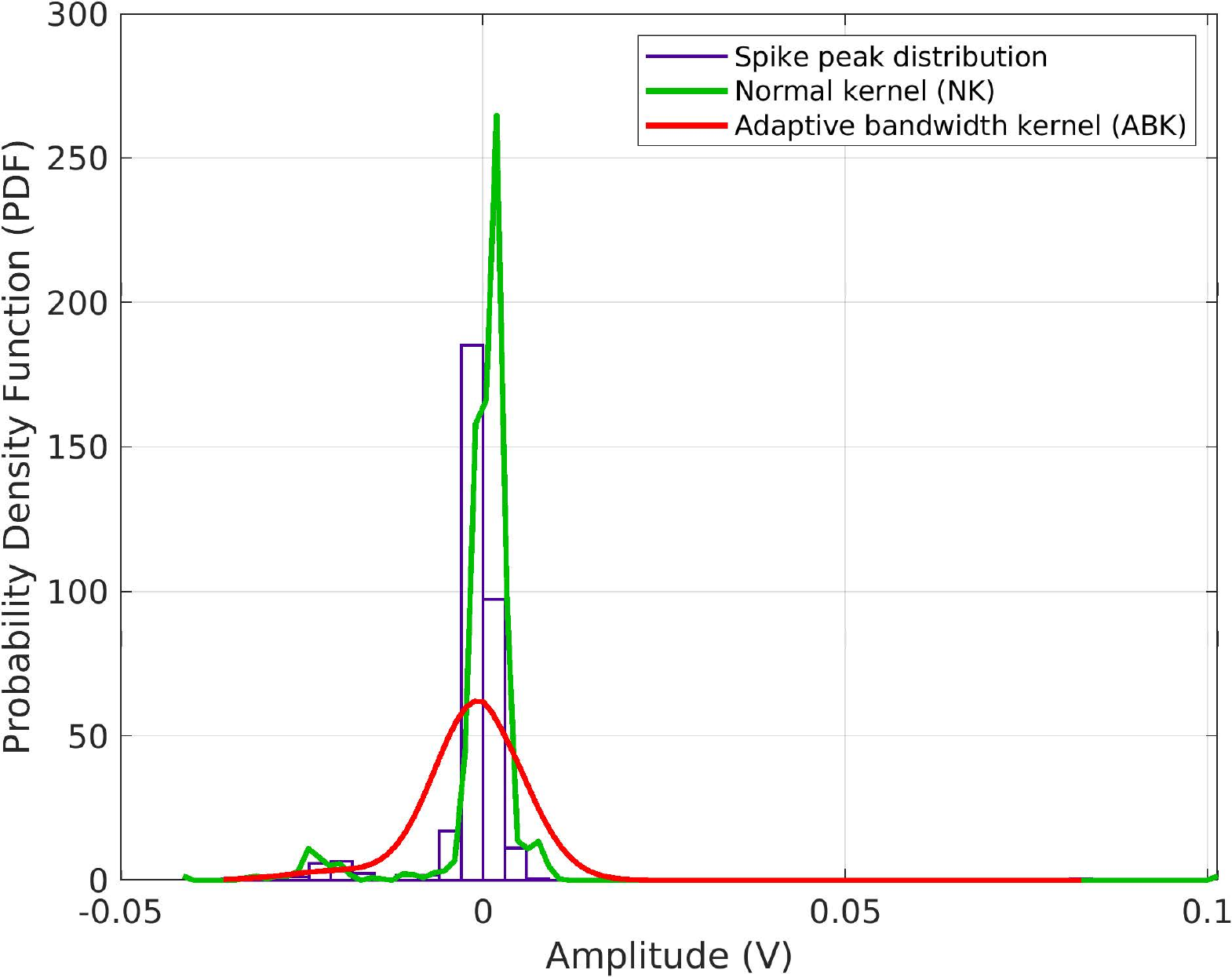}}
    \subfigure[]{\includegraphics[width=0.32\textwidth]{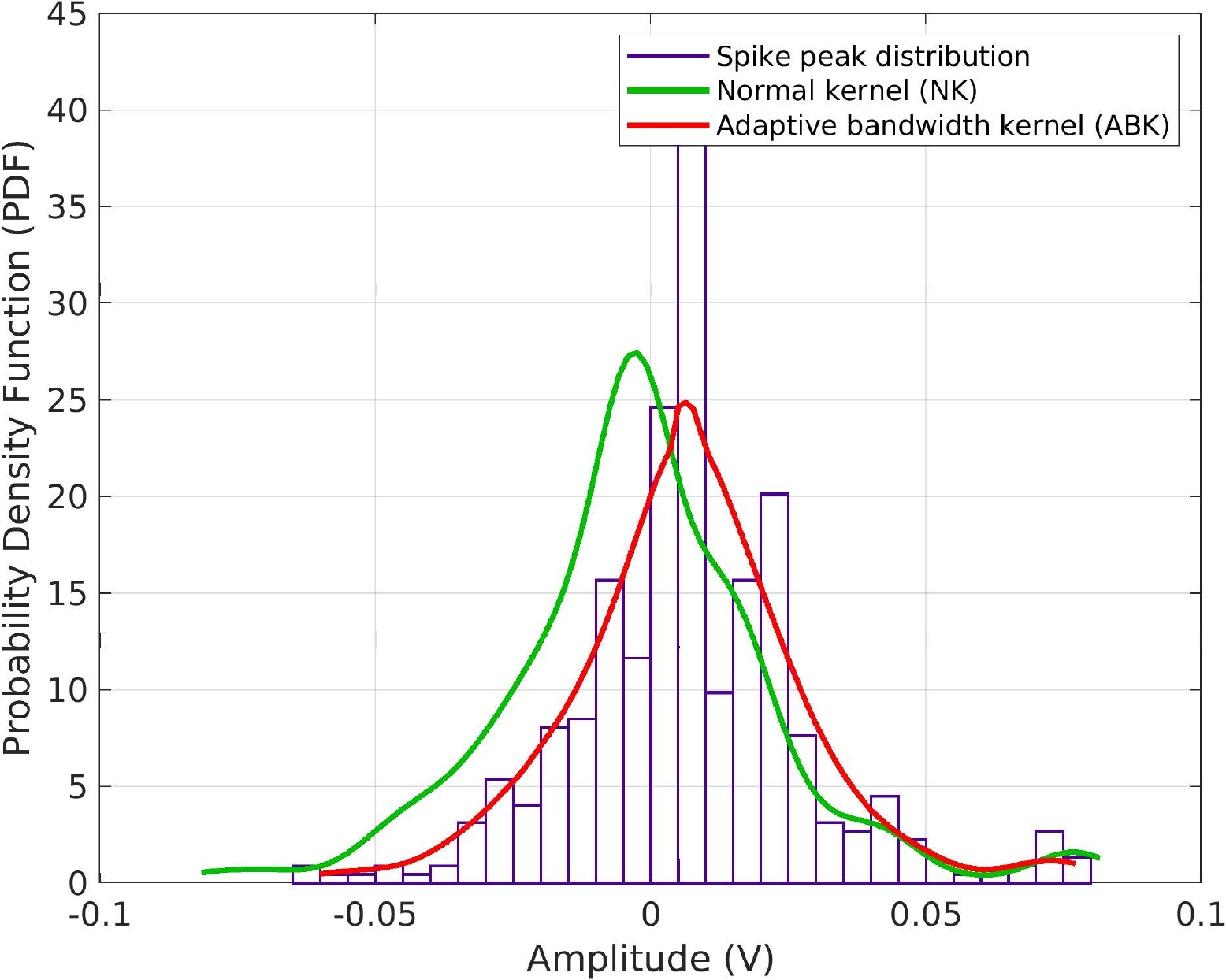}}
    \subfigure[]{\includegraphics[width=0.32\textwidth]{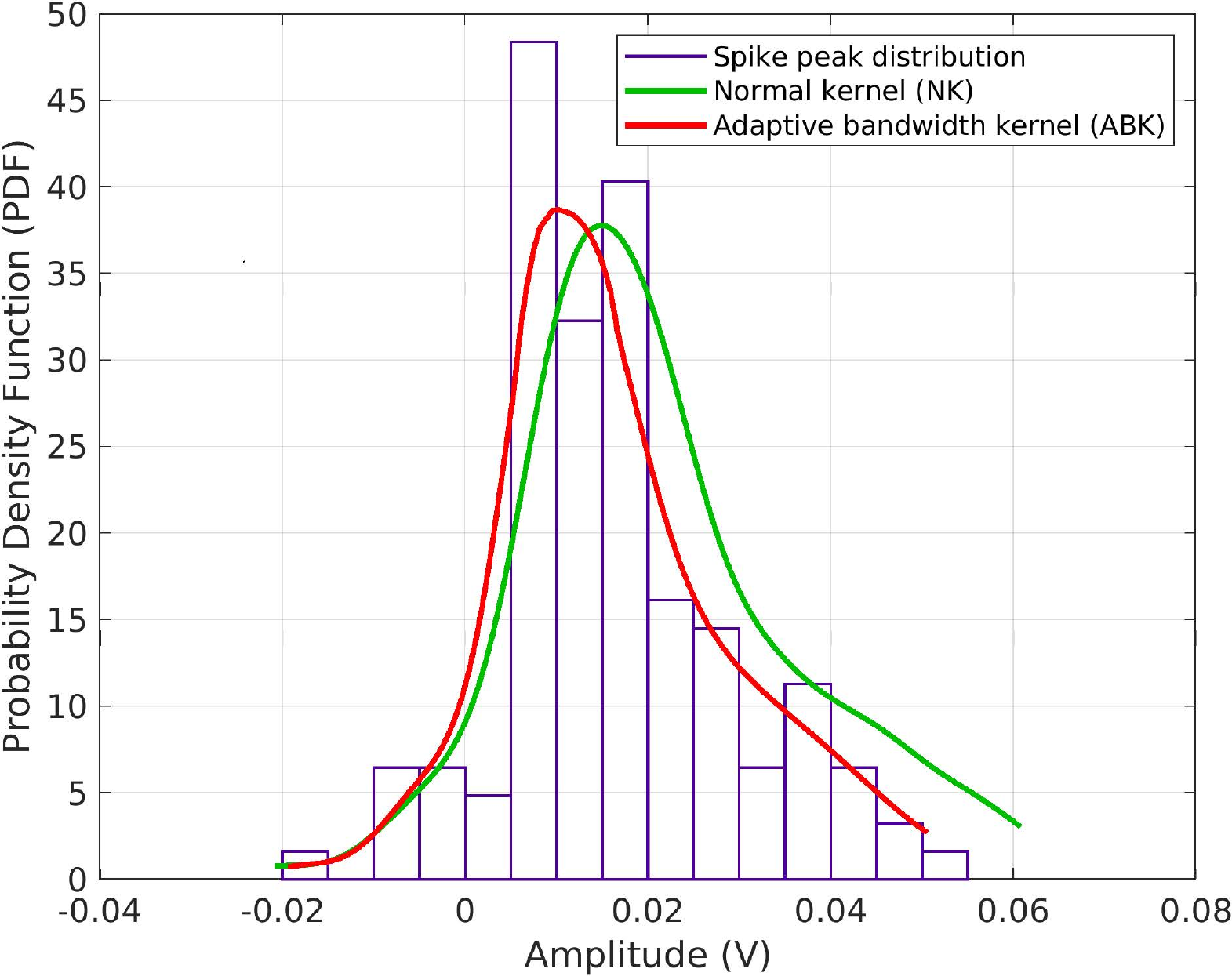}}
    \subfigure[]{\includegraphics[width=0.32\textwidth]{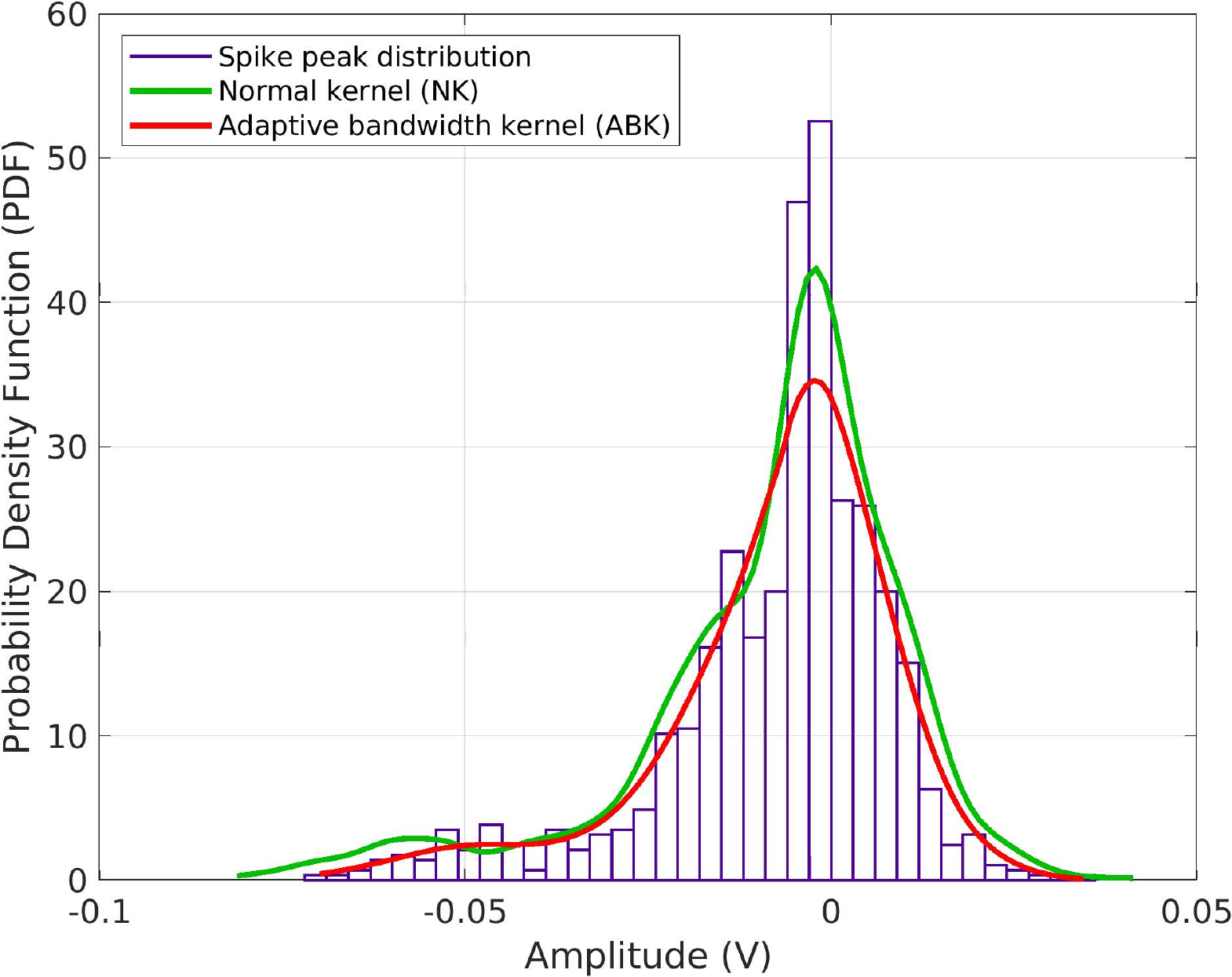}}
    \subfigure[]{\includegraphics[width=0.32\textwidth]{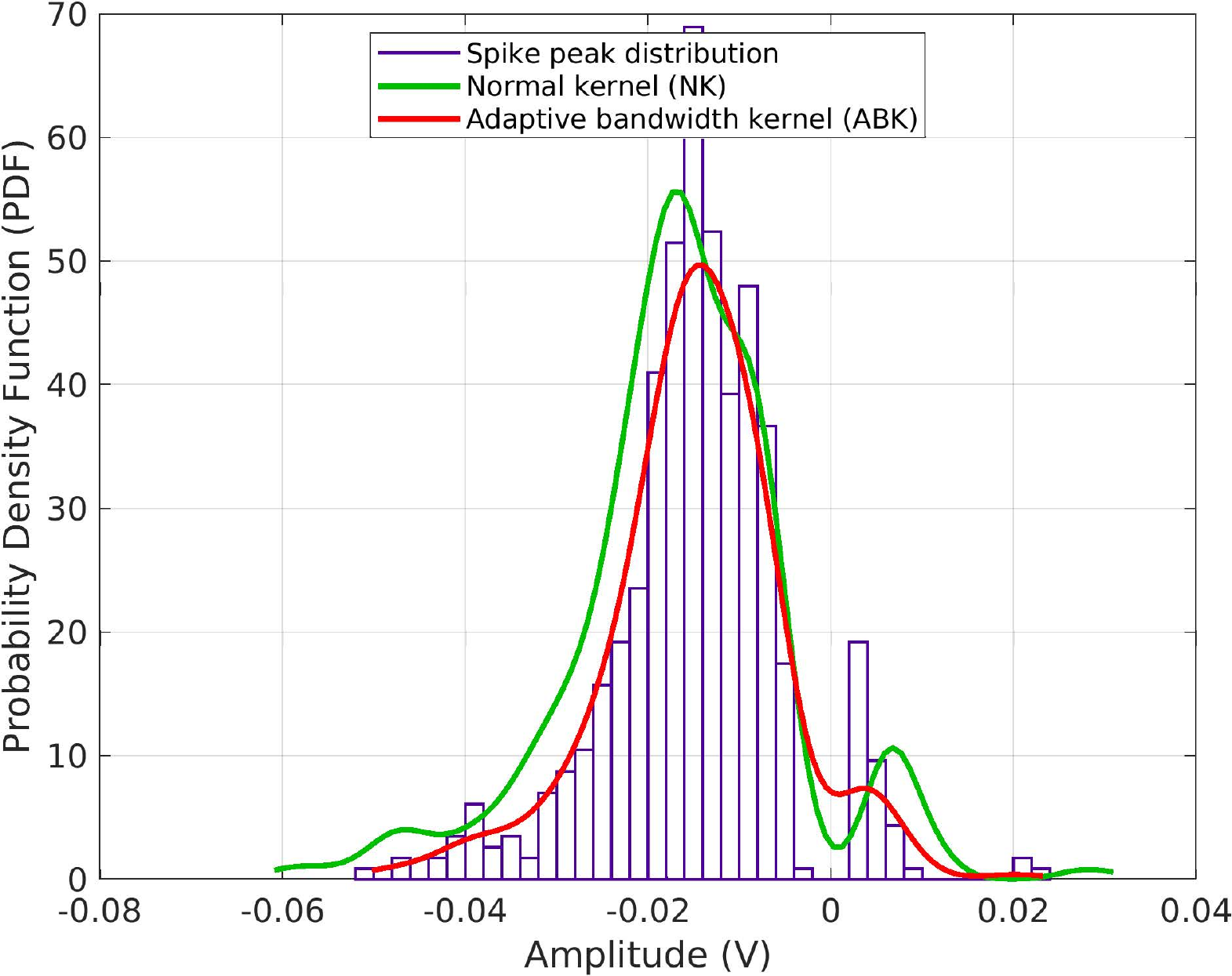}}
    \subfigure[]{\includegraphics[width=0.32\textwidth]{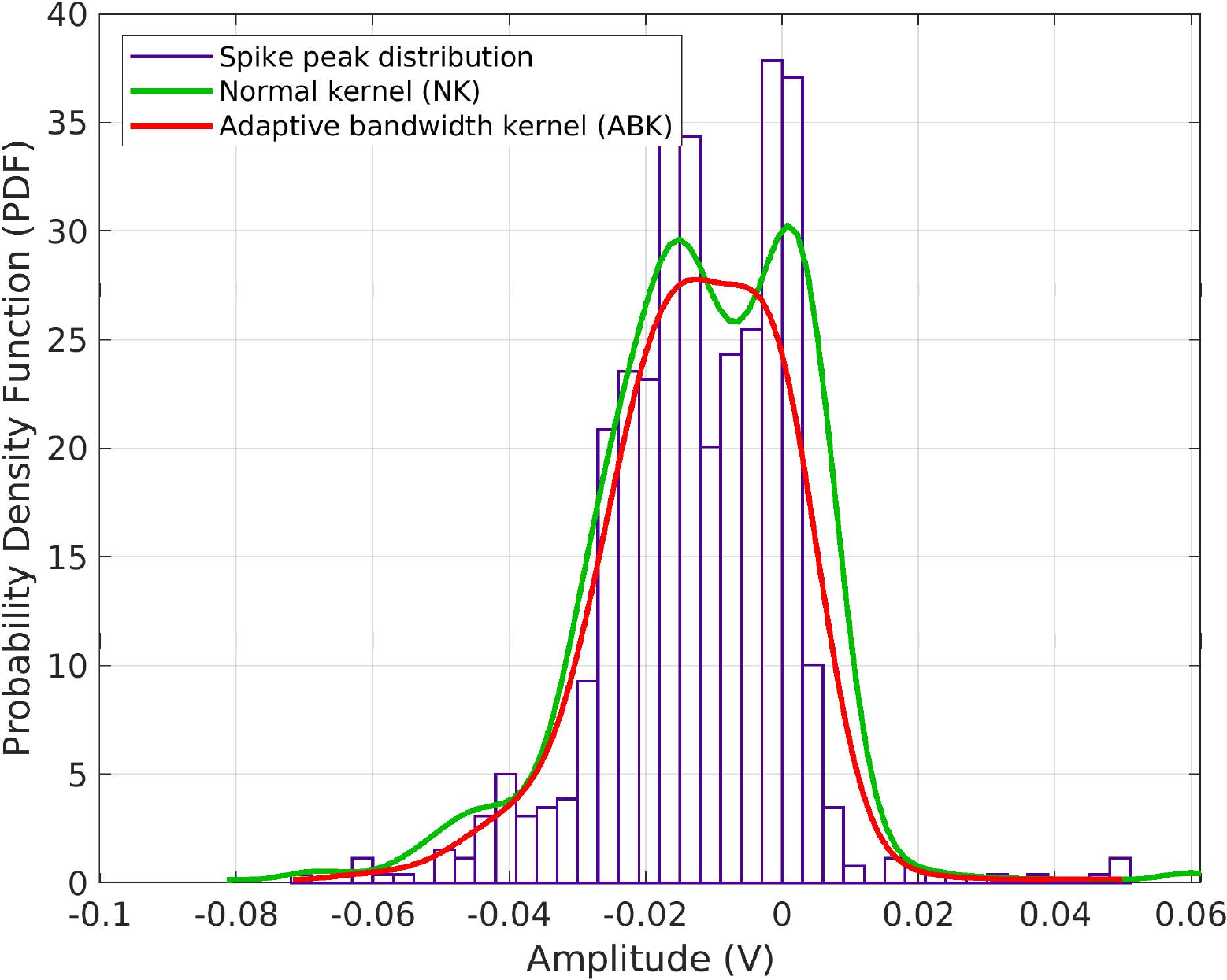}}
    \caption{Distribution of spike event lengths with superimposed  Gaussian and Adaptive bandwidth kernels~\cite{shimazaki2010kernel}. (a,b)~In-line electrode arrangements with a distance of 1~cm. (c,d)~In-line electrode arrangements with a distance of 2~cm. (e,f)~Random electrode arrangements with an approximate distance of 2~cm.}
\label{fig:9}
\end{figure}

\begin{figure}[!htb]
    \centering
    \subfigure[]{\includegraphics[width=0.32\textwidth]{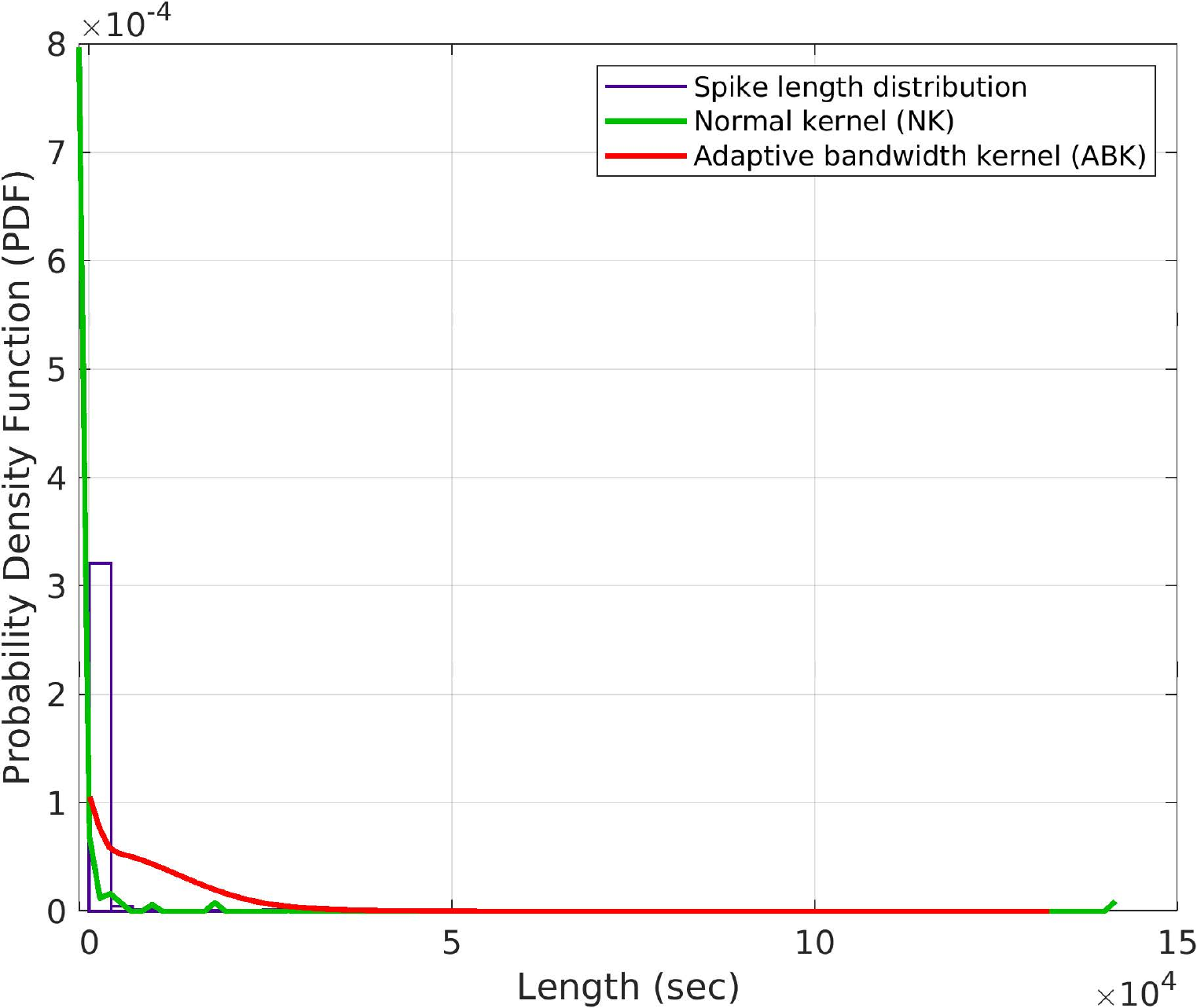}}
    \subfigure[]{\includegraphics[width=0.32\textwidth]{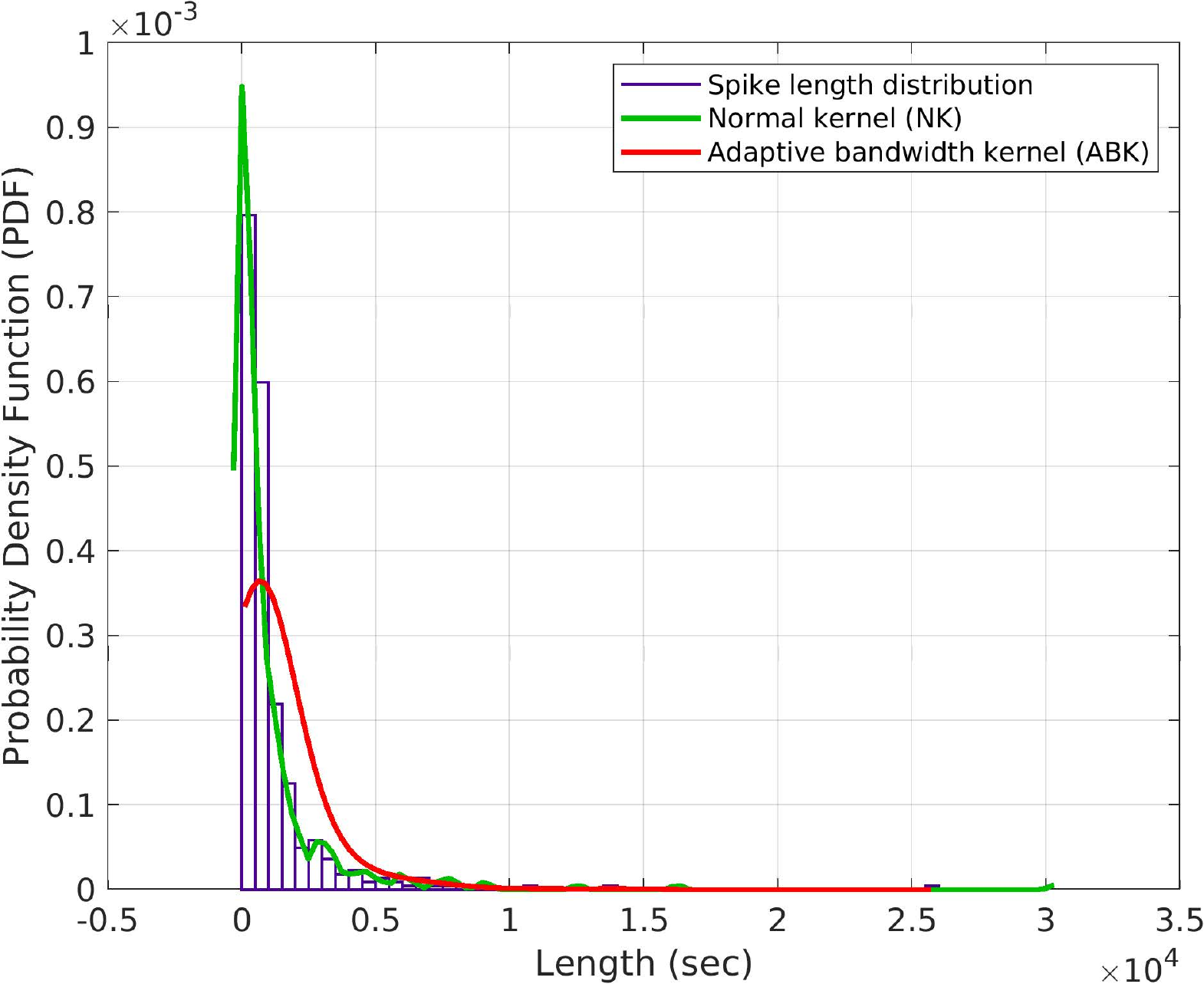}}
    \subfigure[]{\includegraphics[width=0.32\textwidth]{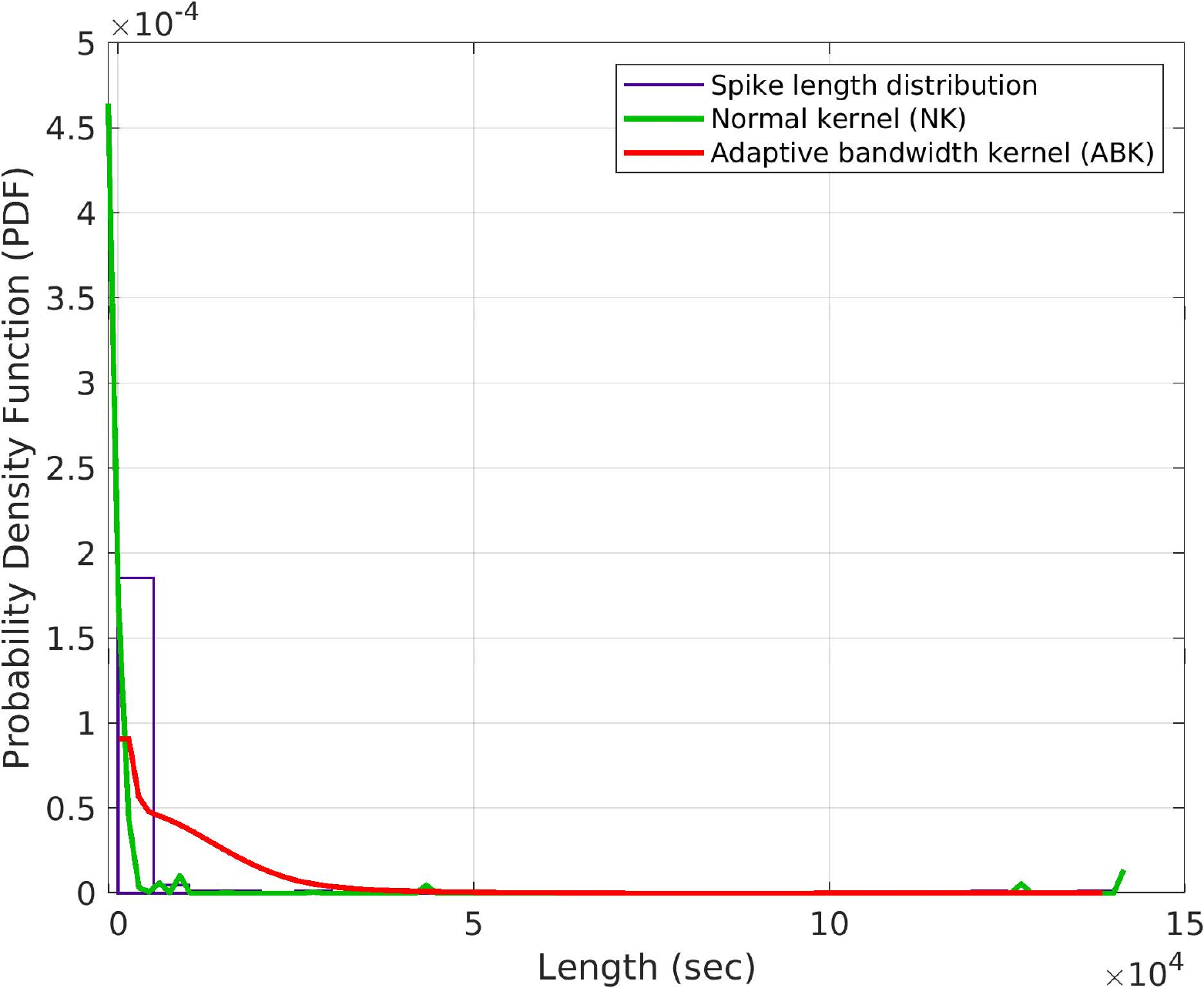}}
    \subfigure[]{\includegraphics[width=0.32\textwidth]{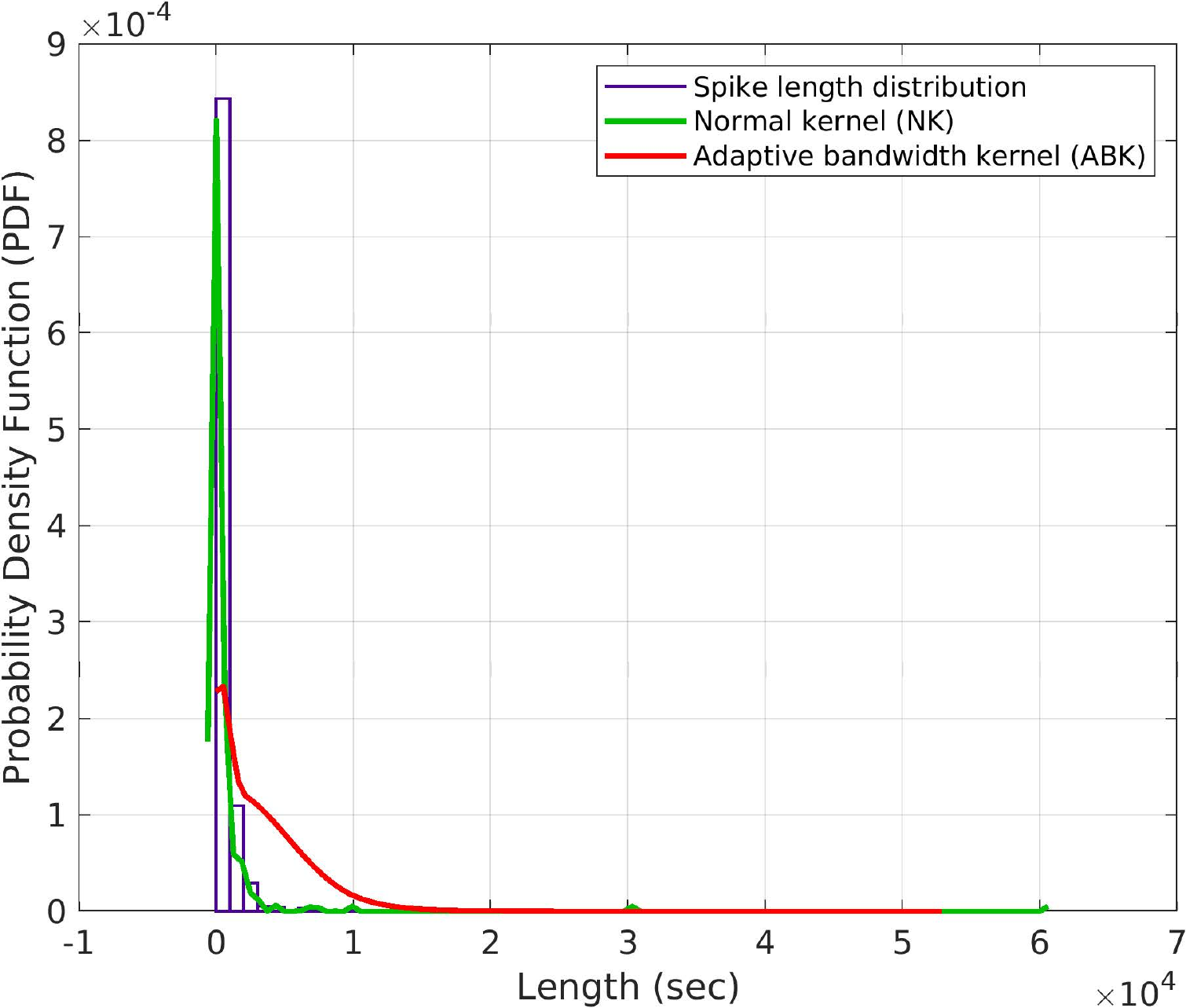}}
    \subfigure[]{\includegraphics[width=0.32\textwidth]{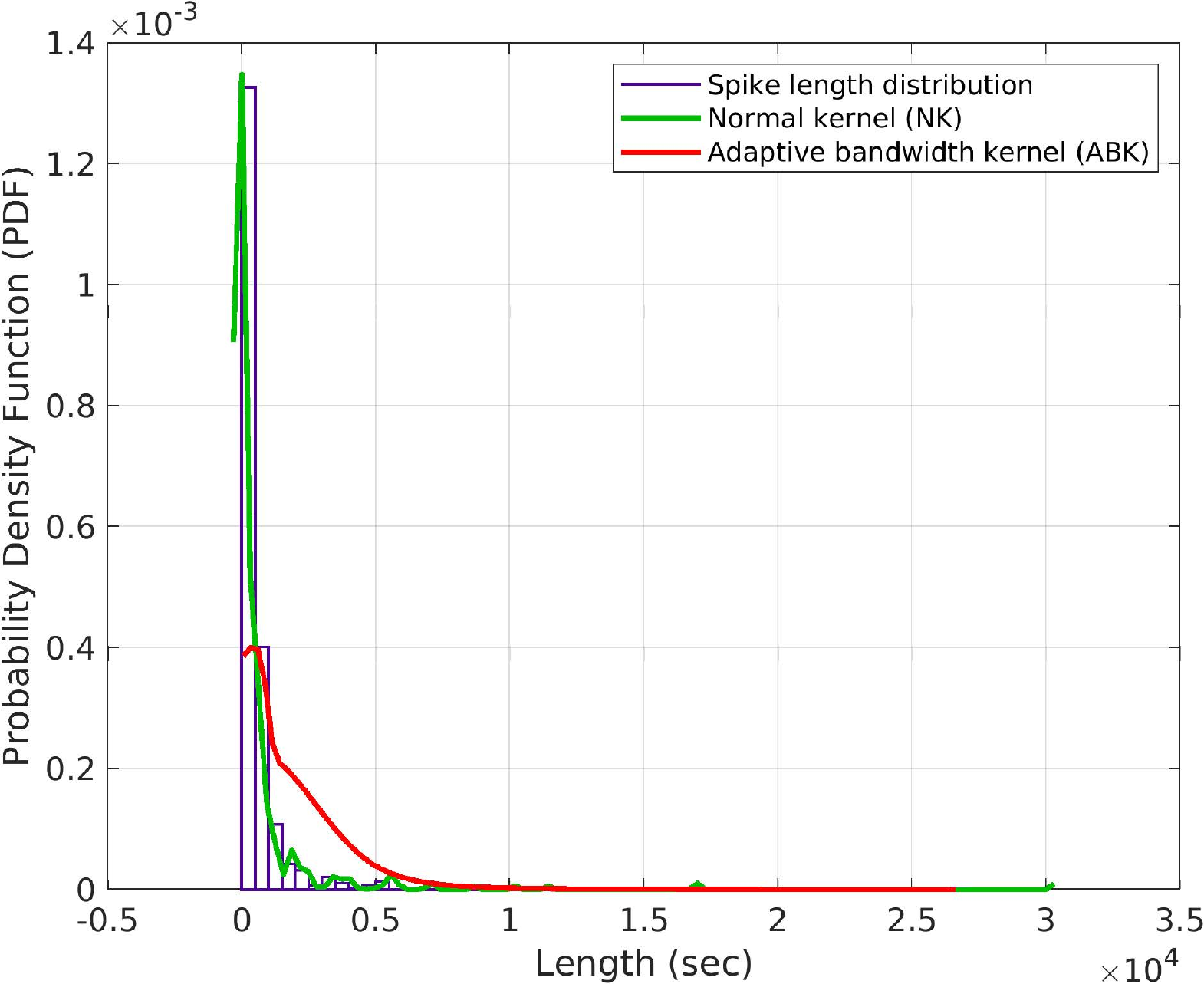}}
    \subfigure[]{\includegraphics[width=0.32\textwidth]{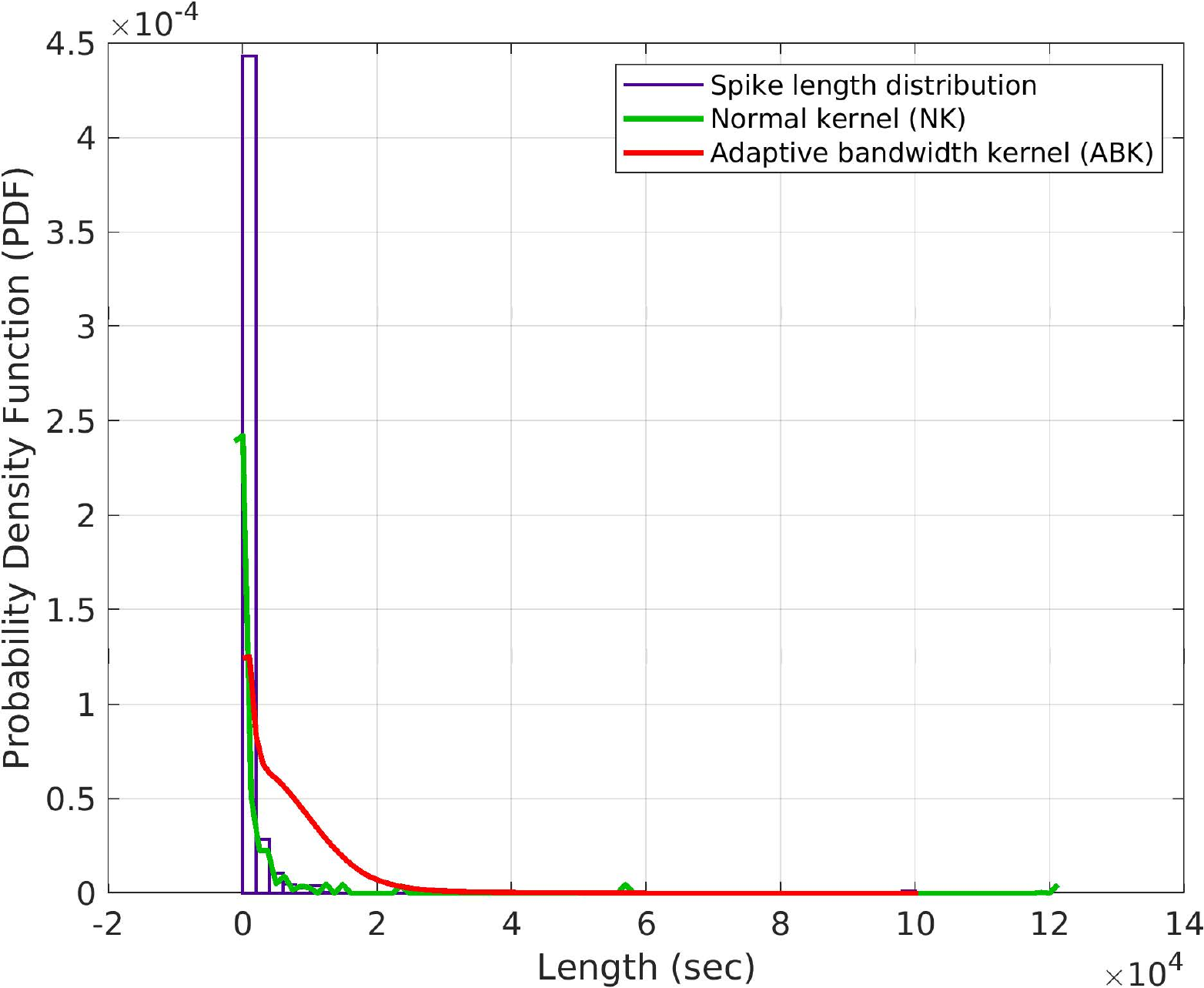}}
    \caption{Distribution of spike maximum amplitudes with superimposed Gaussian and Adaptive bandwidth kernels \cite{shimazaki2010kernel} for 
    (a,b)~in lines electrode placement with a distance of 1~cm. 
    (c,d)~in lines electrode placement with a distance of 2~cm. 
    (e,f)~random electrode placement with an approximate distance of 2~cm.}
\label{fig:10}
\end{figure}

We applied the proposed method to six experiments where the statistical results are shown in Figs.~\ref{fig:9} -- \ref{fig:10} and summarised in Table \ref{tbl:1}. It should be noted that the placement of the electrodes in two experiments was in lines with a distance of 1~cm, in two experiments it was in lines with a distance of 2~cm, and in two experiments it was random with a distance of approximately 2~cm. The implementation in MATLAB R2020a and details of experiments are available at~\cite{dehshibi2020electrical}.

\begin{table}[!htb]
\caption{The dominant value and bandwidth for the spike's length and amplitude in each experiment across all recording channels. The duration and amplitude of spikes are estimated via probability density function (PDF) and adaptive bandwidth kernel (ABK)~\cite{shimazaki2010kernel}. The bold-face \textcolor{blue}{\textbf{blue}} and \textcolor{red}{\textbf{red}} entries indicate the absolute minimum and maximum values, respectively. We considered the absolute value since we have bidirectional changes in potential.}
\label{tbl:1}
\resizebox{\textwidth}{!}{%
\begin{tabular}{ccccccccccc}
\hline
\multirow{3}{*}{} & \multirow{3}{*}{\#Channels} & \multirow{3}{*}{\#Spikes} & \multicolumn{4}{c}{Length (sec)}                            & \multicolumn{4}{c}{Amplitude (V)}                             \\ \cline{4-11} 
                  &                             &                           & \multicolumn{2}{c}{PDF}      & \multicolumn{2}{c}{ABK}      & \multicolumn{2}{c}{PDF}       & \multicolumn{2}{c}{ABK}       \\ \cline{4-11} 
                  &                             &                           & Dominant         & Bandwidth & Dominant         & Bandwidth & Dominant          & Bandwidth & Dominant          & Bandwidth \\ \hline
\#1               & 8                           & 565                       & \textcolor{blue}{\textbf{84.00}}   & 75.61     & \textcolor{blue}{\textbf{84.00}}   & 60.22     & \textcolor{blue}{\textbf{0.00003}}  & 0.00048   & \textcolor{blue}{\textbf{-0.00117}} & 0.00576   \\
\#2               & 5                           & 447                       & 366.80           & 154.31    & 625.60           & 126.47    & 0.00642           & 0.00544   & 0.00642           & 0.00667   \\
\#3               & 4                           & 124                       & 84.00            & 75.61     & 84.00            & 60.22     & 0.00003           & 0.00048   & -0.00117          & 0.00576   \\
\#4               & 5                           & 951                       & 534.12           & 80.09     & 534.12           & 84.80     & -0.00239          & 0.00301   & -0.00239          & 0.00508   \\
\#5               & 5                           & 573                       & 334.25           & 74.52     & 334.25           & 80.9      & \textcolor{red}{\textbf{-0.01536}} & 0.00218   & \textcolor{red}{\textbf{-0.01462}} & 0.00357   \\
\#6               & 15                          & 862                       & \textcolor{red}{\textbf{1014.72}} & 99.53     & \textcolor{red}{\textbf{1014.72}} & 92.67     & -0.00172          & 0.00381   & -0.01277          & 0.00591   \\ \hline
\end{tabular}%
}
\end{table}
\clearpage

These findings are aligned with the previously reported results on electrical activity of \emph{Physarum polycephalum}~\cite{adamatzky2013tactile, adamatzky2018spiking} in which we reported that Physarum spike lengths are in the range of 60-120 seconds. In terms of growth, Physarum is faster than fungi. Therefore, we can now hypothesise that fungal spikes can not be less than 60-120 seconds, with more observations. 

\subsection{Complexity Analysis}

To quantify the complexity of the electrical signalling recorded, we used the following measurements: 

\begin{enumerate}
    \item The Shannon entropy, $H$, is calculated as $H =- \sum_{w \in W} (\nu(w)/\eta \cdot ln (\nu(w)/\eta))$, where $\nu(w)$ is a number of times the neighbourhood configuration $w$ is found in configuration $W$, and $\eta$ is the total number of spike events found in all channels of an experiment.
    \item Simpson's diversity, $S$, is calculated as $S=\sum_{w \in W} (\nu(w)/\eta)^2$. It linearly correlates with Shannon entropy for $H<3$ and the relationships becomes logarithmic for higher values of $H$. The value of $S$ ranges between 0 and 1, where 1 represents infinite diversity and 0, no diversity.
    \item Space filling, $D$, is the ratio of non-zero entries in $W$ to the total length of string.
    \item Expressiveness, $E$, is calculated as the Shannon entropy $H$ divided by space-filling ratio $D$, the expressiveness reflects the `economy of diversity'.
\end{enumerate}

\begin{figure}[!htb]
    \centering
    \subfigure[]{\includegraphics[width=0.43\textwidth]{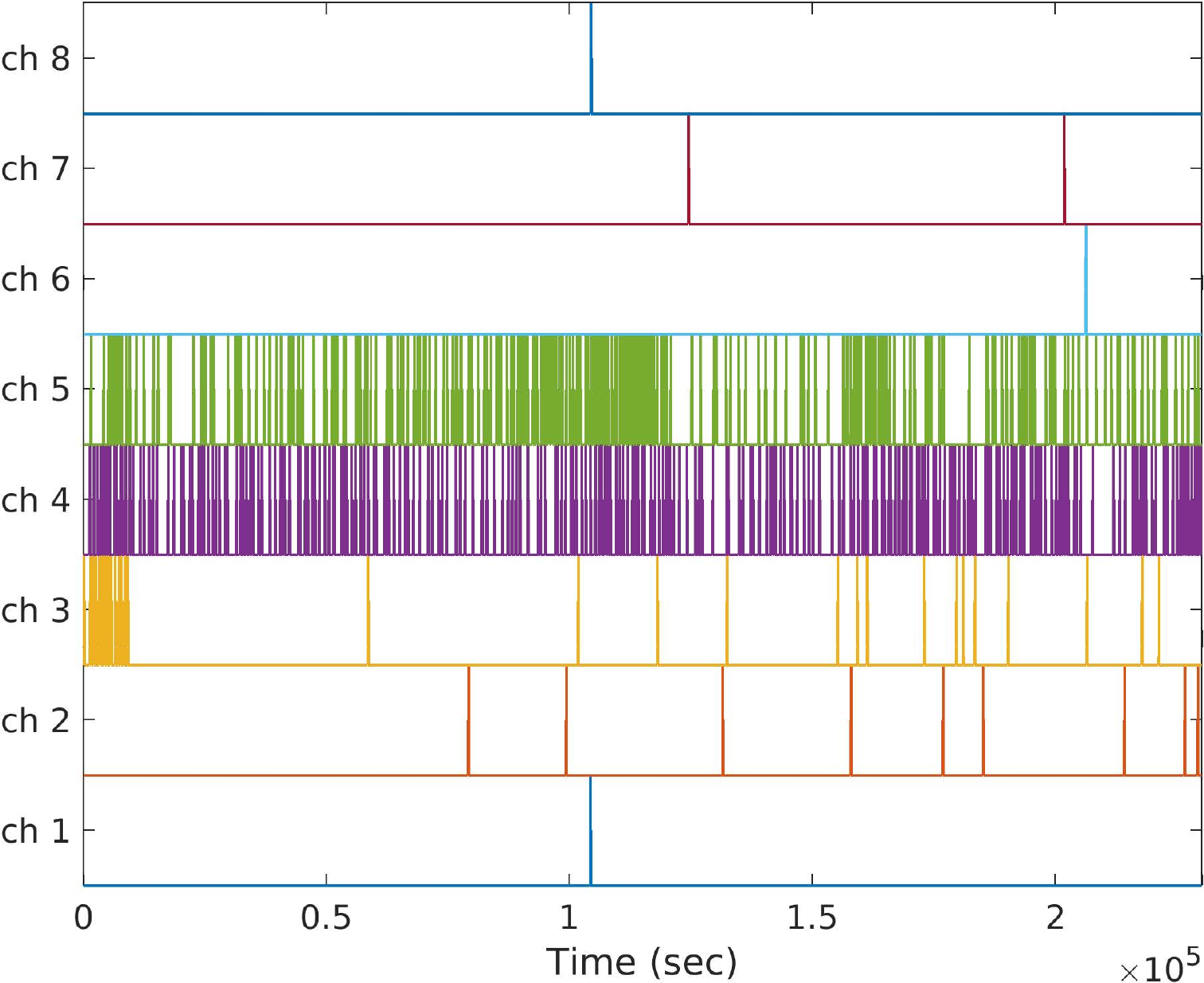}}
    \subfigure[]{\includegraphics[width=0.43\textwidth]{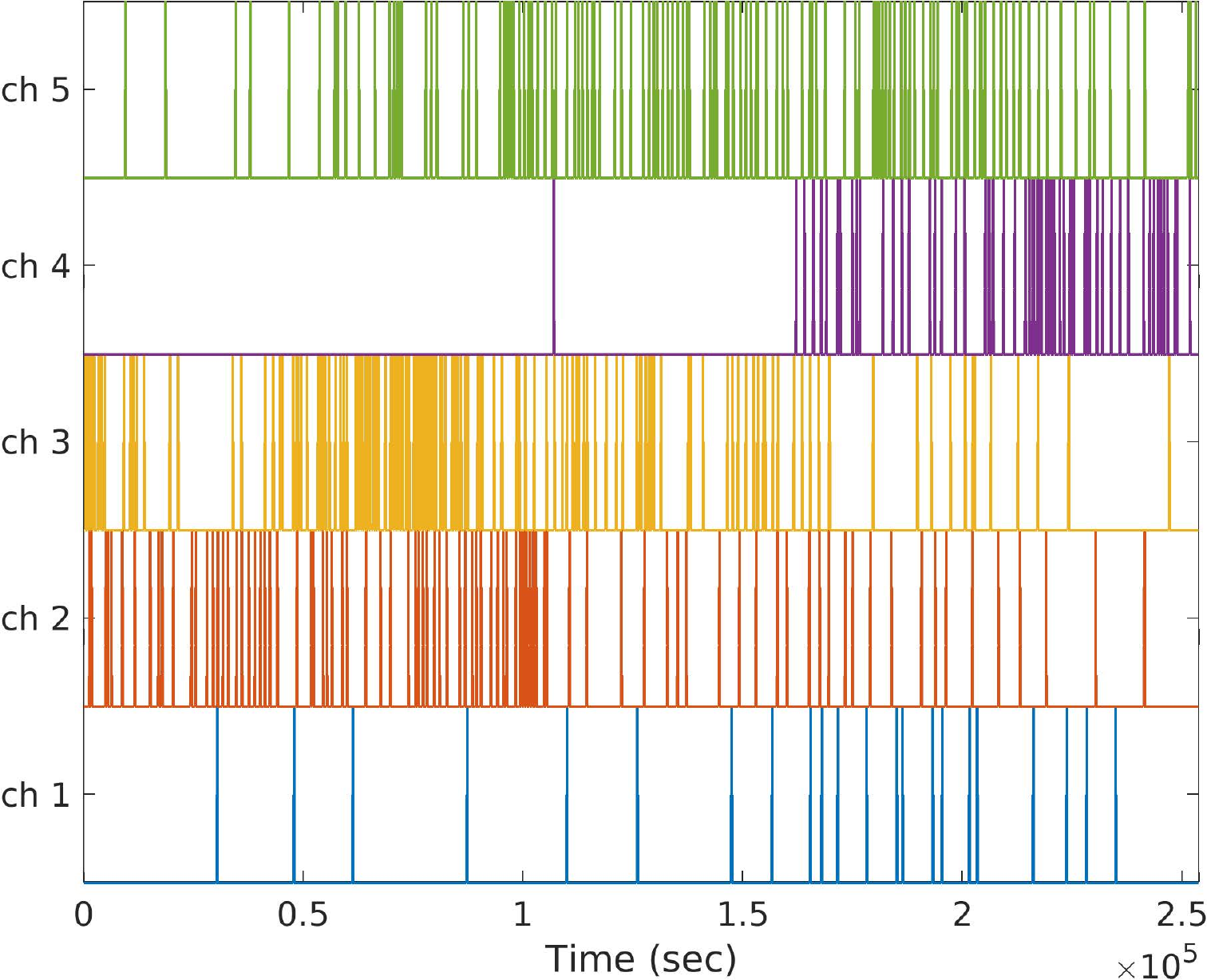}}
    \subfigure[]{\includegraphics[width=0.43\textwidth]{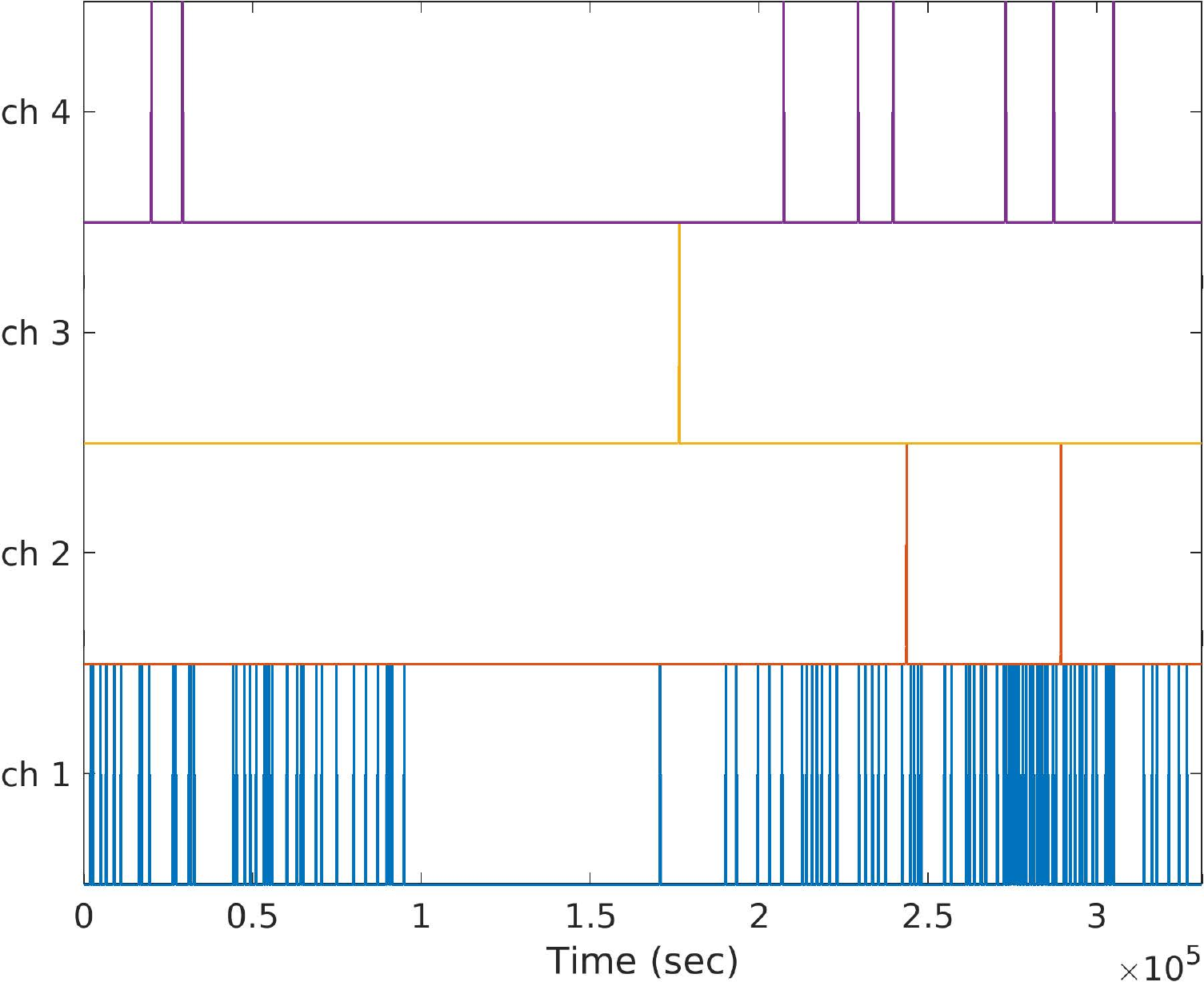}}
    \subfigure[]{\includegraphics[width=0.43\textwidth]{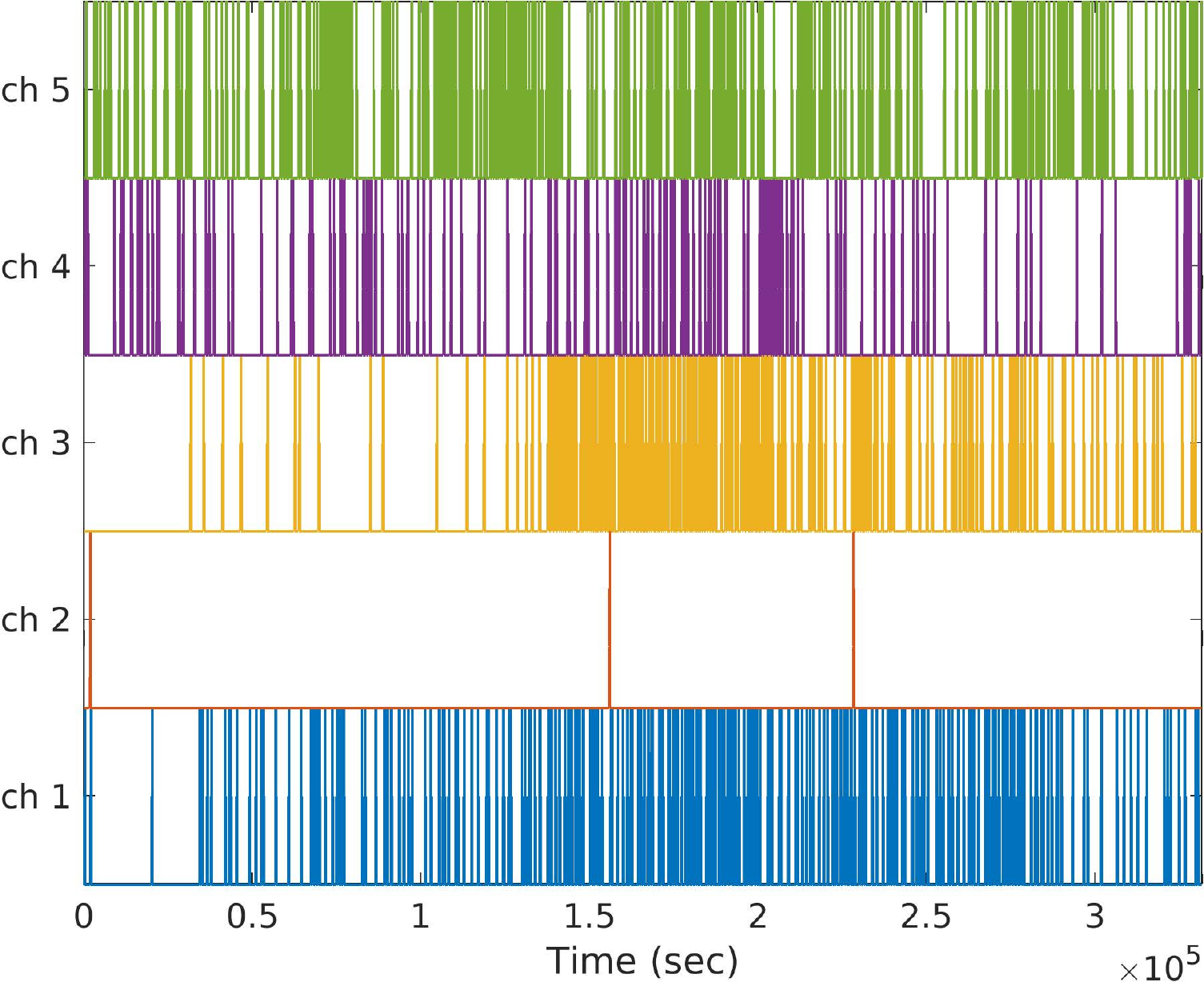}}
    \subfigure[]{\includegraphics[width=0.43\textwidth]{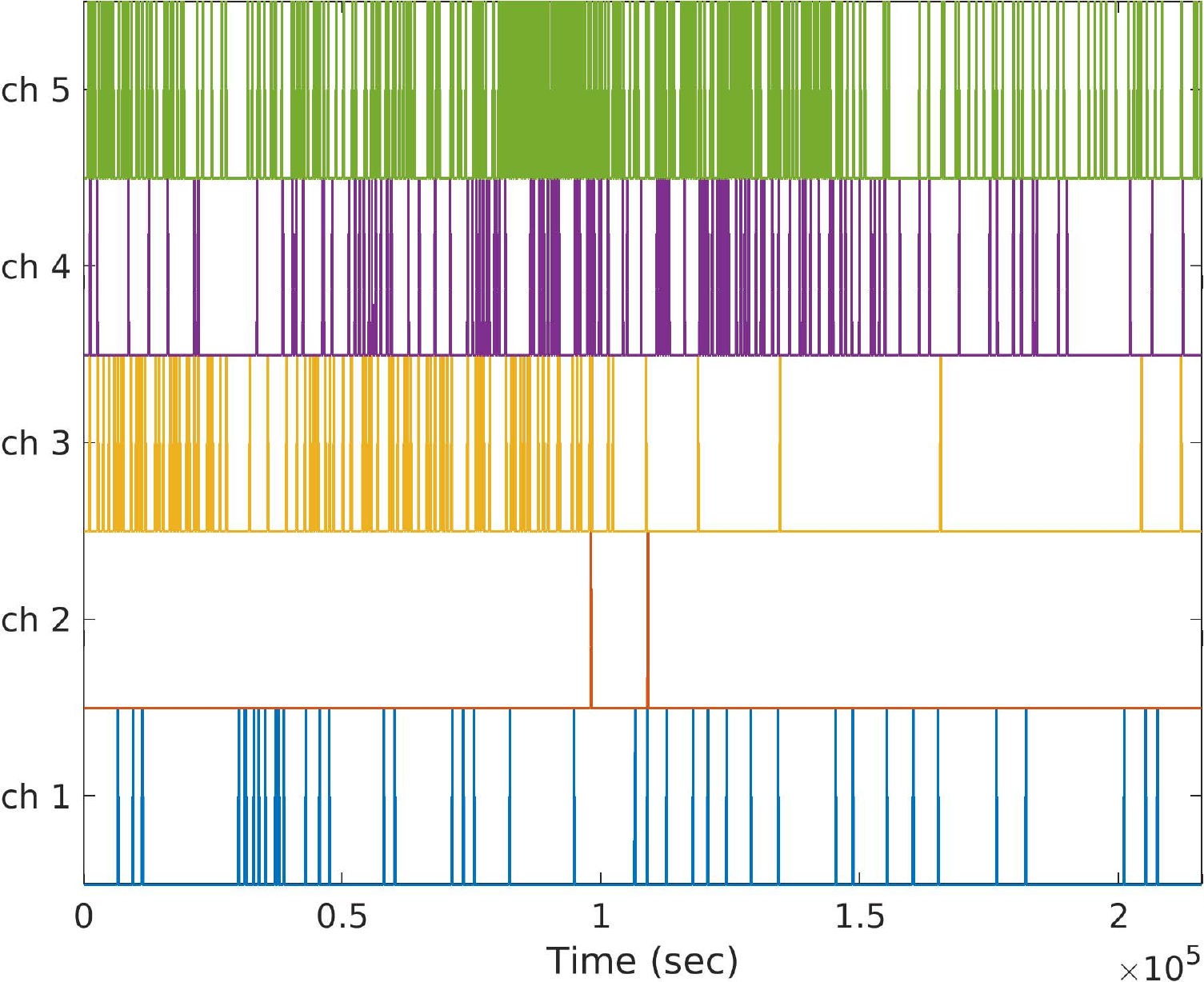}}
    \subfigure[]{\includegraphics[width=0.43\textwidth]{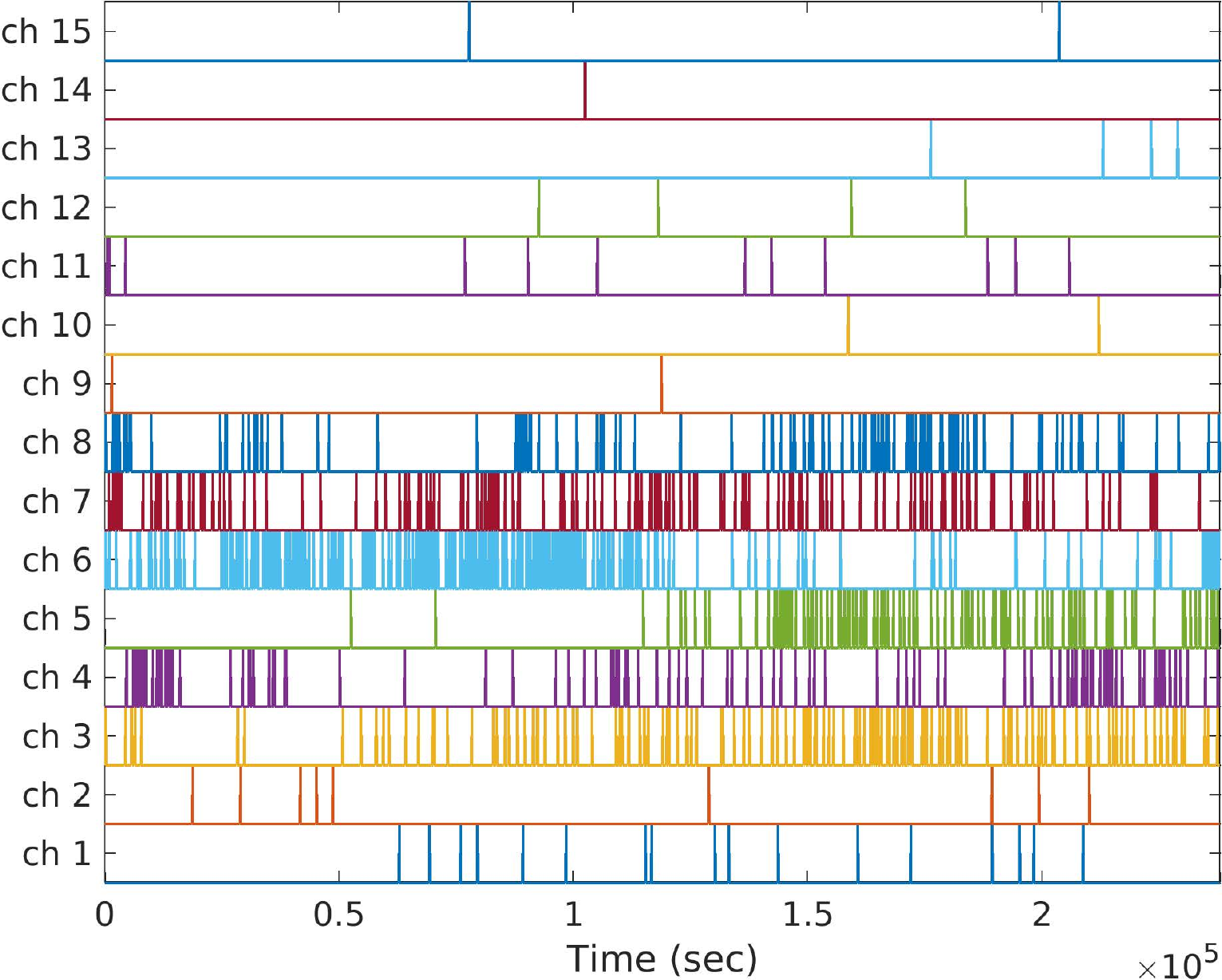}}
    \caption{Barcode-like representation of spike events in different channels for (a,b) in-line electrode arrangements with a distance of 1~cm, (c,d) in-line electrode arrangements with a distance of 2~cm, and (e,f) random electrode arrangements with an approximate distance of 2~cm.} 
\label{fig:11}
\end{figure}
\clearpage

\begin{enumerate}
    \setcounter{enumi}{4}
    \item Lempel--Ziv complexity (compressibility), $LZ$, is evaluated by the size of binary string, $n$, and used to assess temporal signal diversity. Here, we represented the spiking behaviour of mycelium with a binary string where `1s' indicates the presence of a spike and `0s' otherwise (see Fig.~\ref{fig:11}).
    \item Perturbation complexity index $PCI = LZ/H$.
\end{enumerate}

\begin{figure}[!htb]
    \centering
    \subfigure[]{\includegraphics[width=0.43\textwidth]{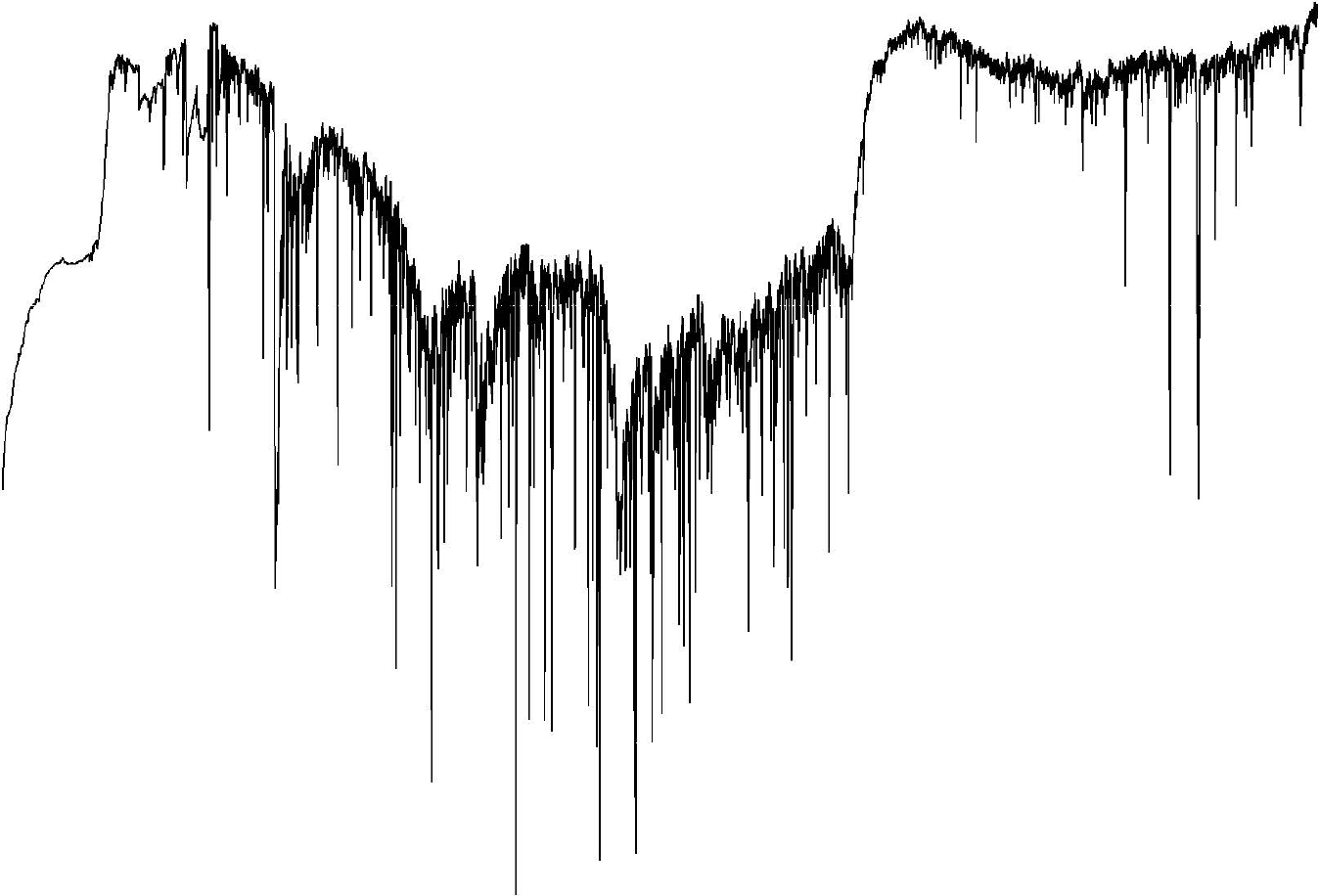}}
    \subfigure[]{\includegraphics[width=0.43\textwidth]{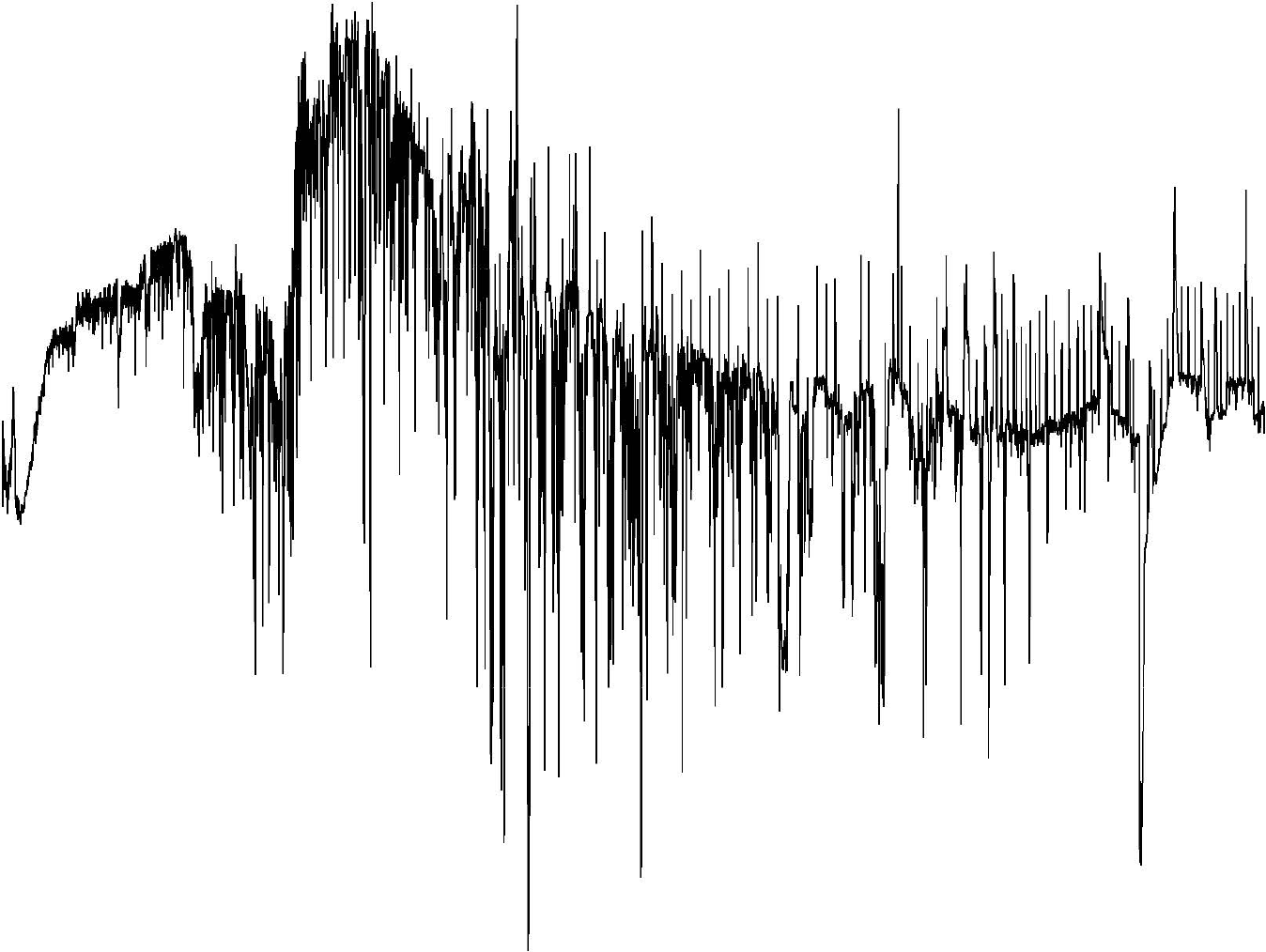}}
    \caption{Two samples from input channels, which are saved in black and white PNG format without axes and annotations.} \label{fig:12}
\end{figure}

To calculate Lempel--Ziv complexity, we saved each signal as a PNG image (see two examples in Fig.~\ref{fig:12}), where the `deflation' algorithm used in PNG lossless compression~\cite{deutsch1996zlib,howard1993design,roelofs1999png} is a variation of the classical LZ77 algorithm~\cite{ziv1977universal}. We employed this approach as the recorded signal is a non-binary string. We take the largest PNG file size to normalise this measurement.

\begin{figure}[!tbp]
    \centering
    \subfigure[]{\includegraphics[width=0.43\textwidth]{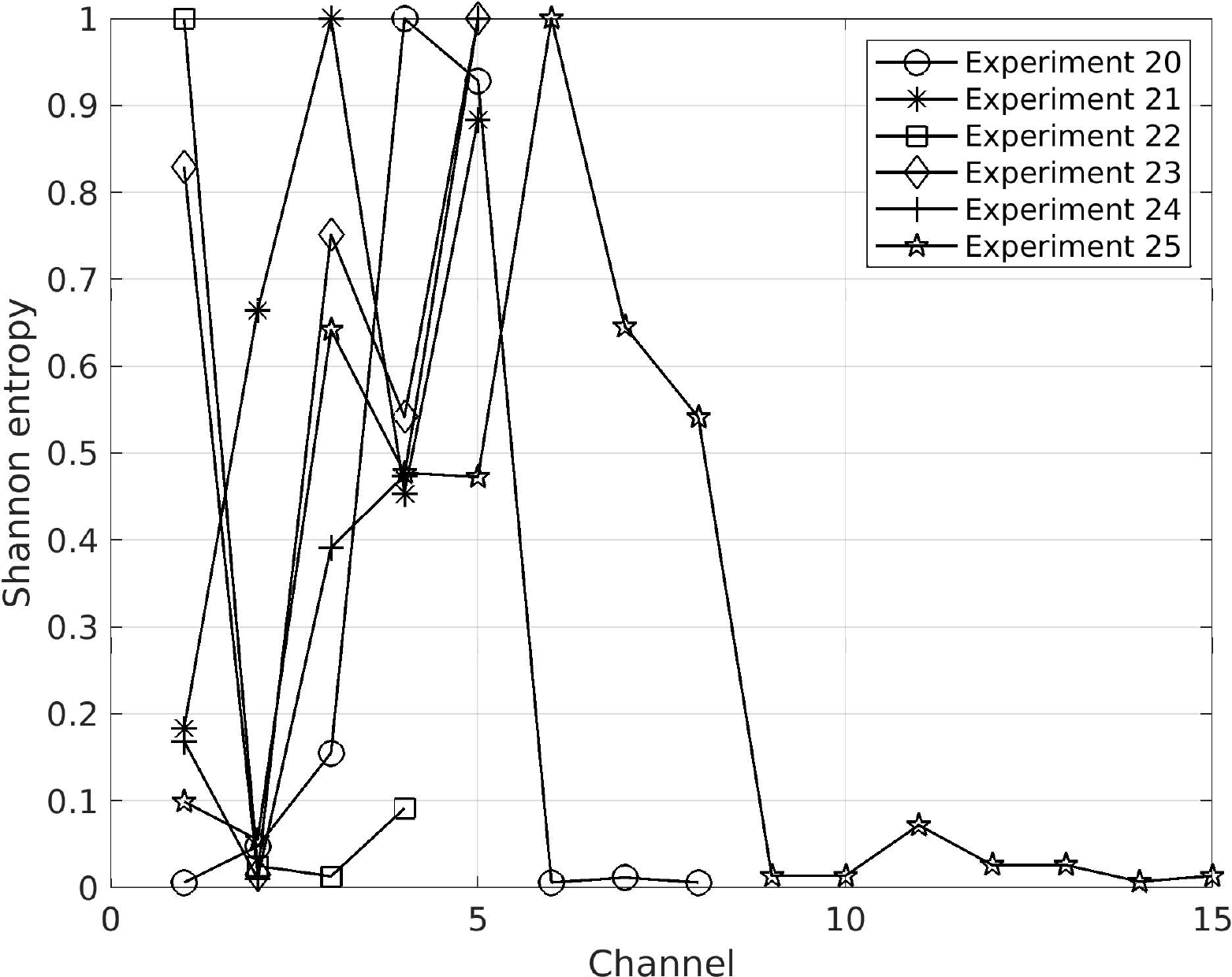}}
    \subfigure[]{\includegraphics[width=0.43\textwidth]{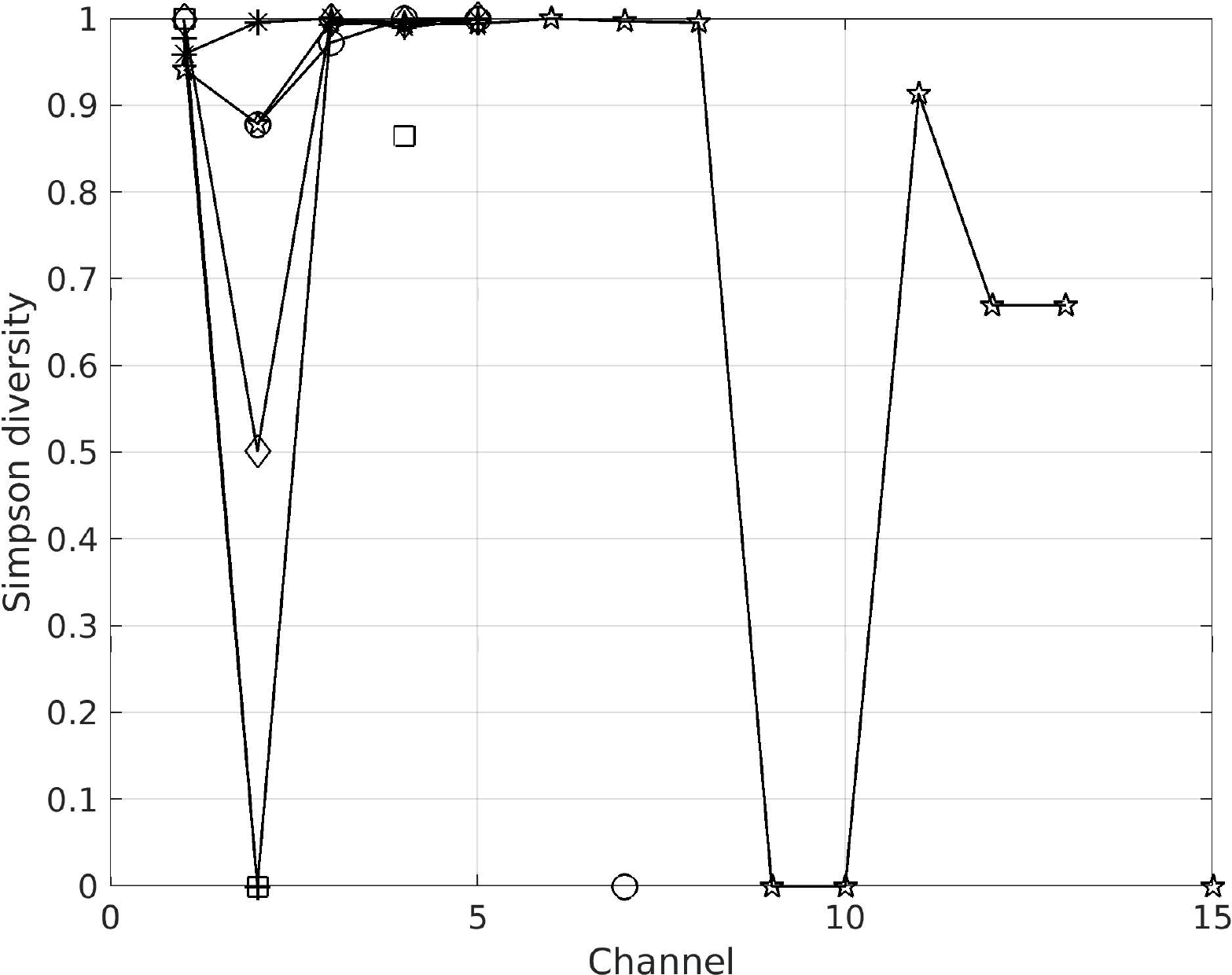}}
    \subfigure[]{\includegraphics[width=0.43\textwidth]{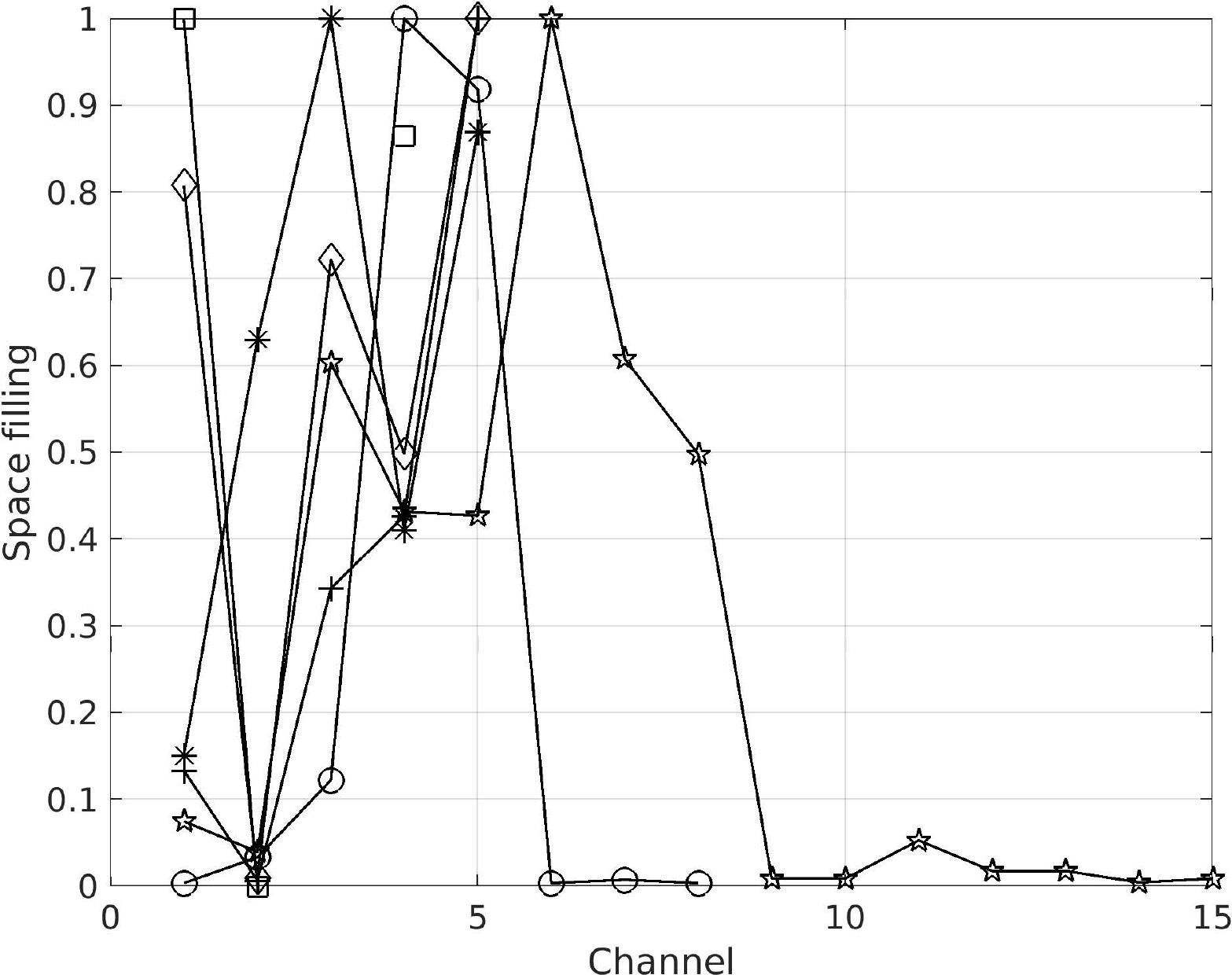}}
    \subfigure[]{\includegraphics[width=0.43\textwidth]{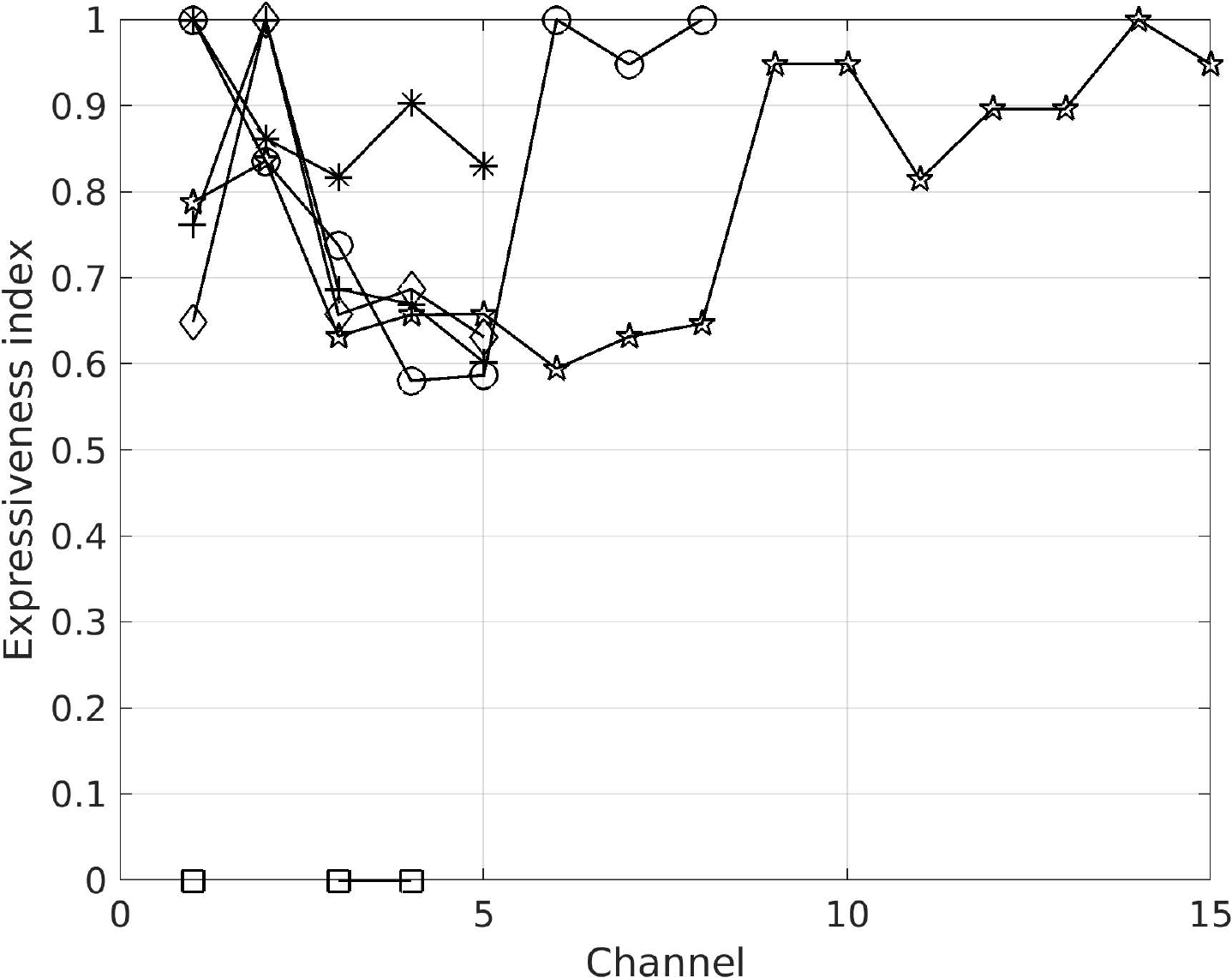}}
    \subfigure[]{\includegraphics[width=0.43\textwidth]{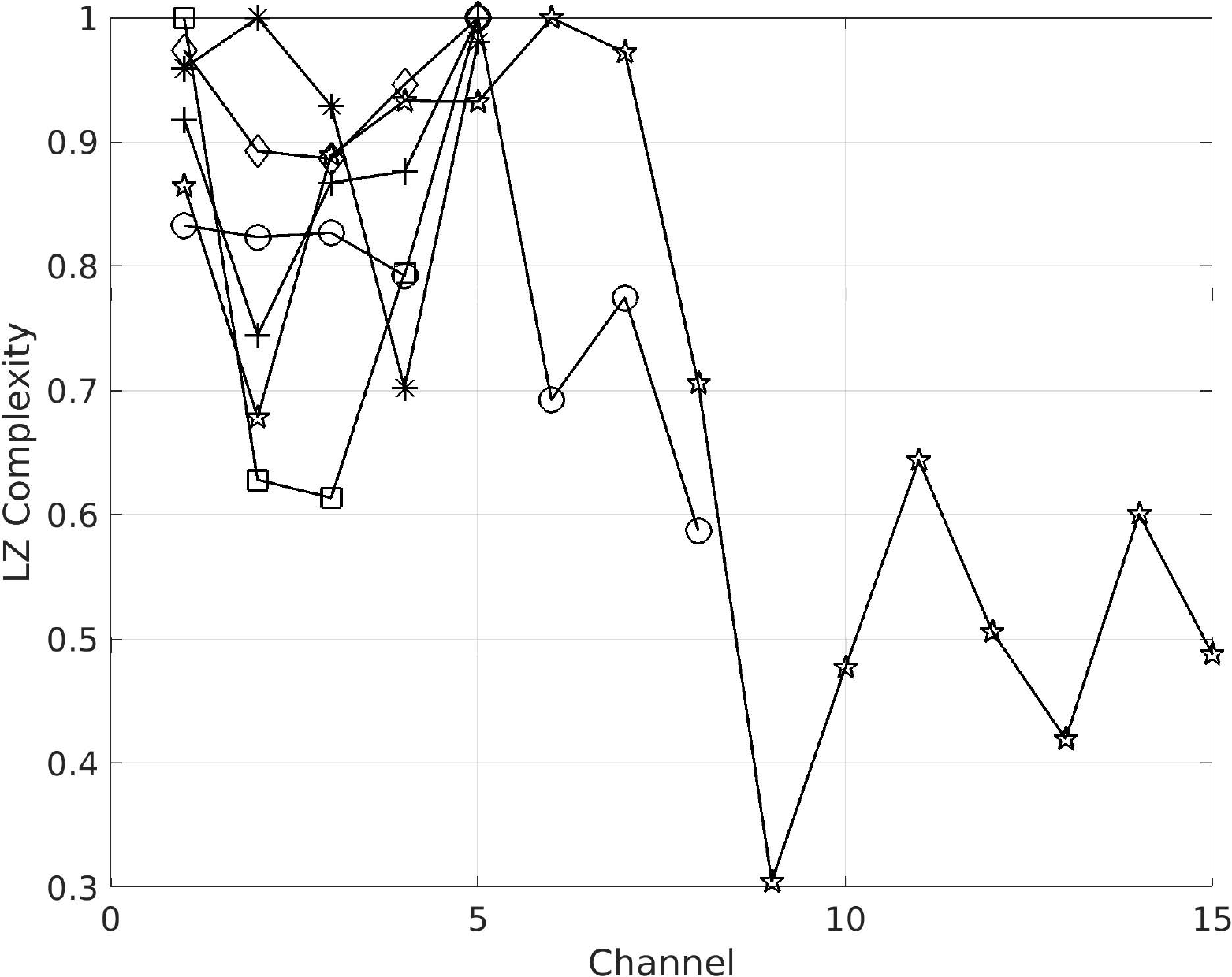}}
    \subfigure[]{\includegraphics[width=0.43\textwidth]{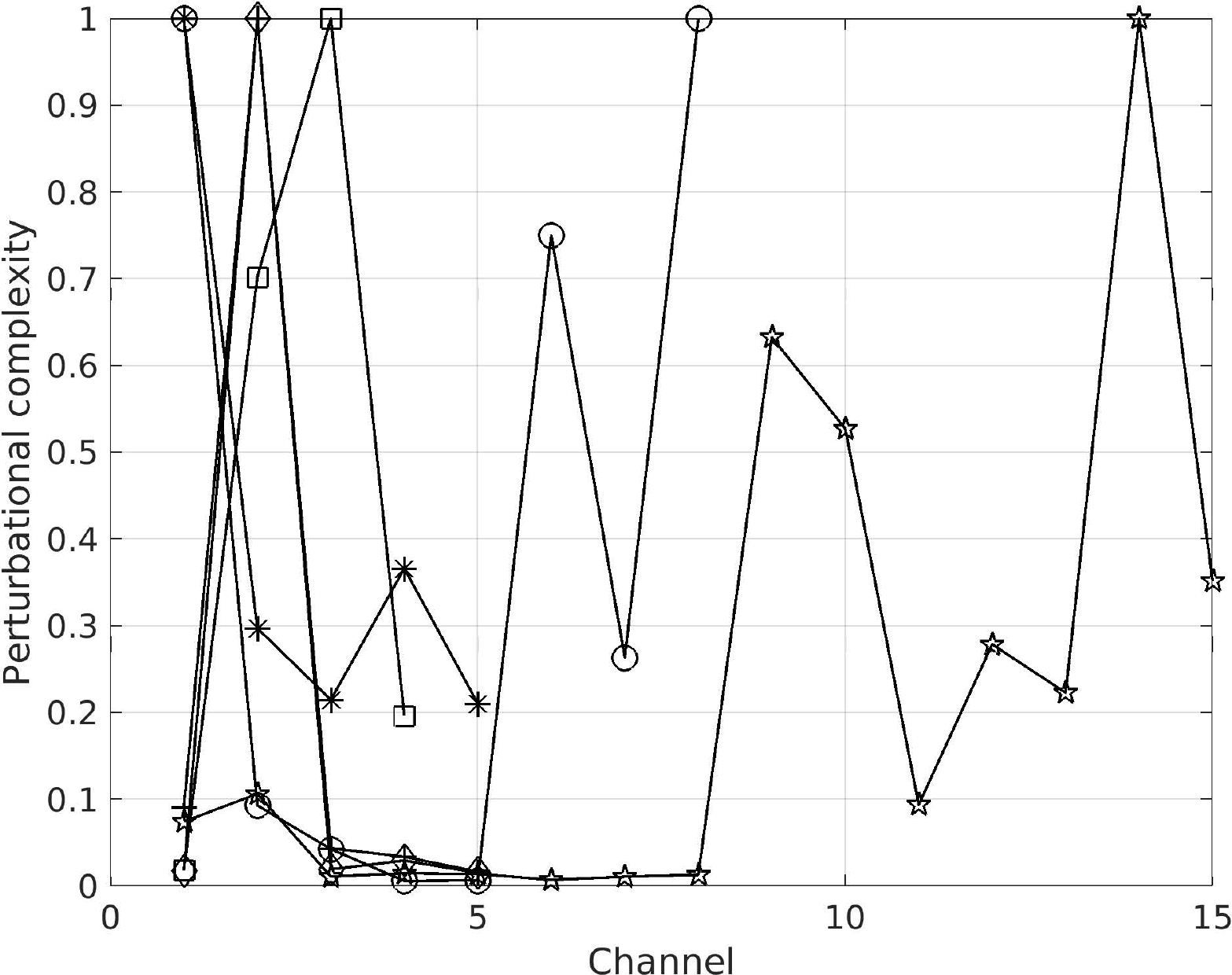}}
    \caption{(a) Shannon entropy, (b) Simpson's diversity, (c)  Space filling, (d) Expressiveness, (e) Lempel–Ziv  complexity, and (f) Perturbation complexity index. All measurements are scaled to the range of $[0,1]$.} \label{fig:13}
\end{figure}

To assess the signal diversity across all channels and observations, we represented each experiment as a matrix with binary entries with a row for each channel and a column for each observation. This binary matrix is then concatenated observation-by-observation to form one binary string. We applied Kolmogorov complexity algorithm \cite{kaspar1987easily} to calculate the across channels Lempel--Ziv complexity, $LZc$. $LZc$ captures temporal signal diversity of single channels as well as spatial signal diversity across channels as the result of the observation-by-observation concatenation of the binarised data matrix. We also normalise $LZc$ by dividing the raw value by the value obtained for the same binary input sequence randomly shuffled. Since the value of $LZ$ for a binary sequence of fixed length is maximal if the sequence is entirely random, the normalised values indicate the level of signal diversity on a scale from 0 to 1. Results of calculating these complexity measurements for all six setups are illustrated in Fig.~\ref{fig:13}, and summarised in Tab.~\ref{tbl:2}. 

\begin{table}[!htb]
\caption{The mean of complexity measurements for six experiments.}
\label{tbl:2}
\resizebox{\textwidth}{!}{%
\begin{tabular}{cccccccccc}
\hline
 &
  \#Channel &
  \#Spike &
  \begin{tabular}[c]{@{}c@{}}Lempel–Ziv\\ complexity\end{tabular} &
  \begin{tabular}[c]{@{}c@{}}Shannon \\ entropy\end{tabular} &
  \begin{tabular}[c]{@{}c@{}}Simpson's\\ diversity\end{tabular} &
  \begin{tabular}[c]{@{}c@{}}Space\\ filling\end{tabular} &
  Kolmogorov &
  PCI &
  Expressiveness \\ \hline
\#1 & 8  & 565 & 0.79 & 45.81  & 0.76 & 30.68$\times 10^{-5}$ & 30.36$\times 10^{-4}$ & 0.365 & 20.8$\times 10^{4}$ \\
\#2 & 5  & 447 & 0.91 & 63.27  & 0.98 & 35.20$\times 10^{-5}$ & 35.78$\times 10^{-4}$ & 0.021 & 18.6$\times 10^{4}$ \\
\#3 & 4  & 124 & 0.75 & 22.57  & 0.61 & 48.10$\times 10^{-5}$ & 10.94$\times 10^{-4}$ & 0.333 & 29.71        \\
\#4 & 5  & 951 & 0.93 & 123.11 & 0.89 & 57.30$\times 10^{-5}$ & 56.05$\times 10^{-4}$ & 0.072 & 23.8$\times 10^{4}$ \\
\#5 & 5  & 573 & 0.88 & 75.75  & 0.79 & 53.02$\times 10^{-5}$ & 52.80$\times 10^{-4}$ & 0.077 & 16.4$\times 10^{4}$ \\
\#6 & 15 & 862 & 0.69 & 39.96  & 0.71 & 24.20$\times 10^{-5}$ & 25.06$\times 10^{-4}$ & 0.207 & 20.4$\times 10^{4}$ \\ \hline
\end{tabular}%
}
\end{table}

\begin{figure}[!htb]
    \centering
    \subfigure[]{\includegraphics[width=0.32\textwidth]{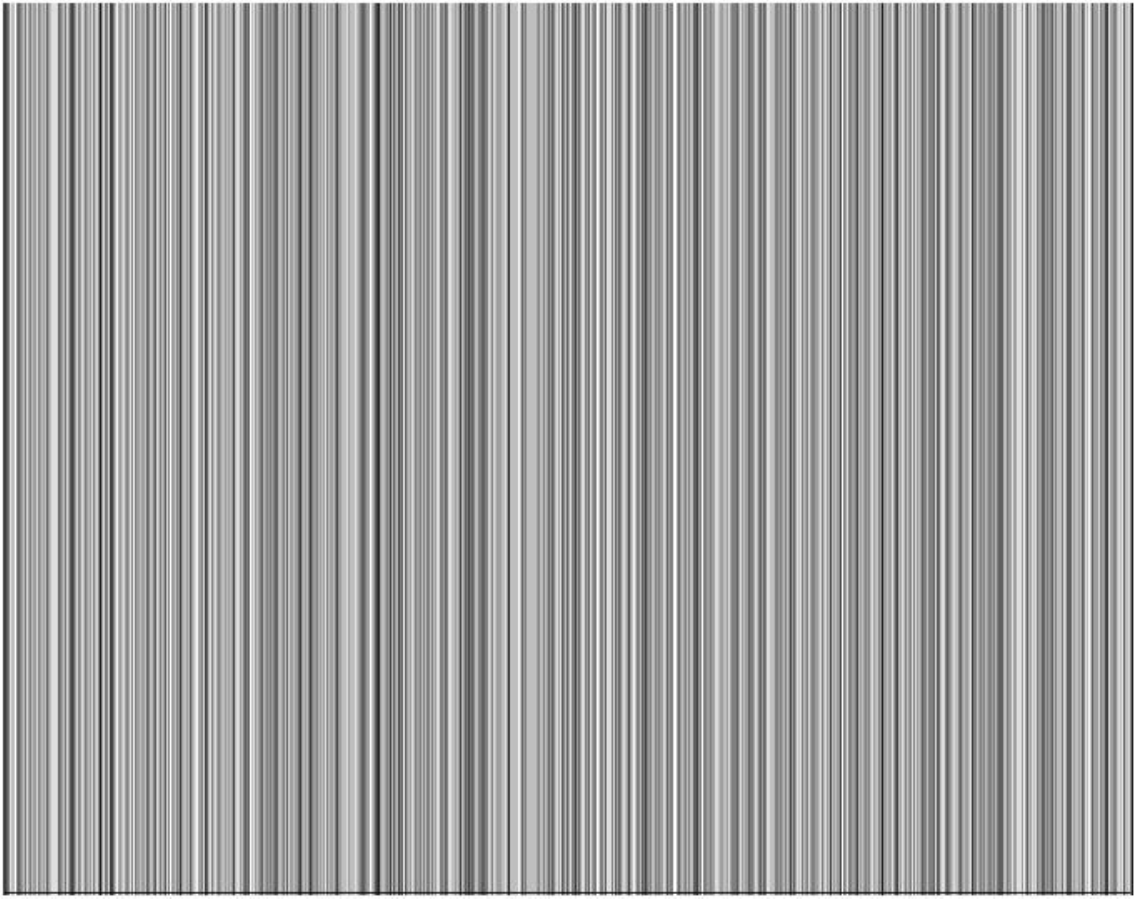}}
    \subfigure[]{\includegraphics[width=0.32\textwidth]{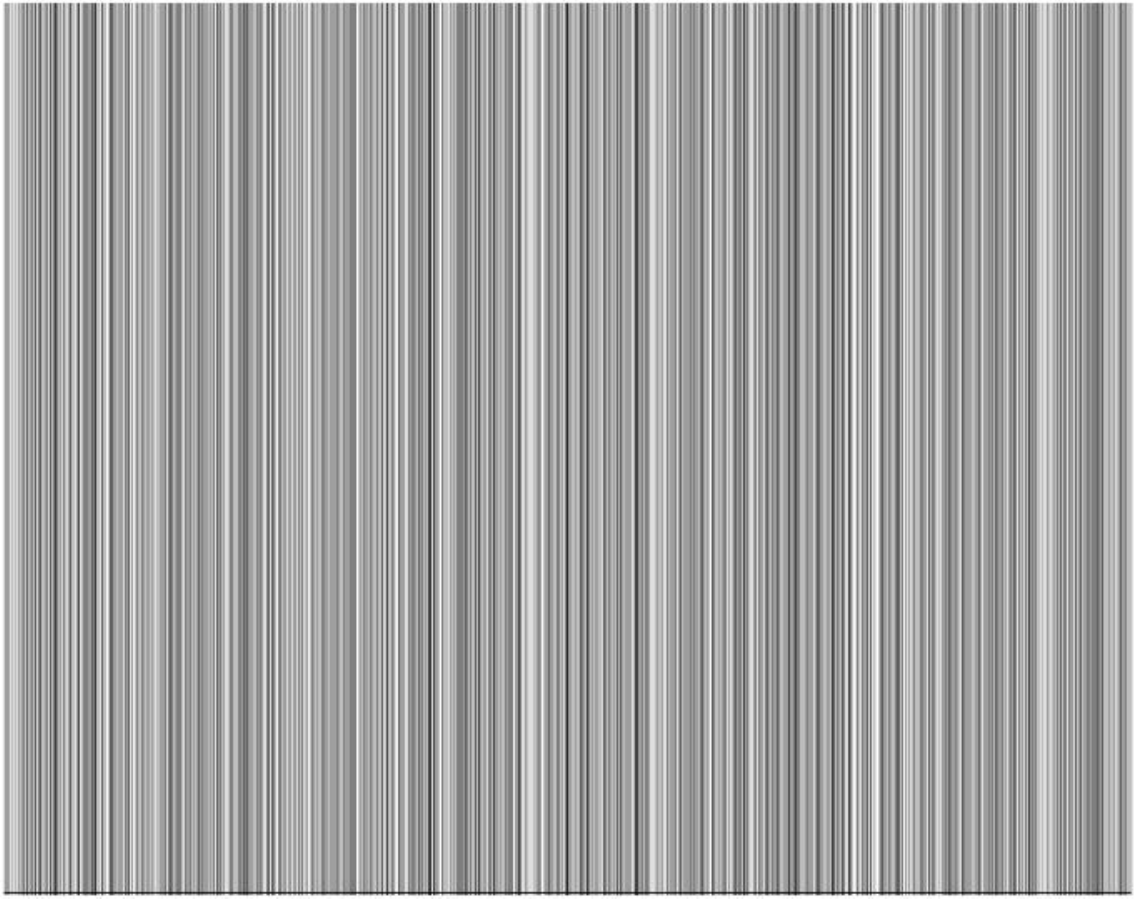}}
    \subfigure[]{\includegraphics[width=0.32\textwidth]{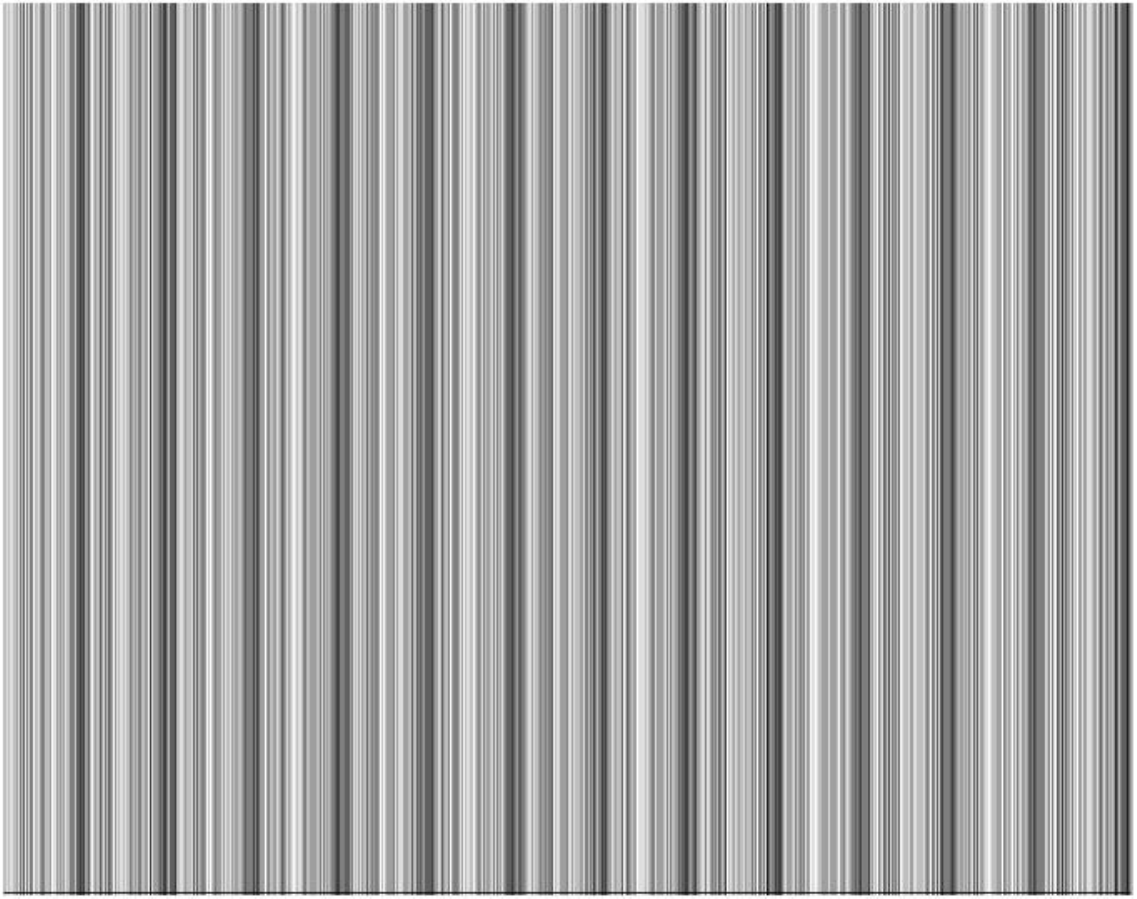}}
    \caption{Binary representation of (a) pieces of news,(b) random sequence of alphanumeric, and (c) periodic sequence of alphanumeric after applying Huffman coding.} \label{fig:14}
\end{figure}

In order to clarify the communication complexity in the mycelium substrate, we also calculated the mentioned complexity measurements for the communications in the forms of (i) pieces of news\footnote{https://www.sciencemag.org/news/2020/07/meet-lizard-man-reptile-loving-biologist-tackling-some-biggest-questions-evolution}, (ii) a random sequence of alphanumeric\footnote{We used available service at https://www.random.org/ }, and (iii) a periodic sequence of alphanumeric converted to binary strings by applying Huffman coding~\cite{huffman1952method} (see barcodes in Fig.~\ref{fig:14}). Results of comparing the complexity of fungi electrical activity with these three forms are reported in Tab.~\ref{tbl:3}.

\begin{table}[!htb]
\caption{The complexity measurements for pieces of news, a random sequence of alphanumeric, a periodic sequence of alphanumeric along with three chunks randomly selected from our experiments.}
\label{tbl:3}
\resizebox{\textwidth}{!}{%
\begin{tabular}{ccccccccc}
\hline
                  & Length & \begin{tabular}[c]{@{}c@{}}Lempel–Ziv\\ complexity\end{tabular} & \begin{tabular}[c]{@{}c@{}}Shannon \\ entropy\end{tabular} & \begin{tabular}[c]{@{}c@{}}Simpson's\\ diversity\end{tabular} & \begin{tabular}[c]{@{}c@{}}Space\\ filling\end{tabular} & Kolmogorov & PCI      & Expressiveness \\ \hline
News              & 36187  & 0.127919                                                        & 4.421728                                                   & 0.999941                                                      & 0.465996                                                & 0.765382   & 0.173096 & 9.49           \\
Random sequence   & 36002  & 0.125465                                                        & 5.770331                                                   & 0.999941                                                      & 0.469835                                                & 1.001850   & 0.173621 & 12.28          \\
Periodic sequence & 36006  & 0.127090                                                        & 3.882058                                                   & 0.999937                                                      & 0.442426                                                & 0.076508   & 0.019708 & 8.77           \\
Chunk 1              & 36000  & 0.067611                                                        & 16.194914                                                  & 0.947368                                                      & 0.000556                                                & 0.006307   & 0.000389 & 29150.84       \\
Chunk 2              & 36000  & 0.007250                                                        & 15.478087                                                  & 0.944444                                                      & 0.000528                                                & 0.006727   & 0.000435 & 29326.90       \\
Chunk 3              & 36000  & 0.068417                                                        & 31.680374                                                  & 0.976190                                                      & 0.001194                                                & 0.012613   & 0.000398 & 26523.10       \\ \hline
\end{tabular}%
}
\end{table}

\section{Discussion} \label{sec:5}

We developed algorithmic framework for exhaustive characterisation of electrical activity of a substrate colonised by mycelium of oyster fungi \emph{Pleurotus djamor}. We evidenced spiking activity of the mycelium. We found that average dominant duration of an action-potential like spike is 402 sec. The spikes amplitudes' depends on the location of the source of electrical activity related to the position of electrodes, thus the amplitudes provide less useful information. The amplitudes vary from 0.5~mV to 6~mV. This is indeed low compared to 50-60~mV of intracellular recording, nevertheless understandable due to the fact the electrodes are inserted not even in mycelium strands but in the substrate colonised by mycelium. The spiking events have been characterised with several complexity measures. Most measures, apart of Kolmogorov complexity shown a low degree of variability between channels (different sites of the recordings). The Kolmogorov complexity of fungal spiking varies from 11$\times 10^{-4}$ to 57$\times 10^{-4}$. This might indicated mycelium sub-networks in different parts of the substrate have been transmitting different information to other parts of the mycelium network. This is somehow echoes experimental results on communication between ants analysed with Kolmogorov complexity: longer paths communicated ants corresponds to higher values of complexity~\cite{ryabko1996using}.

\begin{figure}[!tbp]
    \centering
    \includegraphics[width=0.7\textwidth]{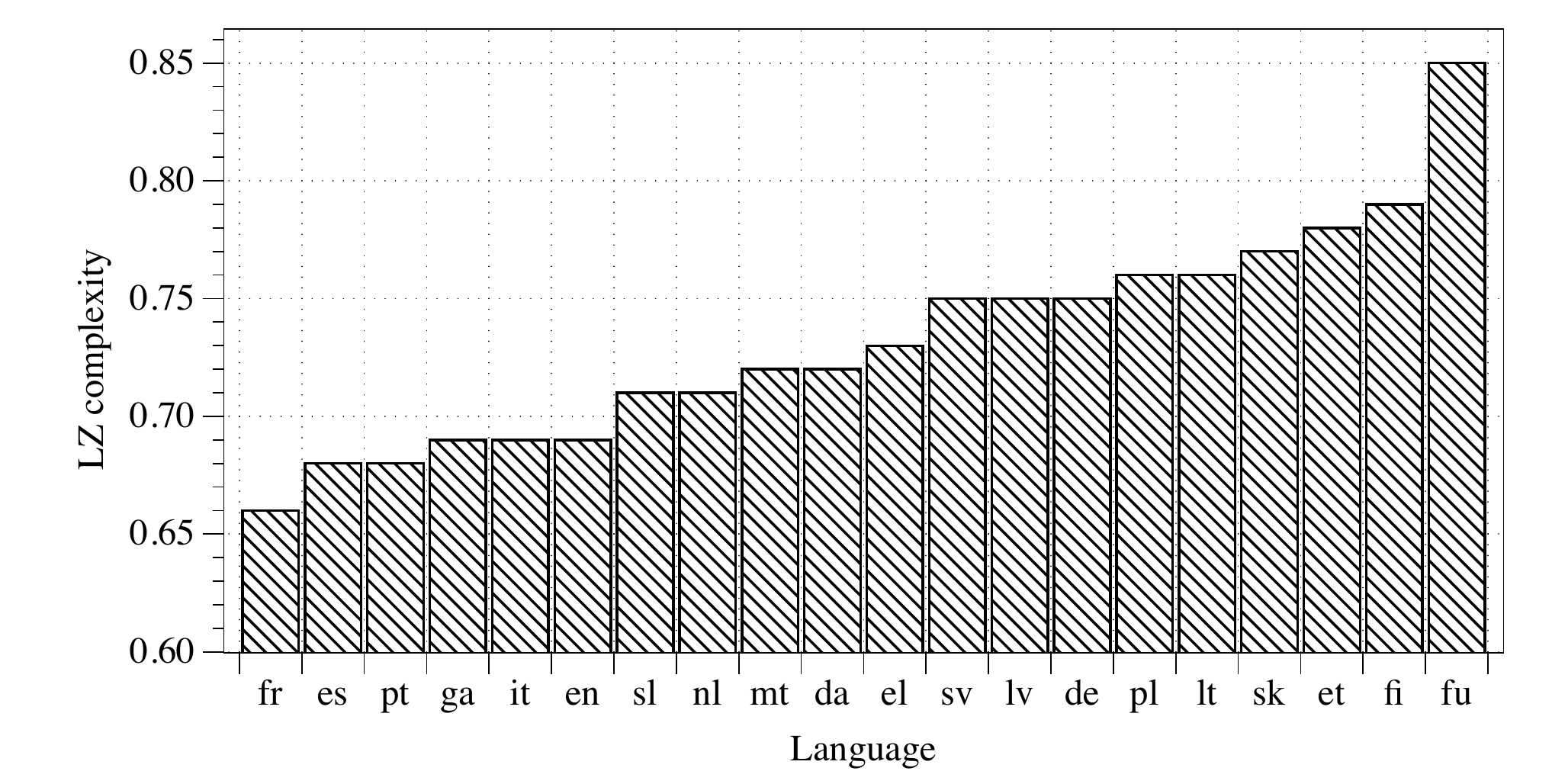}
    \caption{Lempel-Ziv complexity of European languages (data from \cite{sadeniemi2008complexity}) with average complexity of fungal (`fu') electrical activity language added.}
    \label{fig:languagecomplexty}
\end{figure}

LZ complexity of fungal language (Tab.~\ref{tbl:2}) is much higher than of news, random or periodic sequences (Tab.~\ref{tbl:3}). The same can be observed for Shannon entropy. Kolmogorov complexity of the fungal language is much lower than that of news sampler or random or periodic sequences. 
Complexity of European languages based on their compressibility~\cite{sadeniemi2008complexity} is shown in Fig.~\ref{fig:languagecomplexty}, French having lowest LZ complexity 0.66 and Finnish highest LZ complexity 0.79. Fungal language of electrical activity has minimum LZ complexity 0.61 and maximum 0.91 (media 0.85, average 0.83). Thus, we can speculate that a complexity of fungal language is higher than that of human languages (at least for European languages).

\section*{Acknowledgement}
We are grateful to Dr. Anna L. Powell for assisting in collecting experimental data.

This project has received funding from the European Union's Horizon 2020 research and innovation programme FET OPEN ``Challenging current thinking'' under grant agreement No 858132.



\bibliographystyle{plain}
\bibliography{references,geometryFHNbib}

\begin{thebibliography}{10}

\bibitem{ieee2003}
{IEEE} standard on transitions, pulses, and related waveforms.
\newblock {\em IEEE Std 181-2003}, pages 1--60, 2003.

\bibitem{adamatzky2013tactile}
Andrew Adamatzky.
\newblock Tactile bristle sensors made with slime mold.
\newblock {\em IEEE Sensors journal}, 14(2):324--332, 2013.

\bibitem{adamatzky2018spiking}
Andrew Adamatzky.
\newblock On spiking behaviour of oyster fungi {P}leurotus djamor.
\newblock {\em Scientific reports}, 8(1):1--7, 2018.

\bibitem{adamatzky2018towards}
Andrew Adamatzky.
\newblock Towards fungal computer.
\newblock {\em Interface focus}, 8(6):20180029, 2018.

\bibitem{adamatzky2019plant}
Andrew Adamatzky.
\newblock Plant leaf computing.
\newblock {\em Biosystems}, 182:59--64, 2019.

\bibitem{adamatzky2019exploring}
Andrew Adamatzky and Mohammad~Mahdi Dehshibi.
\newblock Exploring tehran with excitable medium.
\newblock In Andrew Adamatzky, Selim Akl, and Georgios~Ch. Sirakoulis, editors,
  {\em From Parallel to Emergent Computing}, chapter~22, pages 475--488. CRC
  Press, 2019.

\bibitem{aidley1998physiology}
David~J Aidley and DJ~Ashley.
\newblock {\em The physiology of excitable cells}, volume~4.
\newblock Cambridge University Press Cambridge, 1998.

\bibitem{belousov1959periodic}
Boris~P {B}elousov.
\newblock A periodic reaction and its mechanism.
\newblock {\em Compilation of Abstracts on Radiation Medicine}, 147(145):1,
  1959.

\bibitem{bingley1966membrane}
MS~Bingley.
\newblock Membrane potentials in amoeba proteus.
\newblock {\em Journal of Experimental Biology}, 45(2):251--267, 1966.

\bibitem{casali2013theoretically}
Adenauer~G Casali, Olivia Gosseries, Mario Rosanova, M{\'e}lanie Boly, Simone
  Sarasso, Karina~R Casali, Silvia Casarotto, Marie-Aur{\'e}lie Bruno, Steven
  Laureys, Giulio Tononi, et~al.
\newblock A theoretically based index of consciousness independent of sensory
  processing and behavior.
\newblock {\em Science translational medicine}, 5(198):198ra105--198ra105,
  2013.

\bibitem{davidenko1992stationary}
Jorge~M Davidenko, Arcady~V Pertsov, Remy Salomonsz, William Baxter, and
  Jos{\'e} Jalife.
\newblock Stationary and drifting spiral waves of excitation in isolated
  cardiac muscle.
\newblock {\em Nature}, 355(6358):349, 1992.

\bibitem{dehshibi2020electrical}
Mohammad~Mahdi Dehshibi and Andrew Adamatzky.
\newblock Supplementary material for ``{E}lectrical activity of fungi: {S}pikes
  detection and complexity analysis".
\newblock \url{https://doi.org/10.5281/zenodo.3997031}, 08 2020.
\newblock (Accessed on 24/08/2020).

\bibitem{deutsch1996zlib}
Peter Deutsch and J~Gailly.
\newblock Zlib compressed data format specification version 3.3.
\newblock Technical report, RFC 1950, May, 1996.

\bibitem{eckert1979ionic}
Roger Eckert and Paul Brehm.
\newblock Ionic mechanisms of excitation in paramecium.
\newblock {\em Annual review of biophysics and bioengineering}, 8(1):353--383,
  1979.

\bibitem{farkas2003human}
I~Farkas, Dirk Helbing, and T~Vicsek.
\newblock Human waves in stadiums.
\newblock {\em Physica A: statistical mechanics and its applications},
  330(1-2):18--24, 2003.

\bibitem{farkas2002social}
Ill{\'e}s Farkas, Dirk Helbing, and Tam{\'a}s Vicsek.
\newblock Social behaviour: Mexican waves in an excitable medium.
\newblock {\em Nature}, 419(6903):131, 2002.

\bibitem{franke2010online}
Felix Franke, Michal Natora, Clemens Boucsein, Matthias~HJ Munk, and Klaus
  Obermayer.
\newblock An online spike detection and spike classification algorithm capable
  of instantaneous resolution of overlapping spikes.
\newblock {\em Journal of computational neuroscience}, 29(1-2):127--148, 2010.

\bibitem{fromm2007electrical}
J{\"o}rg Fromm and Silke Lautner.
\newblock Electrical signals and their physiological significance in plants.
\newblock {\em Plant, cell \& environment}, 30(3):249--257, 2007.

\bibitem{fyhn2004spatial}
Marianne Fyhn, Sturla Molden, Menno~P Witter, Edvard~I Moser, and May-Britt
  Moser.
\newblock Spatial representation in the entorhinal cortex.
\newblock {\em Science}, 305(5688):1258--1264, 2004.

\bibitem{gorbunov1987excitation}
LM~Gorbunov and VI~Kirsanov.
\newblock Excitation of plasma waves by an electromagnetic wave packet.
\newblock {\em Sov. Phys. JETP}, 66(290-294):40, 1987.

\bibitem{gotman1991state}
J~Gotman and LY~Wang.
\newblock State-dependent spike detection: concepts and preliminary results.
\newblock {\em Electroencephalography and clinical Neurophysiology},
  79(1):11--19, 1991.

\bibitem{hall1976optimal}
Charles~A Hall and W~Weston Meyer.
\newblock Optimal error bounds for cubic spline interpolation.
\newblock {\em Journal of Approximation Theory}, 16(2):105--122, 1976.

\bibitem{hansma1979sodium}
Helen~G Hansma.
\newblock Sodium uptake and membrane excitation in paramecium.
\newblock {\em The Journal of cell biology}, 81(2):374--381, 1979.

\bibitem{hodgkin1952quantitative}
Alan~L Hodgkin and Andrew~F Huxley.
\newblock A quantitative description of membrane current and its application to
  conduction and excitation in nerve.
\newblock {\em The Journal of physiology}, 117(4):500--544, 1952.

\bibitem{howard1993design}
Paul~Glor Howard.
\newblock {\em The Design and Analysis of Efficient Lossless Data Compression
  Systems}.
\newblock PhD thesis, Citeseer, 1993.

\bibitem{huffman1952method}
David~A Huffman.
\newblock A method for the construction of minimum-redundancy codes.
\newblock {\em Proceedings of the IRE}, 40(9):1098--1101, 1952.

\bibitem{kaspar1987easily}
F~Kaspar and HG~Schuster.
\newblock Easily calculable measure for the complexity of spatiotemporal
  patterns.
\newblock {\em Physical Review A}, 36(2):842, 1987.

\bibitem{kittel1958excitation}
Ch~Kittel.
\newblock Excitation of spin waves in a ferromagnet by a uniform rf field.
\newblock {\em Physical Review}, 110(6):1295, 1958.

\bibitem{lilly2017element}
Jonathan~M Lilly.
\newblock Element analysis: a wavelet-based method for analysing time-localized
  events in noisy time series.
\newblock {\em Proceedings of the Royal Society A: Mathematical, Physical and
  Engineering Sciences}, 473(2200):20160776, 2017.

\bibitem{lilly2008higher}
Jonathan~M Lilly and Sofia~C Olhede.
\newblock Higher-order properties of analytic wavelets.
\newblock {\em IEEE Transactions on Signal Processing}, 57(1):146--160, 2008.

\bibitem{lilly2012generalized}
Jonathan~M Lilly and Sofia~C Olhede.
\newblock Generalized morse wavelets as a superfamily of analytic wavelets.
\newblock {\em IEEE Transactions on Signal Processing}, 60(11):6036--6041,
  2012.

\bibitem{liu2020robust}
Zuozhi Liu, Xiaotian Wang, and Quan Yuan.
\newblock Robust detection of neural spikes using sparse coding based features.
\newblock {\em Mathematical Biosciences and Engineering}, 17(4):4257, 2020.

\bibitem{marple1999computing}
Lawrence Marple.
\newblock Computing the discrete-time" analytic" signal via fft.
\newblock {\em IEEE Transactions on signal processing}, 47(9):2600--2603, 1999.

\bibitem{masi2015electrical}
Elisa Masi, Marzena Ciszak, Luisa Santopolo, Arcangela Frascella, Luciana
  Giovannetti, Emmanuela Marchi, Carlo Viti, and Stefano Mancuso.
\newblock Electrical spiking in bacterial biofilms.
\newblock {\em Journal of The Royal Society Interface}, 12(102):20141036, 2015.

\bibitem{mcgillviray1987transhyphal}
Ann~M. McGillviray and Neil~A.R. Gow.
\newblock The transhyphal electrical current of \emph{{N}euruspua crassa} is
  carried principally by protons.
\newblock {\em Microbiology}, 133(10):2875--2881, 1987.

\bibitem{nelson2012excitable}
Phillip~G Nelson and Melvyn Lieberman.
\newblock {\em Excitable cells in tissue culture}.
\newblock Springer Science \& Business Media, 2012.

\bibitem{nenadic2004spike}
Zoran Nenadic and Joel~W Burdick.
\newblock Spike detection using the continuous wavelet transform.
\newblock {\em IEEE transactions on Biomedical Engineering}, 52(1):74--87,
  2004.

\bibitem{obeid2004evaluation}
Iyad Obeid and Patrick~D Wolf.
\newblock Evaluation of spike-detection algorithms fora brain-machine interface
  application.
\newblock {\em IEEE Transactions on Biomedical Engineering}, 51(6):905--911,
  2004.

\bibitem{quiroga2004unsupervised}
R~Quian Quiroga, Zoltan Nadasdy, and Yoram Ben-Shaul.
\newblock Unsupervised spike detection and sorting with wavelets and
  superparamagnetic clustering.
\newblock {\em Neural computation}, 16(8):1661--1687, 2004.

\bibitem{quiroga2009explicit}
Rodrigo~Quian Quiroga, Alexander Kraskov, Christof Koch, and Itzhak Fried.
\newblock Explicit encoding of multimodal percepts by single neurons in the
  human brain.
\newblock {\em Current Biology}, 19(15):1308--1313, 2009.

\bibitem{racz2020spike}
Melinda R{\'a}cz, Csaba Liber, Erik N{\'e}meth, Rich{\'a}rd Fi{\'a}th,
  J{\'a}nos Rokai, Istv{\'a}n Harmati, Istv{\'a}n Ulbert, and Gergely
  M{\'a}rton.
\newblock Spike detection and sorting with deep learning.
\newblock {\em Journal of Neural Engineering}, 17(1):016038, 2020.

\bibitem{roelofs1999png}
Greg Roelofs and Richard Koman.
\newblock {\em {PNG}: the definitive guide}.
\newblock O'Reilly \& Associates, Inc., 1999.

\bibitem{ryabko1996using}
Boris Ryabko and Zhanna Reznikova.
\newblock Using shannon entropy and kolmogorov complexity to study the
  communicative system and cognitive capacities in ants.
\newblock {\em Complexity}, 2(2):37--42, 1996.

\bibitem{sablok2020interictal}
Shlok Sablok, Githali Gururaj, Naushaba Shaikh, I~Shiksha, and Antara~Roy
  Choudhary.
\newblock Interictal spike detection in eeg using time series classification.
\newblock In {\em 2020 4th International Conference on Intelligent Computing
  and Control Systems (ICICCS)}, pages 644--647. IEEE, 2020.

\bibitem{sadeniemi2008complexity}
Markus Sadeniemi, Kimmo Kettunen, Tiina Lindh-Knuutila, and Timo Honkela.
\newblock Complexity of european union languages: A comparative approach.
\newblock {\em Journal of Quantitative Linguistics}, 15(2):185--211, 2008.

\bibitem{schartner2017increased}
Michael~M Schartner, Robin~L Carhart-Harris, Adam~B Barrett, Anil~K Seth, and
  Suresh~D Muthukumaraswamy.
\newblock Increased spontaneous meg signal diversity for psychoactive doses of
  ketamine, lsd and psilocybin.
\newblock {\em Scientific reports}, 7:46421, 2017.

\bibitem{shimazaki2010kernel}
Hideaki Shimazaki and Shigeru Shinomoto.
\newblock Kernel bandwidth optimization in spike rate estimation.
\newblock {\em Journal of computational neuroscience}, 29(1-2):171--182, 2010.

\bibitem{slonczewski1999excitation}
JC~Slonczewski.
\newblock Excitation of spin waves by an electric current.
\newblock {\em Journal of Magnetism and Magnetic Materials}, 195(2):L261--L268,
  1999.

\bibitem{taghipour2016complexity}
Nassim Taghipour, Hamid Haj~Seyyed Javadi, Mohammad~Mahdi Dehshibi, and Andrew
  Adamatzky.
\newblock On complexity of persian orthography: L-systems approach.
\newblock {\em Complex Systems}, 25(2):127--156, 2016.

\bibitem{trainito2019extracellular}
Caterina Trainito, Constantin von Nicolai, Earl~K Miller, and Markus Siegel.
\newblock Extracellular spike waveform dissociates four functionally distinct
  cell classes in primate cortex.
\newblock {\em Current Biology}, 29(18):2973--2982, 2019.

\bibitem{trebacz2006electrical}
Kazimierz Trebacz, Halina Dziubinska, and Elzbieta Krol.
\newblock Electrical signals in long-distance communication in plants.
\newblock In {\em Communication in plants}, pages 277--290. Springer, 2006.

\bibitem{tsoi1998excitation}
M~Tsoi, AGM Jansen, J~Bass, W-C Chiang, M~Seck, V~Tsoi, and P~Wyder.
\newblock Excitation of a magnetic multilayer by an electric current.
\newblock {\em Physical Review Letters}, 80(19):4281, 1998.

\bibitem{wang2020novel}
Zimeng Wang, Duanpo Wu, Fang Dong, Jiuwen Cao, Tiejia Jiang, and Junbiao Liu.
\newblock A novel spike detection algorithm based on multi-channel of bect eeg
  signals.
\newblock {\em IEEE Transactions on Circuits and Systems II: Express Briefs},
  2020.

\bibitem{wilson2002spike}
Scott~B Wilson and Ronald Emerson.
\newblock Spike detection: a review and comparison of algorithms.
\newblock {\em Clinical Neurophysiology}, 113(12):1873--1881, 2002.

\bibitem{wilson1999spike}
Scott~B Wilson, Christine~A Turner, Ronald~G Emerson, and Mark~L Scheuer.
\newblock Spike detection ii: automatic, perception-based detection and
  clustering.
\newblock {\em Clinical neurophysiology}, 110(3):404--411, 1999.

\bibitem{zhabotinsky1964periodic}
AM~{Z}habotinsky.
\newblock Periodic processes of malonic acid oxidation in a liquid phase.
\newblock {\em Biofizika}, 9(306-311):11, 1964.

\bibitem{zhabotinsky2007belousov}
Anatol~M Zhabotinsky.
\newblock Belousov-zhabotinsky reaction.
\newblock {\em Scholarpedia}, 2(9):1435, 2007.

\bibitem{zimmermann2013electrical}
Matthias~R Zimmermann and Axel Mith{\"o}fer.
\newblock Electrical long-distance signaling in plants.
\newblock In {\em Long-Distance Systemic Signaling and Communication in
  Plants}, pages 291--308. Springer, 2013.

\bibitem{ziv1977universal}
Jacob Ziv and Abraham Lempel.
\newblock A universal algorithm for sequential data compression.
\newblock {\em IEEE Transactions on information theory}, 23(3):337--343, 1977.

\end{thebibliography}
\end{document}